\documentclass[12pt]{article}
\usepackage{amssymb}
\usepackage{amsfonts}
\usepackage{amsmath}
\usepackage[mathscr]{eucal}
\usepackage{amssymb}
\usepackage{amsthm}
\usepackage{bbold}
\usepackage{bm}
\usepackage{graphicx}
\usepackage{caption}

\numberwithin{equation}{section}
\newcommand{\be}{\begin{equation}}
\newcommand{\ee}{\end{equation}}
\newcommand{\bea}{\begin{eqnarray}}
\newcommand{\eea}{\end{eqnarray}}
\newcommand{\nn}{\nonumber \\}
\newcommand{\lb}{\label}
\newcommand{\p}[1]{(\ref{#1})}

\usepackage{MnSymbol}

\newcounter{rown}

\begin{document}

\begin{titlepage}

\vspace*{0.2cm}

\renewcommand{\thefootnote}{\star}
\begin{center}

{\LARGE\bf  From $\mathcal{N}{=}\,4$ Galilean superparticle }\\

\vspace{0.5cm}

{\LARGE\bf to three-dimensional }\\

\vspace{0.5cm}

{\LARGE\bf non-relativistic $\mathcal{N}{=}\,4$ superfields}\\

\vspace{1.5cm}

{\large\bf Sergey\,Fedoruk}${}^{a}$,\,\,\,{\large\bf Evgeny\,Ivanov}${}^a$,\,\,\,{\large\bf Jerzy\,\,Lukierski}${}^b$

\vspace{1cm}

${}^a${\it Bogoliubov Laboratory of Theoretical Physics, JINR,  Dubna, Moscow region, Russia} \\
\vspace{0.2cm}
{\tt fedoruk@theor.jinr.ru, \, eivanov@theor.jinr.ru }\\

\vspace{0.6cm}

${}^b${\it Institute for Theoretical Physics,  University of Wroc{\l}aw, Poland} \\
\vspace{0.2cm}
{\tt jerzy.lukierski@ift.uni.wroc.pl}
\vspace{0.7cm}

\end{center}


\vspace{1.2cm}

\begin{abstract}
\noindent
We consider the general $\mathcal{N}{=}\,4,$ $d{=}\,3$ Galilean superalgebra with arbitrary central charges
and study its dynamical realizations. Using the nonlinear realization techniques,
we introduce a class of actions
for $\mathcal{N}{=}\,4$ three-dimensional non-relativistic superparticle, such that
they are linear in the central charge Maurer-Cartan one-forms.
As a prerequisite to the quantization, we analyze the phase space constraints structure of our model for various choices
of the central charges.
The first class constraints generate gauge transformations, involving fermionic $\kappa$-gauge transformations.
The quantization of the model gives rise to the collection of free $\mathcal{N}{=}\,4$, $d{=}\,3$ Galilean superfields,
 which can be further employed, e.g.,  for description of three-dimensional non-relativistic
 $\mathcal{N}{=}\,4$ supersymmetric theories.

\end{abstract}

\vspace{1.5cm}

\phantom{Keywords: }  superparticles, quantization methods

\newpage

\end{titlepage}

\setcounter{equation}{0}
\section{Introduction}

\quad\, In recent years, one can observe the growth of interest in the non-relativistic (NR)
field-theoretic models, in particular those describing NR gravity and NR supergravity, e.g., in the framework
of the so-called Newton-Cartan geometry \cite{DuHor,AndBRS1,AndBRS2,fli10,BerRos}.
Until present, the NR supersymmetric  framework \cite{AndBRS2,fli10,BerRos}
has been basically developed for $D=2{+}1$-dimensional case,\footnote{We denote by $D$ the number of space and time dimensions,
and by $d$ the number of space dimensions, i.e. $D= d+1$.} which corresponds to the exotic version of
Galilean symmetry with two central charges \cite{Levy,Grig,LukSZ,JacNa,fli16}. In this paper we address the next
 physically interesting case of  $\mathcal{N}{=}\,4,$ $D=3{+}1$  supersymmetric extension of Galilean
symmetries. Due to the distinguished role of $\mathcal{N}{=}\,4$ supersymmetric Yang-Mills theory
(see e.g. \cite{AharGMOO}), this kind of extended supersymmetry merits as well an attention in the NR case.

Similarly as in the relativistic case, one can study the NR  $\mathcal{N}{=}\,4,$ $D=3{+}1$ supersymmetric theories by following several paths:
\begin{description}
\item[i)]
One can start with NR  $\mathcal{N}{=}\,4,$ $d=3$ Galilean superalgebras (for arbitrary $\mathcal{N}$ see \cite{fli8})
 and then construct their superspace and superfield realizations.
In this way we obtain the universal tool for constructing NR supersymmetric field theories.
\item[ii)] The NR field theories can be reproduced  by performing the non-relativistic contraction
$c\to\infty$ ($c$ is the speed of  light) in the known relativistic non-supersymmetric, as well as supersymmetric field
theory  models (see, e.g., \cite{fli10,JenK,BerRT}).
One of the advantages of such a method is the possibility to derive the proper NR
contractions of relativistic action integrals.
\item[iii)]
For a definite type of (super)symmetric framework one can consider the dynamics associated with particles,
fields, string, $p$-branes, etc. An important role in such a list is played by the free classical and first-quantized
(super)particle models, with the property that their first quantization leads to the classical (super)field realizations (see, e.g., \cite{Cas,Schw}).
In our case, we will look for the free superparticle models invariant under $\mathcal{N}{=}\,4,$ $d=3$ Galilean supersymmetry.
One can mention that in the relativistic case this way of deriving free superfields  from the classical and first-quantized
superparticles with extended $\mathcal{N}{=}\,4$, $D=4$ Poincar\'e supersymmetry  was already proposed in \cite{Fryd}.\footnote{
Here we will deal only with those Galilean supersymmetries which, after quantization, leads to
the superfields depending on spinorial Grassmann variables. One could introduce as well
relativistic \cite{BarCasLus,BerMar,HowePPT} and Galilean \cite{BarCasGom,DuHor94}
world-line supersymmetries with vectorial Grassmann variables. However, this option is out of the scope of our study.
}
\end{description}

In this paper we will follow the path {\bf iii)}. We will consider the most general
NR $\mathcal{N}{=}\,4,$ $d{=}\,3$ Galilean superalgebra,
introduce the corresponding $\mathcal{N}{=}\,4$ Galilean supergroup and its cosets, construct the relevant nonlinear realizations and use
the associated Maurer-Cartan (MC) one-forms   to build NR $\mathcal{N}{=}\,4$ superparticle models. They will be subsequently quantized
to obtain the NR superfields providing realizations of $\mathcal{N}{=}\,4$ Galilean supersymmetries. Note that in such a setting
the original coset parameters are treated as the $D{=}1$ world-line fields. However, the whole formalism could be equally applied along
the lines of path  {\bf i)}, with the coset parameters treated as independent NR superspace coordinates.

As a prelude to our considerations, we will describe the Galilean symmetries and their supersymmetrizations
in a short historical survey.

The Galilean theories describe the low energy, non-relativistic dynamical systems
\footnote{The Galilean symmetries are used also in the description of light cone quantization of relativistic
theories \cite{fli1}-\cite{fli3}.
In this paper, we shall not deal  with this application of Galilean symmetries.},
which can also be obtained as non-relativistic limit ($c\to \infty$) of the corresponding relativistic theories
(see, e.g., \cite{fli4}--\cite{fli9}).
Such a contraction limit,  applied to $D=4$ Poincar\'{e} algebra
($P_\mu, M_{\mu\nu}; P_\mu=(P_0,\,{P}_i$),  $M_{\mu\nu}=(M_i =\frac{1}{2} \varepsilon_{ijk} M_{jk},  N_i= M_{i0})$),  after shifting
and rescaling
\begin{equation}\label{eqfli1.1}
P_0 = m_0 c + \frac{\mathbf{H}}{c}\,,\qquad P_i=\mathbf{P}_i\,, \qquad \qquad N_i = c\, \mathbf{B}_i \,,\qquad M_i=\mathbf{J}_i\,,
\end{equation}
where $\mathbf{H}$ stands  for non-relativistic Hamiltonian and $\mathbf{B}_i$ for the Galilean boosts,
yields  ``quantum'' $d=3$ Galilean
algebra  \cite{fli11}\footnote{The Galilean algebra  (\ref{eqfli1.2}) with central charge $\mathbf{M}\neq 0$ is called Bargmann algebra.
In $d=2$ one can introduce also second central charge through the modified commutator $[B_i,B_j]=\epsilon_{ij}\rho$ \cite{Levy,Grig,LukSZ,JacNa}.}
\begin{equation}\label{eqfli1.2}
\begin{array}{c}
\left[ \mathbf{J}_i, \mathbf{J}_j \right] = i\varepsilon_{ijk} \mathbf{J}_k\,,
 \\  \\
\left[\mathbf{J}_i, \mathbf{P}_j \right] = i\varepsilon_{ijk} \mathbf{P}_k\,, \qquad
\left[ \mathbf{J}_i, \mathbf{B}_j \right] = i\varepsilon_{ijk} \mathbf{B}_k\,,
\\ \\
\left[ \mathbf{H}, \mathbf{B}_i \right] = -i\mathbf{P}_i\,, \qquad \left[ \mathbf{P}_i, \mathbf{B}_j \right] = -i\mathbf{M} \, \delta_{ij}\,.
\end{array}
\end{equation}
The central charge $\mathbf{M}=m_0$ describes a non-relativistic mass which can be identified with the relativistic rest mass.

Because bosons and fermions occur in both relativistic and non-relativistic settings,
one can consider the non-relativistic supersymmetry as well.
The first proposal for supersymmetrization of Galilei algebra (\ref{eqfli1.2}) was given in \cite{fli12},
where $\mathcal{N}{=}\,1$ and $\mathcal{N}{=}\,2,$ $d{=}\,3$ Galilean superalgebras were presented.
The $\mathcal{N}{=}\,1,$ $d{=}\,3$ Galilean superalgebra is an extension of relations (\ref{eqfli1.2})
by complex NR  $\mathrm{USp}(2)\simeq \mathrm{SU}(2)$ supercharges
$\mathbf{S}_\alpha, \bar{\mathbf{S}}^\alpha := (\mathbf{S}_\alpha)^\dagger$ ($A \to A^\dagger$ denotes
Hermitian conjugation) which satisfy the relations\footnote{We list only non-vanishing (anti)commutators.}
\begin{equation}\label{N1superG}
\{ \mathbf{S}_\alpha, \bar{\mathbf{S}}^\beta \} = \mathbf{M} \, \delta_{\alpha}^{\beta} \,,
\qquad
[ \mathbf{J}_i, \mathbf{S}_\alpha ] = (\sigma_i)_\alpha{}^{\beta} \mathbf{S}_\beta\,,
\qquad
[ \mathbf{J}_i, \bar{\mathbf{S}}^\alpha ] = -\bar{\mathbf{S}}^\beta(\sigma_i)_\beta{}^{\alpha} \,.
\end{equation}
Passing to the $\mathcal{N}{=}\,2$  $d{=}\,3$  Galilean superalgebra \cite{fli12}
is accomplished by adding to the $\mathcal{N}{=}\,1$ Galilean superalgebra generators
$(\mathbf{J}_i, \mathbf{P}_i, \mathbf{B}_i, \mathbf{H}; \mathbf{S}_\alpha, \bar{\mathbf{S}}^\alpha, \mathbf{M})$
the second pair of complex $\mathrm{SU}(2)$ supercharges $\mathbf{Q}_\alpha, \bar{\mathbf{Q}}^\alpha := (\mathbf{Q}_\alpha)^\dagger$,
subject to the following relations:
\begin{equation}\label{eqfli1.3}
\begin{array}{l}
\{ \mathbf{Q}_\alpha, \bar{\mathbf{Q}}^\beta \} = 2\delta_\alpha^{\  \beta} \mathbf{H}\,,
\\ [7pt]
\{ \mathbf{Q}_\alpha, \bar{\mathbf{S}}^\beta \} = 2(\sigma_i )_{\alpha}{}^{\beta}\mathbf{P}_i + 2i {\mathbf{Y}} \delta_{\alpha}^{\ \beta}\,,
\\  [7pt]
[\mathbf{J}_i, \mathbf{Q}_\alpha ]= (\sigma_i)_\alpha{}^\beta \mathbf{Q}_\beta \,, \qquad
[ \mathbf{B}_i, \mathbf{Q}_\alpha ] = (\sigma_i)_\alpha^{\ \beta} \mathbf{S}_\beta \,, \\  [7pt]
[ \mathbf{J}_i, \bar{\mathbf{Q}}^\alpha ] = -\bar{\mathbf{Q}}^\beta(\sigma_i)_\beta{}^{\alpha}\,,\qquad
[ \mathbf{B}_i, \bar{\mathbf{Q}}^\alpha ] = -\bar{\mathbf{S}}^\beta(\sigma_i)_\beta{}^{\alpha}\,.
\end{array}
\end{equation}

In the relations (\ref{eqfli1.3}), (\ref{N1superG}),  besides the central charge $\mathbf{M}$, there appears the new
central charge  $\mathbf{Y}$. The $\mathcal{N}{=}\,2,$  $d{=}\,3$  Galilean superalgebra can be derived
from $\mathcal{N}{=}\,2$, $D{=}\,4$ Poincar\'{e} superalgebra ($a=1,2$)
\begin{equation}\label{N2salg}
\begin{array}{l}
\{ {Q}^a_\alpha, \bar{{Q}}_{\dot\beta b} \} = 2(\sigma^\mu)_{\alpha\dot\beta}P_\mu\,\delta_b^a\,,
\\ [7pt]
\{ {Q}^a_\alpha, {Q}^b_\beta \} = 2\epsilon^{ab}\epsilon_{\alpha\beta}\,Z\,,
\\  [7pt]
\{ \bar{{Q}}_{\dot\alpha a}, \bar{{Q}}_{\dot\beta b} \} = -2\epsilon_{ab}\epsilon_{\dot\alpha\dot\beta}\,\bar Z
\end{array}
\end{equation}
(plus the commutation relations with Poincar\'e and internal R-symmetry $\mathrm{U}(2)$ generators)
by taking the $c \to \infty$ contraction limit with $\mathbf{M}=m_0\,$.
In general, the $\mathcal{N}{=}\,2,$  $D{=}\,4$ Poincar\'{e} superalgebra
is endowed with one complex central charge $Z$ or, equivalently, two real central charges,  $Z=X+iY\,$.
\footnote{Since $\mathcal{N}{=}\,2$ $D{=}\,4$ Poincar\'{e} superalgebra is covariant under the phase transformation of
Weyl supercharges $(Q^r_\alpha, \bar{Q}^r_{\dot{\alpha}}; r=1,2)$
$$
Q_\alpha^{r} \to e ^{\frac{i}{2} \alpha}Q_\alpha^{r}
 \qquad Z  \to e^{i\alpha} Z \qquad (Z= e^{-i\alpha} |z|)
$$
one could think that  one real central charge is enough in ${\cal N}=2$ case.
However, as was found  by studying concrete dynamical models
\cite{fli13,fli14}, it is the complex ${\cal N}=2$ central charge $Z= X_1 +i X_2$ what actually matters. It
amounts to two physical real central charges: the topological magnetic charge $X_1$ and the non-topological electric charge $X_2$. Only
if these charges take constant eigenvalues, i.e. are numerical, they can be rotated to the single central charge by the phase transformations just mentioned.}.
Before taking the NR limit $c \to \infty$, the Galilean supercharges $\mathbf{Q}_\alpha$ and $\mathbf{S}^\alpha$
(see (\ref{N1superG}), (\ref{eqfli1.3})) should be identified with the following linear combinations
of two $\mathcal{N}{=}\,2$ Weyl supercharges in (\ref{N2salg})
\begin{equation}\label{N2QS}
\mathbf{Q}_\alpha=c^{1/2}Q^+_\alpha\,,\quad
\bar{\mathbf{Q}}^\alpha=c^{1/2}\bar Q^+_{\dot\alpha}\,,\qquad
\mathbf{S}_\alpha=c^{-1/2}Q^-_\alpha\,,\quad
\bar{\mathbf{S}}_\alpha=c^{-1/2}\bar Q^-_{\dot\alpha}\,,
\end{equation}
where
\begin{equation}\label{N2Q+-}
Q^\pm_\alpha=\frac{1}{\sqrt{2}}\left(Q^1_\alpha \mp \epsilon_{\alpha\beta}\bar Q_{\dot\beta\,2}\right)\,,\qquad
\bar Q^\pm_{\dot\alpha}=\frac{1}{\sqrt{2}}\left(\bar Q_{\dot\alpha\,1} \mp \epsilon_{\dot\alpha\dot\beta}Q^2_{\beta}\right)
\end{equation}
and $\bar Q^\pm_{\dot\alpha}=(Q^\pm_\alpha)^\dagger$.
Also, we should postulate the following $c$-dependence of the central charges in (\ref{N2salg})
\begin{equation}\label{XY-lim}
X=-m_0 c + \frac{\mathbf{X}}{c} + \mathcal{O}\left(\frac{1}{c^2}\right)\,,\qquad
Y=\mathbf{Y} + \mathcal{O}\left(\frac{1}{c}\right)\,.
\end{equation}
If $\mathbf{X}$ is finite in the contraction limit,   it merely generates
the shift $\mathbf{H}\to \mathbf{H}+\mathbf{X}$ in the relations of the
$\mathcal{N}{=}\,2,$ $d{=}\,3$ Galilean  superalgebra (see the first relation in (\ref{eqfli1.3})).

The Galilean ${\cal N}=2$ superalgebra and its dynamical realizations were studied in several papers,
but mostly for the case of two ($d{=}\,2$) space dimensions \cite{fli15,fli16,AndBRS1,AndBRS2,fli10,fli17}.

In the present paper we consider $\mathcal{N}{=}\,4,$ $d{=}\,3$ Galilean superalgebra  with all possible central charges.
It will be obtained by the $c\to \infty$ contraction procedure from
the general $\mathcal{N}{=}\,4,$ $D{=}\,4$ relativistic Poincar\'{e} superalgebra \cite{fli18} which involves 6
complex central charges $Z^{AB}=-Z^{BA}$ ($A,B=1,2,3,4$). Correspondingly,  the  $\mathcal{N}{=}\,4,$ $d{=}\,3$
Galilean superalgebra obtained in the $c \to \infty$ limit involves 12 real central
charges\footnote{In fact, the NR $\mathcal{N}{=}\,4$ Galilean superalgebra involves 13 central charges if we take into account
the Bargmann central charge $\mathbf{M}=m_0$ obtained from the leading terms in the asymptotic expansion of $P_0$ and $X$ in $c$
(see (\ref{eqfli1.1}) and (\ref{XY-lim})).}.
If these central charges are numerical, then, using a suitable redefinition of supercharges by an unitary $4\times 4$ matrix,
one can cast the antisymmetric $4{\times}4$ complex matrix of six central charges $Z^{AB} = - Z^{BA}$ ($A,B = 1,2,3,4$)
into a quasi-diagonal Jordanian form \cite{fli19,fli20}
\begin{equation}\label{eqfli1.5}
Z^{AB} = \begin{pmatrix}
\begin{matrix}
0 & -Z_1
\\
Z_1 & 0
\end{matrix} & 0
\\
0
&
\begin{matrix}
0 & -Z_2
\\
Z_2 & 0
\end{matrix}
\end{pmatrix} \quad  \longleftrightarrow  \quad
Z^{AB} = {\rm diag} (\epsilon Z_1 , \epsilon Z_2),
\end{equation}
where $\left(\epsilon^{ab}\right) = \left(
\begin{matrix}
0 & -1
\\
1 & 0
\end{matrix}\right)=-i\sigma_2$ is antisymmetric matrix\footnote{The transformation
of the general case with 6 complex central charges $Z^{AB}$ to the case with central charges given by matrix (\ref{eqfli1.5})
is straightforward  if
$Z^{AB}$ encompass constant  mass-like  parameters. If $Z^{AB}$ are operators, the map (\ref{eqfli1.5}) is valid
only if  the real operators $\hat{\alpha}_{AB}$ and $| Z_{A', B'}|$ defined as $Z_{AB} = \exp (i \hat{\alpha}_{AB}) |Z_{AB}|$
mutually commute.}.
The choice (\ref{eqfli1.5}) breaks $\mathrm{U}(4)$ internal symmetry of $\mathcal{N}{=}\,4,$ $D{=}\,4$ Poincar\'{e} superalgebra
down to $\mathrm{USp} (2)\otimes \mathrm{USp}(2) \simeq \mathrm{SU}(2) \otimes \mathrm{SU}(2) \simeq \mathrm{O}(4)$ \footnote{Modulo chiral $\mathrm{U}(1)$, see below.};
if we further put $Z=Z_1 = Z_2$ we obtain $\mathcal{N}{=}\,4,$ $D{=}\,4$ Poincar\'{e} superalgebra with one complex central charge
and unbroken $\mathrm{USp}(4) \simeq \mathrm{O}(5) $ internal symmetry.

Such a structure of the internal sectors survives in the non-relativistic limit; one can therefore consider
 $\mathcal{N}{=}\,4,$ $d{=}\,3$ Galilean supersymmetric theories with the internal sectors $\mathrm{USp}(2)\otimes \mathrm{USp}(2)$
(four real Galilean central charges) or $\mathrm{USp}(4)$ (a pair of real Galilean central charges).\footnote{Further
we denote these two non-relativistic superalgebras as $S\hat{g}(3;4|4)$ and $S\hat{g}(3;4|2)$, where
$S\hat{g}(d;N|n)$ stands for N-extended d-dimensional Galilean superalgebra with $n$ real central charges.
The corresponding supergroups will be denoted $\mathrm{SG}(d; N|n)$.}

In the most general $\mathcal{N}{=}\,4$ case, when we deal with six complex central charges,
the central charge $4{\times}4$ matrix can be written as follows
\begin{equation}\label{Z-matr}
Z^{AB}=\left(
\begin{array}{cc}
Z^{ab} & Z^{a\tilde b} \\
Z^{\tilde a b} & Z^{\tilde a\tilde b} \\
\end{array}
\right),\qquad Z^{ab}=\epsilon^{ab}\,Z_1\,,\qquad  Z^{\tilde a\tilde b}=\epsilon^{\tilde a\tilde b}\,Z_2\,,
\qquad Z^{\tilde a b}=-Z^{b\tilde a}
\,,
\end{equation}
where $a,b=1,2$ ($\tilde a,\tilde b=1,2$) are the left (right) $\mathrm{USp}(2)\simeq\mathrm{SU}(2)$
spinor indices. The four complex central charges $Z^{a \tilde b}$ constitute complex $\mathrm{O}(4,\mathbb{C})$
isovector $Z_M= \frac{1}{2i}\,(\sigma_M)_{\tilde a b}Z^{b \tilde a}$, where
$(\sigma_M)_{\tilde a b}$ are $D=4$ Euclidean Pauli matrices $\sigma_M=(\sigma_i,i\mathbb{1}_2)$.
If $Z_M=0$ ({\it i.e.}, the central charge matrix is reduced to \p{Z-matr}) we deal with the decomposition of $\mathcal{N}{=}\,4$ Galilean superalgebra into the direct sum of two
$\mathcal{N}{=}\,2$ Galilean superalgebras, each possessing $\mathrm{USp}(2)$ automorphism;
if $Z_M\neq 0$ the decomposition of $\mathcal{N}{=}\,4$ Galilean supersymmetry into such a sum
of two $\mathcal{N}{=}\,2$ superalgebras is not possible.
As we will see, in the absence of central charges
the full compact internal R-symmetry in the NR case is $\mathrm{U}(1) \otimes \mathrm{USp}(4)$
as opposed to  $\mathrm{U}(4)$ of the relativistic ${\cal N}=4, D=4$ superalgebra. If the central charges take numerical values, the presence
of off-diagonal supercharges (\ref{Z-matr}) provides the breaking of $\mathrm{USp}(2){\otimes}\mathrm{USp}(2)\simeq\mathrm{O}(4)
\subset \mathrm{USp}(4)$ internal symmetry (still preserved by the diagonal central charges)
down to the exact $\mathrm{O}(3)$ or $\mathrm{O}(2)$ internal symmetries which form diagonal subgroups in the product
$\mathrm{O}(3){\otimes}\mathrm{O}(3)=\mathrm{O}(4)$.

The central charges, besides bringing in the mass parameters, are also capable to simplify the formulation of
$\mathcal{N}\geqslant 2$ supersymmetric gauge theories.
In particular, recall that $\mathcal{N}{=}\,4,$ $D{=}\,4$ Yang-Mills theory with one central
charge  and internal symmetry broken to $\mathrm{O}(5)$, contrary to $\mathcal{N}{=}\,4,$ $D{=}\,4$ supersymmetric Yang-Mills theory
with $\mathrm{SU}(4)$  R-symmetry and without central charges, permits an off-shell superspace formulation which does not
require harmonic variables \cite{fli21,fli22}.

The plan of the paper is as follows. In Sect.~2, following \cite{fli8},  we derive the general $\mathcal{N}{=}\,4$, $d{=}\,3$
Galilean superalgebra, which contains 12 independent real central charges and the additional thirteenth Bargmann central charge
describing the rest mass.
As in \cite{fli4,fli5,fli6,fli7}, in this derivation we employ the NR contraction
$c \to \infty$ of  relativistic $\mathcal{N}{=}\,4,$ $D{=}\,4$ Poincar\'{e} superalgebra.
In Sect.~3 we calculate the MC one-forms on the coset ${\mathcal{G}}/{\mathcal{H}}$,
where ${\mathcal{G}} = \mathrm{SG}(3;4|12)$ (see footnote 9) and stability subgroup $\mathcal{H}$ is
given by $\mathrm{SU}(2)\simeq \mathrm{O}(3)$  and $\mathrm{USp}(4)$ generators.
In Sect.~4 we study the $\mathcal{G}$-invariant actions linear in MC one-forms associated with central charges.
For different choices of the central charges these actions describe various models of $\mathcal{N}{=}\,4,$ $d{=}\,3$
Galilean superparticles. We consider the  phase superspace formulation of these superparticle models and
present complete set of
first and second class constraints. The first class fermionic constraints generate the non-relativistic
$\mathcal{N}{=}\,4$ $\kappa$-gauge transformations which act in the non-physical part of the Grassmann coordinate sector.
In Sect.~5 we quantize the model.
Using super Schr\"{o}dinger  realization of quantum phase superspace  algebra, we obtain as
the quantum solutions of the model a set of free $\mathcal{N}{=}\,4,$ $d{=}\,3$ Galilean superfields.
In Sect.~6 we present an outlook, in particular, we describe briefly the alternative ways of constructing the $\mathcal{N}{=}\,4$ Galilean
superparticle models.
Concluding, we hope that our paper will contribute to the issue of superfield description of the interacting non-relativistic
$\mathcal{N}{=}\,4,$ $d{=}\,3$  supersymmetric field theories.\footnote{For examples of supersymmetric extensions of QED and
Yang-Mills Galilean theories see \cite{fli23}-\cite{fli25}.}

\section{General Galilean $\mathcal{N}{=}\,4,$ $d{=}\,3$ superalgebra with central charges}

\quad\, The $\mathcal{N}{=}\,4,$ $D{=}\,4$ Poincar\'{e} superalgebra is spanned by the following generators
\footnote{
We define $D=4$ sigma-matrices as follows: $(\sigma^\mu)_{\alpha\dot\alpha}=({1}_2,\vec{\sigma})_{\alpha\dot\alpha}$,
$(\tilde\sigma^\mu)^{\dot\alpha\alpha}=\epsilon^{\alpha\beta}\epsilon^{\dot\alpha\dot\beta}(\sigma^\mu)_{\beta\dot\beta}=
({1}_2,-\vec{\sigma})^{\dot\alpha\alpha}$, $\sigma^{\mu\nu}=i\,\sigma^{[\mu}\tilde\sigma^{\nu]}$,
$\tilde\sigma^{\mu\nu}=i\,\tilde\sigma^{[\mu}\sigma^{\nu]}$. Always in this paper we use weight coefficient in (anti)symmetrization:
$A_{(\mu}B_{\nu)}=\frac12\,(A_{\mu}B_{\nu}+A_{\nu}B_{\mu})$, $A_{[\mu}B_{\nu]}=\frac12\,(A_{\mu}B_{\nu}-A_{\nu}B_{\mu})$.
The $D=4$ metric tensor is taken as $\eta_{\mu\nu}={\rm diag}(+1,-1,-1,-1)$. }
\begin{description}
\item[i)] $4$ two-component complex Weyl supercharges
$Q_{\alpha}^{\,A}$, $\bar{Q}_{\dot{\alpha}\, A} =(Q_{\alpha}^{\,A})^\dagger$ ($\alpha=1,2$, $A=1,...,4$);
\item[ii)] Poincar\'{e} algebra generators $P_\mu=\left( P_0,P_i\right)$,
$M_{\mu\nu}=\left( M_i=\frac12\,\epsilon_{ijk}M_{jk}, N_i=M_{i0}\right)$, $i=1,2,3$;
\item[iii)] 16 internal Hermitian  $\mathrm{U}(4)$ R-symmetry generators $G^{A}{}_{B}$, $(G^{A}{}_{B})^\dagger = G^{B}{}_{A}$,
where antisymmetric 6 generators $G_{(a)}^{A}{}_{B} = \frac12\left(G^{A}{}_{B}-G^{B}{}_{A}\right)$
are $\mathrm{O}(4)$ generators and 10 symmetric ones $G_{(s)}^{A}{}_{B} = \frac12\left(G^{A}{}_{B}+G^{B}{}_{A}\right)$
describe the coset $\mathrm{U}(4)/\mathrm{O}(4)$. The axial $\mathrm{U}(1)$ generator $A=G_{(s)}^{A}{}_{A}$
can be separated out, i.e. $\mathrm{U}(4)=\mathrm{SU}(4)\otimes \mathrm{U}(1)$,
where $\mathrm{SU}(4)$ generators $T^{A}_{B}=G^{A}{}_{B}-\frac14\,\delta^{A}{}_{B} A$ are traceless, $T^{A}_{A}=0$,
and satisfy the relation
\begin{equation}\label{suN}
[T^{A}_{B},T^{C}_{D}]=\delta^{C}_{B}\,T^{A}_{D} - \delta^{A}_{D}\,T^{C}_{B}  \,;
\end{equation}
\item[iv)] The set of 6 complex central charges $Z^{AB}=-Z^{BA}$, $\bar Z_{AB} =(Z^{AB})^\dagger$,
or equivalently the set of 12 real central charges $\mathcal{X}^{AB}=-\mathcal{X}^{BA}$, $\mathcal{Y}^{AB}=-\mathcal{Y}^{BA}$, where
\begin{equation}\label{Z-XY}
Z^{AB}=\mathcal{X}^{AB}+i\,\mathcal{Y}^{AB} \,.
\end{equation}
Each of the real central charges $\mathcal{X}^{AB}$ and $\mathcal{Y}^{AB}$ transform in the adjoint of
$\mathrm{O}(4)\propto G_{(a)}^{A}{}_{B}\,$. The $\mathrm{U}(4)/\mathrm{O}(4)$ coset generators properly mix
$\mathcal{X}^{AB}$ and $\mathcal{Y}^{AB}\,.$
\end{description}

The general $\mathcal{N}{=}\,4,$ $D{=}\,4$ super Poincar\'{e} anticommutation relations are
\begin{equation}\label{QbQ-224}
\{Q_{\alpha}^{\,A}, \bar{Q}_{\dot{\beta} B} \} =
2(\sigma^\mu)_{\alpha \dot{\beta}} \, P_\mu  \,
\delta_{B}^{\, A}\,,
\end{equation}
\begin{equation}\label{QQ-224}
\begin{array}{l}
\{Q_{\alpha}^{\,A}, {Q}_{\beta}^{\,B} \}=2\epsilon_{\alpha\beta}Z^{AB}=2\epsilon_{\alpha\beta}\left(\mathcal{X}^{AB}+i\,\mathcal{Y}^{AB}\right),\\ [6pt]
\{\bar{Q}_{\dot{\alpha} A}, \bar{Q}_{\dot{\beta} B} \}=2\epsilon_{\dot\alpha\dot\beta}\bar Z_{AB}=
2\epsilon_{\dot\alpha\dot\beta}\left(\mathcal{X}^{AB}-i\,\mathcal{Y}^{AB}\right) \,,
\end{array}
\end{equation}
and the remaining non-zero commutation relations read
\begin{equation}\label{QM-224}
[M_{\mu\nu}, Q_{\alpha}^{\,A}] = - \frac{1}{2}\, (\sigma_{\mu\nu})_{\alpha}{}^{\beta} Q_{\beta}^{\,A}\,,\qquad
[M_{\mu\nu}, \bar{Q}_{\dot{\alpha} A}] =  \frac{1}{2}\, (\tilde{\sigma}_{\mu\nu})^{\dot{\beta}}{}_{\dot{\alpha}} \, \bar{Q}_{\dot{\beta} A}\,,
\end{equation}
\begin{equation}
[ P_\mu , Q_{\alpha}^{\,A}] = [ P_\mu , \bar{Q}_{\dot{\alpha} A}] =0\,,
\end{equation}
\begin{equation}\label{I-Q}
[ T^{A}_{B}, Q_{\alpha}^{\, C} ] = \delta^{C}_{B} \, Q_{\alpha}^{\, A}
- \frac{1}{4} \, \delta^{A}_{B}\,  Q_{\alpha}^{\, C}\,,
\qquad
[T^{A}_{B}, \bar{Q}_{\dot{\alpha}C} ] = -\delta^{A}_{C} \,\bar{Q}_{\dot{\alpha}B}
+ \frac{1}{4} \, \delta^{A}_{B}\,  \bar{Q}_{\dot{\alpha}C}\,,
\end{equation}
\begin{equation}\label{A-QS}
[ A , Q_{\alpha}^{A}] = \alpha\, Q_{\alpha}^{A}\,, \qquad [ A , \bar{Q}_{\dot{\alpha}A}] =  -\alpha \,\bar{Q}_{\dot{\alpha}A}\,,
\end{equation}
\begin{equation}\label{A-Zbz}
[ A , Z^{AB}] = 2\alpha\, Z^{AB}\,, \qquad [ A , \bar Z_{AB}] =  -2\alpha \,\bar Z_{AB}\,.
\end{equation}
Here $\alpha$ is some real parameter.
If we choose $\alpha =1\,$, it defines  the chirality of supercharges (see (\ref{A-QS})) and so identifies $A$ as the generator
of axial symmetry.\footnote{In the cases of $\mathcal{N}{=}\,4,$ $D{=}\,4$ superconformal symmetry
and the corresponding $d{=}\,3$ Galilean superconformal algebra the generator $A$ fully decouples, so implying $\alpha = 0$ (see \cite{FL11}).}

In order to perform the non-relativistic contraction of
$\mathcal{N}{=}\,4,$ $D{=}\,4$ Poincar\'{e} superalgebra to the limit describing
$\mathcal{N}{=}\,4,$ $d{=}\,3$ Galilean superalgebra one should rewrite the superalgebra (\ref{QbQ-224})-(\ref{A-QS})
in the  new fermionic Weyl basis
\footnote{
The basis  (\ref{Q-pm}) was introduced in \cite{Luk86} and further used in \cite{fli4,fli6,FL11}
in order to obtain Galilean limit for $D=4$ Majorana supercharges.}
\begin{equation}\label{Q-pm}
Q^{\pm A}_{\alpha} =
\frac{1}{\sqrt{2}}\,\Big(Q_{\alpha}^{A} \pm\epsilon_{\alpha \beta}\,\Omega^{AB}\, \bar{Q}_{\dot{\beta}B}\,\Big)\,,
\qquad
\bar{Q}^{\pm }_{\dot{\alpha}A}=\frac{1}{\sqrt{2}}\,\Big(\bar{Q}_{\dot{\alpha} A} \mp \epsilon_{\dot{\alpha} \dot{\beta}}\,
\Omega_{AB} \, {Q}^{B}_{\beta} \,\Big)\,,
\end{equation}
where the real $4{\times}4$ matrix $\Omega^{AB}=-\Omega^{BA}$ is
the symplectic metric,
$\Omega_{AB}=-\Omega^{AB}$, $\Omega^{AC}\Omega_{CB}=\delta^A_B\,$.
In this paper we choose the following explicit form of $\Omega$:
\begin{equation}\label{Om}
\Omega_{AB}=
\left(
\begin{array}{cc}
\epsilon_{ab} & 0 \\
0 & \epsilon_{\tilde a \tilde b} \\
\end{array}
\right)
,\qquad
\Omega^{AB}= \left(
\begin{array}{cc}
\epsilon^{ab} & 0 \\
0 & \epsilon^{\tilde a \tilde b} \\
\end{array}
\right),
\end{equation}
where $\epsilon_{ab}=
\left(
\begin{array}{cc}
0 & 1 \\
-1 & 0 \\
\end{array}
\right)
$,
$\epsilon^{ab}=
\left(
\begin{array}{cc}
0 & -1 \\
1 & 0 \\
\end{array}
\right)
$, $a=1,2$, $\tilde a=1,2$.

The relations (\ref{Q-pm}) break manifest Lorentz symmetry $\mathrm{O}(3,1)$ to $\mathrm{O}(3)$
(spinorial scalar product $a_{\alpha}b_{\dot\alpha}$ is $\mathrm{U}(2)$-invariant) and the
internal symmetry $\mathrm{U}(4)$ is broken to its subgroups which depend on the choice of central charges \cite{fli20}.

The supercharges (\ref{Q-pm}) by definition satisfy the subsidiary symplectic-Majorana conditions \cite{KugoT}
\begin{equation}\label{MW-cond}
({Q}^{\pm A}_{\ {\alpha}})^\dagger
=\bar{Q}^{\pm}_{\dot{\alpha}A}
= \mp\, \epsilon_{\dot\alpha\dot\beta}
\Omega_{AB}\, Q^{\pm B}_{\beta} \,.
\end{equation}

The full set of supercharges $\left( {Q}^{\pm a}_{\ {\alpha}},\bar{Q}^{\pm}_{\dot{\alpha}a};
{Q}^{\pm \tilde a}_{\ {\alpha}},\bar{Q}^{\pm}_{\dot{\alpha}\tilde a}\right)$
can be split into the holomorphic sector
$\left( {Q}^{\pm a}_{\ {\alpha}},{Q}^{\pm \tilde a}_{\ {\alpha}}\right)$
and the antiholomorphic one
$\left( \bar{Q}^{\pm}_{\dot{\alpha}a},\bar{Q}^{\pm}_{\dot{\alpha}\tilde a}\right)$;
these both sectors are related by the subsidiary conditions (\ref{MW-cond}),
thus revealing the quaternionic structure of the pairs of complex supercharges
related by Hermitian conjugation (see \cite{LukNow,FL11}).
Due to the  constraints  (\ref{MW-cond}) one can choose as unconstrained sets
of linearly independent supercharges the generators from  either holomorphic or antiholomorphic sectors.
The $\mathcal{N}{=}\,4$ superalgebra spanned by the generators from holomorphic sector
is however not self-conjugate. In order to define the complete self-conjugated
Hermitian basis one should choose the full set of pairs of supercharges, which are related by Hermitian conjugation.
For the choice (\ref{Om}) of the matrix $\Omega$ these self-conjugated pairs are
\begin{equation}\label{Q-pm-H}
{Q}^{\pm}_{\ {\alpha}}={Q}^{\pm 1}_{\ {\alpha}}\,,\quad \bar{Q}^{\pm}_{\dot{\alpha}}=\bar{Q}^{\pm}_{\dot{\alpha}1}\,;\qquad
\tilde{Q}^{\pm}_{\ {\alpha}}={Q}^{\pm \tilde 1}_{\ {\alpha}}\,,\quad
\bar{\tilde{Q}}^{\pm}_{\dot{\alpha}}=\bar{Q}^{\pm}_{\dot{\alpha}\tilde 1} \,.
\end{equation}

In this paper we will use the supercharges belonging to the holomorphic sector,
i.e. $Q^{\pm A}_{\alpha}$ ($A=(a,\tilde a)=1,...,4$).
They transform linearly under the $\mathrm{USp}(4)\simeq \mathrm{O}(5)$ R-symmetry,
which defines the compact R-symmetry sector of   $\mathcal{N}{=}\,4,$ $d{=}\,3$ supersymmetry
with one central charge $\mathcal{Z}$ corresponding to the following choice of $4{\times}4$ central charge matrix (\ref{Z-matr})
\begin{equation}\label{Z-1}
Z^{AB}= \mathcal{Z}\,\Omega^{AB}\,.
\end{equation}

In the holomorphic basis the non-vanishing relations (\ref{QbQ-224}), (\ref{QQ-224}) can be represented as
\begin{eqnarray}\label{QQ-pm}
\{ Q^{\pm A}_{\alpha} , Q^{\pm B}_{\beta} \}&=&2\epsilon_{{\alpha} {\beta}}\,\Big(
\pm\,P_0\,\Omega^{AB}   +X^{AB}\Big)\,, \\ [5pt]
\{ Q^{+ A}_{\alpha}, Q^{- B}_{\beta} \}&=&2\,\Big(
P_i(\sigma_i )_{{\alpha} {\beta}}\,\Omega^{AB}   +i\epsilon_{{\alpha} {\beta}}
Y^{AB}\Big)\,,\label{QQ-+-}
\end{eqnarray}
where $(\sigma_i )_{\alpha\beta}=(\sigma_i )_{\beta\alpha}=\epsilon_{\alpha\gamma}(\sigma_i )_{\beta\dot\gamma}$, $i=1,2,3$
and
\begin{equation}\label{Re-Im-XY}
X^{AB}:= \frac{1}{2} \left(Z^{AB}-\Omega^{AC}\bar Z_{CD} \Omega^{DB}\right),\qquad
Y^{AB} := \frac{1}{2i} \left(Z^{AB}+\Omega^{AC}\bar Z_{CD} \Omega^{DB}\right),
\end{equation}
\begin{equation}\label{Re-XY-g}
\left(X^{AB}\right)^\dagger   =
\Omega_{AC} \Omega_{BD}\,X^{CD}\,, \qquad
\left(Y^{AB}\right)^\dagger   =
\Omega_{AC} \Omega_{BD}\,Y^{CD}\,.
\end{equation}

The relations inverse to \p{Re-Im-XY} are
\begin{equation}\label{Re-Im-XY1}
Z^{AB} = X^{AB} + i Y^{AB}\,, \quad \bar{Z}_{AB} = X_{AB} - i Y_{AB}\,.
\end{equation}
Note that $ X^{AB}$ and $Y^{AB}$ do not coincide with $ {\cal X}^{AB}$ and $\cal{Y}^{AB}$ defined in \p{Z-XY}: while
the latter are real in the ordinary sense, the former are subject to the pseudo-reality conditions  \p{Re-XY-g}. The first
kind of reality is preserved by the subgroup $\mathrm{O}(4) \subset \mathrm{SU}(4)$ with
both $ {\cal X}^{AB}$ and $\cal{Y}^{AB}$ in the adjoint representation of $ \mathrm{O}(4)$; the pseudoreality is consistent with the subgroup
$\mathrm{USp}(4) \simeq \mathrm{O}(5) \subset \mathrm{SU}(4)$, such that each of  $ X^{AB}$ and $Y^{AB}$ encompasses
an $ \mathrm{O}(5)$ vector and an $ \mathrm{O}(5)$ scalar.

If we choose the central charges in accord with (\ref{Z-matr})
the anticommutators  (\ref{QQ-pm}), (\ref{QQ-+-}) take the form
\begin{equation}\label{QQ-pm-2}
\begin{array}{rcl}
\{ Q^{\pm a}_{\alpha} , Q^{\pm b}_{\beta} \}&=&2\,
\epsilon^{ab} \, \epsilon_{{\alpha} {\beta}}\, \Big( \pm P_0 + \mathrm{Re} \left(Z_1\right)\Big)\,, \\  [6pt]
\{ Q^{\pm \tilde a}_{\alpha}, Q^{\pm \tilde b}_{\beta} \}&=&2\,
\epsilon^{\tilde a\tilde b} \, \epsilon_{{\alpha} {\beta}}\, \Big( \pm P_0 + \mathrm{Re} \left(Z_2\right)\Big)\,,\\  [6pt]
\{ Q^{\pm a}_{\alpha}, Q^{\pm \tilde b}_{\beta} \}&=& 2\, \epsilon_{\alpha\beta} \,X^{a \tilde b}\,,
\end{array}
\end{equation}
\begin{equation}\label{QQ-+--2}
\begin{array}{rcl}
\{ Q^{+ a}_{\alpha}, Q^{- b}_{\beta} \}&=&2\,
\epsilon^{ab} \,  \Big( (\sigma_i )_{{\alpha} {\beta}}P_i + i\epsilon_{{\alpha} {\beta}}\,\mathrm{Im} \left(Z_1\right)\Big)\,, \\  [6pt]
\{ Q^{+ \tilde a}_{\alpha}, Q^{- \tilde b}_{\beta} \}&=&2\,
\epsilon^{\tilde a\tilde b} \,  \Big( (\sigma_i )_{{\alpha} {\beta}}P_i +
i\epsilon_{{\alpha} {\beta}}\,\mathrm{Im} \left(Z_2\right)\Big)\,, \\  [6pt]
\{ Q^{+ a}_{\alpha}, Q^{- \tilde b}_{\beta} \}&=&-\,\{ Q^{+ \tilde a}_{\alpha}, Q^{- b}_{\beta} \} \ = \ 2i\, \epsilon_{\alpha \beta}\,
Y^{a\tilde b}\,,
\end{array}
\end{equation}
where
\begin{equation}\label{Re-Im}
X^{a \tilde b}:= \frac{1}{2} \left(Z^{a\tilde b}+\epsilon^{ac}\epsilon^{\tilde c \tilde d}\bar Z_{c\tilde d} \right),\qquad
Y^{a\tilde b} := \frac{1}{2i} \left(Z^{a\tilde b}-\epsilon^{ac}\epsilon^{\tilde c \tilde d}\bar Z_{c\tilde d} \right).
\end{equation}
It can be pointed out that $X^{a \tilde b}$ and $Y^{a\tilde b}$ are ``real'' with respect
to the symplectic pseudoreality conditions similar to (\ref{MW-cond}) and following from \p{Re-XY-g},
$\left(X^{a \tilde b}\right)^\dagger = \epsilon_{ac}\epsilon_{\tilde b \tilde d}\,X^{c \tilde d}\,$,
$\left(Y^{a \tilde b}\right)^\dagger = \epsilon_{ac}\epsilon_{\tilde b \tilde d}\,Y^{c \tilde d}\,$.

The commutators (\ref{QM-224}) for the generators  (\ref{Q-pm}) can be rewritten as follows
\begin{equation}\label{QM-noncov}
[M_{ij}, Q^{\pm A}_{\alpha}] = - \frac{1}{2}\, \epsilon_{ijk}(\sigma_{k})_{\alpha}{}^{\beta}  Q^{\pm A}_{\beta}\,,\qquad
[M_{i0}, Q^{\pm A}_{\alpha}] = \frac{i}{2}\, (\sigma_{i})_{\alpha}{}^{\beta}  Q^{\mp A}_{\beta}\,.
\end{equation}

Further, let us decompose the generators of internal symmetry $\mathrm{SU}(4)\,$ as
\begin{equation}\label{T-pm}
T^{\pm A}_{\ \, B}\equiv T^A_B \mp \Omega^{AC}\, T^D_C\,\Omega_{DB}\,.
\end{equation}
The projections (\ref{T-pm}) of $T^A_B$ satisfy the relations
\begin{equation}\label{rel-T-pm}
T^{+ C}_{\ \,A}\Omega_{BC}=2\, T^{C}_{(A}\, \Omega_{B)C}\,,\qquad
T^{- C}_{\ \,A}\Omega_{BC}=2\, T^{C}_{[A}\, \Omega_{B]C}\,,
\end{equation}
\begin{equation}\label{rel-T-pm-1}
T^{\pm A}_{\ \,B}=\mp\, \Omega^{AC}T^{\pm D}_{\ \,C}\, \Omega_{DB}\,.
\end{equation}
The constraint (\ref{rel-T-pm-1}) amounts to the conditions for the generators $T^{\pm A}_{\ \, B}\,$,
\begin{equation}\label{rel-T-pm-cons}
T^{+ C}_{\ \,[A}\Omega_{B]C}=0\,,\qquad
T^{- C}_{\ \,(A}\Omega_{B)C}=0\,.
\end{equation}
 So, the set of generators $T^{+}$ contains 10 independent generators which are symmetric in their indices
\begin{equation}\label{rel-T-+}
T^{+}_{AB}=\Omega_{AC}T^{+ C}_{\ \,B}=T^{+}_{(AB)}\,.
\end{equation}
The set $T^{-}$ involves 5 independent operators forming a traceless antisymmetric matrix
\begin{equation}\label{rel-T--}
T^{-}_{AB}=\Omega_{AC}T^{- C}_{\ \,B}=T^{-}_{[AB]}\,,\qquad \Omega^{AB}T^{-}_{AB}=0\,.
\end{equation}
Using  (\ref{suN}), we find that operators (\ref{rel-T-+}), (\ref{rel-T--}) satisfy the following algebra
\begin{eqnarray}\label{rel-TTpm}
\left[T^{\pm}_{AB}, T^{\pm}_{CD}\right]&=&-\left(\Omega_{BC}T^{+}_{AD}+\Omega_{AD}T^{+}_{BC}
\pm \Omega_{AC}T^{+}_{BD} \pm \Omega_{BD}T^{+}_{AC} \right) ,\\ [6pt]
\left[T^{+}_{AB}, T^{-}_{CD} \right]&=& -\left(\Omega_{BC}T^{-}_{AD}-\Omega_{AD}T^{-}_{BC}
+ \Omega_{AC}T^{-}_{BD} - \Omega_{BD}T^{-}_{AC} \right). \label{rel-TT+-}
\end{eqnarray}
Thus the original internal SU(4) symmetry generators $T^{A}_{B}$, decomposed according to the relations (\ref{T-pm}),
do split into the ones generating USp(4) and the coset SU(4)$/$USp(4):
\begin{equation}\label{mosluke3.22}
h^{(3)} = \left(\,T^{+ A}_{\ \,B}\,\right) \in {usp}(4)\,,
\qquad \qquad
k^{(3)} = \left(\,T^{- A}_{\ \,B}\,\right)\in {su}(4) {\ominus} {usp}(4)\,.
\end{equation}
This decomposition of the $su(4)$ algebra  provides an example of symmetric Riemannian pair ($h^{(3)}, k^{(3)}$):
\begin{equation}\label{h-k}
[h^{(3)}, h^{(3)}] \subset h^{(3)}\,, \qquad
[h^{(3)}, k^{(3)}] \subset k^{(3)}\,, \qquad
[k^{(3)}, k^{(3)}] \subset h^{(3)}\,.
\end{equation}

The commutators (\ref{I-Q}) are rewritten in the new basis as
\begin{equation}\label{GCA-TQ-pm}
[{T}^{+ A}_{\ \, B},  Q^{\pm C}_{\alpha} ] = \left(\mathscr{U}^A_B\right){}^C_D\,  Q^{\pm D}_{\alpha}\,,
\qquad
[{T}^{- A}_{\ \, B}, Q^{\pm C}_{\alpha} ] = \left(\tau^A_B\right){}^C_D\, Q^{\mp D}_{\alpha}\,,
\end{equation}
where the $4{\times} 4$ matrix $\left(\mathscr{U}^A_B\right)\,$,
\begin{equation}\label{U-Usp}
\left(\mathscr{U}^A_B\right){}^C_D = \delta^A_D\delta^C_B +\Omega^{AC}\Omega_{BD}\,,
\end{equation}
defines the fundamental  $4\times 4$ representations of the $\mathrm{USp}(4)$ algebra given by the supercharges $Q^{\pm C}_{\alpha}$.
The matrices
\begin{equation}\label{tau-Usp}
\left(\tau^A_B\right){}^C_D = \delta^A_D\delta^C_B -\Omega^{AC}\Omega_{BD} - \frac{1}{2}\,\delta^A_B\delta^C_D
\end{equation}
enlarge the matrices $\left(\mathscr{U}^A_B\right)$ to the fundamental representations of $SU(4)$ algebra which interchange the $+$
and $-$ projections.

Let us make a comment on the case of $\alpha\neq 0$ in \p{A-QS}, (\ref{A-Zbz}). Choosing $\alpha=1\,$, one finds
\be
[A, Q^{\pm B}_\alpha] = Q^{\mp B}_\alpha\,; \qquad [A, X^{AB}] = 2i \, Y^{AB}\,, \quad [A, Y^{AB}] = -2i \, X^{AB}\,. \lb{alphanot0}
\ee

Now we are prepared to define the $\mathcal{N}{=}\,4,$  $d{=}\,3$ Galilean superalgebra by making use of the NR contraction procedure.
One rescales the relativistic supercharges as
\begin{equation}\label{QS-tr}
Q^{+ A}_{\alpha} = c^{-1/2} \, \mathbf{Q}^{A}_{\alpha}\,,
\qquad Q^{- A}_{\alpha} = c^{1/2} \, \mathbf{S}^{A}_{\alpha}\,.
\end{equation}
The physical rescaling of the bosonic generators of the algebra $\mathrm{o}(1,3)\oplus \mathrm{u}(4)$,
where $(P_\mu,M_{\mu\nu})\in \mathrm{o}(1,3)$ and $(T^{\pm A}_{\ \, B}, \,A) \in \mathrm{u}(4)$, is performed as follows
\begin{equation}\label{resc-o42}
\begin{array}{lll}
&& P_{0} = m_0c+c^{-1}\,\mathbf{H}\,, \qquad P_{i} = \mathbf{P}_{i}\,, \\[6pt]
&& M_{ij}=  \mathbf{J}_{ij}\,,\qquad  M_{i0}=  c\,\mathbf{B}_{i}\,, \\[6pt]
&& A = c\, \mathbf{A}_0\,, \qquad T^{+ A}_{\ \, B} = \mathbf{T}^{+ A}_{\ \, B}\,,
\qquad T^{- A}_{\ \, B} = c\,\mathbf{T}^{- A}_{\ \, B} \,.
\end{array}
\end{equation}
where $m_0$ is the relativistic rest  mass.
The rescaling of the central charges is given by the formulas (see also (\ref{XY-lim}))
\begin{equation}\label{resc-CCh}
X^{AB}= -m_0 c\,\Omega^{AB}+c^{-1}\mathbf{X}^{AB}\,,\qquad
Y^{AB}=\mathbf{Y}^{AB}\,,
\end{equation}
where $X^{AB}$, $Y^{AB}$ are defined in (\ref{Re-Im-XY})  and the operators
$\mathbf{X}^{AB}=-\mathbf{X}^{BA}$,
$\mathbf{Y}^{AB}=-\mathbf{Y}^{BA}$ satisfy the symplectic pseudoreality conditions .

We will firstly perform the $c\to \infty$ contraction for a simple choice of the central charge matrix.
\\

\noindent {\bf i) Jordanian quasi-diagonal form of central charge matrix}
\\

Let us consider the special case with central charge matrix in the reduced form (\ref{eqfli1.5})
\begin{equation}\label{Z-Om}
Z^{AB}= \left(
\begin{array}{cc}
Z_1 \epsilon^{ab} & 0 \\
0 & Z_2 \epsilon^{\tilde a \tilde b} \\
\end{array}
\right),\qquad
\bar Z_{AB}=
\left(
\begin{array}{cc}
-\bar Z_1 \epsilon_{ab} & 0 \\
0 & -\bar Z_2 \epsilon_{\tilde a \tilde b} \\
\end{array}
\right),
\end{equation}
where the central charge matrix (\ref{Z-1}) is recovered at $\mathcal{Z}=Z_1=Z_2$.
The rescaling  (\ref{resc-CCh}) takes the more explicit form for this choice
\begin{eqnarray}\label{resc-CCh2a}
&\mathrm{Re} \left(Z_{1}\right)= -m_0 c+c^{-1}\mathbf{X}_{1}\,,\qquad \mathrm{Re} \left(Z_{2}\right)= -m_0 c+c^{-1}\mathbf{X}_{2}\,,\\ [6pt]
&\mathrm{Im} \left(Z_{1}\right)=\mathbf{Y}_{1}\,,\qquad  \mathrm{Im} \left(Z_{2}\right)=\mathbf{Y}_{2}\,.\label{resc-CCh2b}
\end{eqnarray}
Substituting these expressions, as well as \p{QS-tr} and \p{resc-o42}, into the superalgebra relations (\ref{QQ-pm-2}),
(\ref{QQ-+--2}) with $Z^{a\tilde b}=Z^{\tilde ab}=0\,$,
and making there the  $c\to\infty$ contraction, we obtain
\begin{equation}\label{contr-QQ-pm-2}
\begin{array}{rcl}
\{ \mathbf{Q}^{a}_{\alpha} , \mathbf{Q}^{b}_{\beta} \}&=&2\,
\epsilon^{ab} \, \epsilon_{{\alpha} {\beta}}\, \Big( \mathbf{H} + \mathbf{X}_{1}\Big)\,, \\  [6pt]
\{ \mathbf{Q}^{\tilde a}_{\alpha}, \mathbf{Q}^{\tilde b}_{\beta} \}&=&2\,
\epsilon^{\tilde a\tilde b} \, \epsilon_{{\alpha} {\beta}}\, \Big( \mathbf{H} + \mathbf{X}_{2}\Big)\,,\\  [6pt]
\{ \mathbf{Q}^{a}_{\alpha}, \mathbf{Q}^{\tilde b}_{\beta} \}&=&0\,,
\end{array}
\end{equation}
\begin{equation}\label{contr-SS-pm-2}
\begin{array}{rcl}
\{ \mathbf{S}^{a}_{\alpha} , \mathbf{S}^{b}_{\beta} \}&=&-4\,
\epsilon^{ab} \, \epsilon_{{\alpha} {\beta}}\, m_0\,, \\  [6pt]
\{ \mathbf{S}^{\tilde a}_{\alpha}, \mathbf{S}^{\tilde b}_{\beta} \}&=&-4\,
\epsilon^{\tilde a\tilde b} \, \epsilon_{{\alpha} {\beta}}\, m_0\,,\\  [6pt]
\{ \mathbf{S}^{a}_{\alpha}, \mathbf{S}^{\tilde b}_{\beta} \}&=&0\,,
\end{array}
\end{equation}
\begin{equation}\label{contr-QQ-+--2}
\begin{array}{rcl}
\{ \mathbf{Q}^{a}_{\alpha}, \mathbf{S}^{b}_{\beta} \}&=&2\,
\epsilon^{ab} \,  \Big( (\sigma_i )_{{\alpha} {\beta}}\mathbf{P}_i + i\epsilon_{{\alpha} {\beta}}\,\mathbf{Y}_{1}\Big)\,, \\  [6pt]
\{ \mathbf{Q}^{\tilde a}_{\alpha}, \mathbf{S}^{\tilde b}_{\beta} \}&=&2\,
\epsilon^{\tilde a\tilde b} \,  \Big( (\sigma_i )_{{\alpha} {\beta}}\mathbf{P}_i +
i\epsilon_{{\alpha} {\beta}}\,\mathbf{Y}_{2}\Big)\,, \\  [6pt]
\{ \mathbf{Q}^{a}_{\alpha}, \mathbf{S}^{\tilde b}_{\beta} \}&=&\{ \mathbf{Q}^{\tilde a}_{\alpha}, \mathbf{S}^{b}_{\beta} \} \ = \ 0\,.
\end{array}
\end{equation}
Here $\mathbf{Q}^{A}_{\alpha}=\left(\mathbf{Q}^{a}_{\alpha},\mathbf{Q}^{\tilde a}_{\alpha}\right)$,
$\mathbf{S}^{A}_{\alpha}=\left(\mathbf{S}^{a}_{\alpha},\mathbf{S}^{\tilde a}_{\alpha}\right)$, and the indices are chosen
so that  $a=1,2$ correspond to $A=1,2$ and $\tilde a=1,2$ to $A=3,4$.
\\

\noindent {\bf ii) General central charge matrix}
\\

In the general case with non-zero off-diagonal central charges $\mathbf{X}^{a\tilde b} = - \mathbf{X}^{\tilde b a}$
and $\mathbf{Y}^{a\tilde b} = - \mathbf{Y}^{\tilde b a}$,
the last lines in \p{contr-QQ-pm-2} and \p{contr-QQ-+--2} are replaced, respectively, by the relations
\bea
&& \{ \mathbf{Q}^{a}_{\alpha}, \mathbf{Q}^{\tilde b}_{\beta} \} \ = \ 2\epsilon_{\alpha\beta} \mathbf{X}^{a \tilde b}\,, \nn
&& \{ \mathbf{Q}^{a}_{\alpha}, \mathbf{S}^{\tilde b}_{\beta} \} \ = \ \{ \mathbf{Q}^{\tilde b}_{\beta},
\mathbf{S}^{a}_{\alpha} \} \ = \ 2i \epsilon_{\alpha \beta} \mathbf{Y}^{a \tilde b}\,. \lb{VectZdop}
\eea

It is easy to check that the rescaling (\ref{QS-tr}) preserves the symplectic-Majorana conditions (\ref{Q-pm})
and in the limit $c\to\infty$ one obtains the following Galilean form of $\mathcal{N}{=}\,4$ symplectic-Majorana conditions
\footnote{We recall that product $A_{\alpha}B_{\dot\alpha}$ of two NR spinors
$A_{\alpha}$, $B_{\dot\alpha}\equiv (\overline{B_{\alpha}})$ is
$\mathrm{U}(2)$ invariant, i.e. $\delta_{\alpha\dot\beta}$
is the $\mathrm{U}(2)$-invariant metric.}
\begin{equation}\label{Q-pm-2}
\begin{array}{ll}
\left(\mathbf{Q}^{a}_{\alpha}\right)^\dagger  \equiv \bar{\mathbf{Q}}_{\dot\alpha a} =
- \epsilon_{\dot\alpha\dot\beta} \epsilon_{ab}\mathbf{Q}^{b}_{\beta}\,, \quad &
\left(\mathbf{Q}^{\tilde a}_{\alpha}\right)^\dagger  \equiv \bar{\mathbf{Q}}_{\dot\alpha \tilde a} =
- \epsilon_{\dot\alpha\dot\beta} \epsilon_{\tilde a\tilde b}\mathbf{Q}^{\tilde b}_{\beta}\,, \\  [6pt]
\left(\mathbf{S}^{a}_{\alpha}\right)^\dagger  \equiv \bar{\mathbf{S}}_{\dot\alpha a} =
\epsilon_{\dot\alpha\dot\beta} \epsilon_{ab}\mathbf{S}^{b}_{\beta}\,, \quad &
\left(\mathbf{S}^{\tilde a}_{\alpha}\right)^\dagger  \equiv \bar{\mathbf{S}}_{\dot\alpha \tilde a} =
\epsilon_{\dot\alpha\dot\beta} \epsilon_{\tilde a\tilde b}\mathbf{S}^{\tilde b}_{\beta}\,.
\end{array}
\end{equation}
Due to (\ref{Re-XY-g}), off-diagonal central charges satisfy the following pseudo-reality conditions
\begin{equation}\label{XY-pm-2}
\left(\mathbf{X}^{a \tilde b}\right)^\dagger   =
\epsilon_{ac} \epsilon_{\tilde b\tilde d}\,\mathbf{X}^{c \tilde d}\,, \qquad
\left(\mathbf{Y}^{a \tilde b}\right)^\dagger   =
\epsilon_{ac} \epsilon_{\tilde b\tilde d}\,\mathbf{Y}^{c \tilde d}\,.
\end{equation}

We point out  that the constant $m_0$ can be considered as an additional thirteenth central charge, i.e. in  fact
the superalgebra (\ref{contr-QQ-pm-2})-(\ref{contr-QQ-+--2}) contains thirteen central charges
($m_0$, $\mathbf{X}_{1}$, $\mathbf{X}_{2}$, $\mathbf{Y}_{1}$, $\mathbf{Y}_{2}$, $\mathbf{X}^{a\tilde b}$, $\mathbf{Y}^{a\tilde b}$)
as compared to twelve in $D{=}\,4$ relativistic case.
\\

\noindent {\bf iii) Internal symmetry sectors}
\\

After $c\to\infty$ contraction (\ref{QM-noncov}), the covariance relations of the supercharges  with respect to NR $\mathrm{O}(3)$ rotations $\mathbf{J}_{ij}$
and Galilean boosts $\mathbf{B}_{i}$ are written as
\begin{equation}\label{CA-N-JQ}
[\mathbf{J}_{ij}, \mathbf{Q}^{A}_{\alpha}] = - {\displaystyle\frac{1}{2}}\,\epsilon_{ijk} (\sigma_{k})_{\alpha}{}^{\beta} \,
\mathbf{Q}^{A}_{\beta}\,, \qquad [\mathbf{J}_{ij}, \mathbf{S}^{A}_{\alpha}] = - {\displaystyle\frac{1}{2}}\,\epsilon_{ijk} (\sigma_{k})_{\alpha}{}^{\beta} \, \mathbf{S}^{A}_{\beta}\,,
\end{equation}
\begin{equation}\label{CA-N-BQ}
[\mathbf{B}_{i}, \mathbf{Q}^{A}_{\alpha}] =  {\displaystyle\frac{i}{2}}\, (\sigma_{i})_{\alpha}{}^{\beta} \,
\mathbf{S}^{A}_{\beta}\,, \qquad [\mathbf{B}_{i}, \mathbf{S}^{A}_{\beta}] = 0 \,.
\end{equation}

Using substitutions (\ref{QS-tr}), (\ref{resc-o42}), the contraction of the relations (\ref{GCA-TQ-pm})
leads also to the covariance relations of supercharges with respect to the internal symmetry
generators $\mathbf{h}{}^{(3)} = (\,\mathbf{T}^{+ A}_{\ \,B}\,)$,
$\mathbf{k}{}^{(3)} = (\,\mathbf{T}^{- A}_{\ \,B}\,)$ and axial charge $\mathbf{A}=\mathbf{A}_0|_{\alpha=1}$
\begin{equation}\label{GCA-N-QT-pl}
[\mathbf{T}^{+ A}_{\ \, B}, \mathbf{Q}^{C}_{\alpha} ] = \left(\mathscr{U}^A_B\right){}^C_D\, \mathbf{Q}^{D}_{\alpha}\,,
\qquad
[\mathbf{T}^{+ A}_{\ \, B}, \mathbf{S}^{C}_{\alpha} ] = \left(\mathscr{U}^A_B\right){}^C_D\, \mathbf{S}^{D}_{\alpha}\,,
\end{equation}
\begin{equation}\label{GCA-N-QT-min}
[\mathbf{T}^{- A}_{\ \, B}, \mathbf{Q}^{C}_{\alpha} ] = \left(\tau^A_B\right){}^C_D\, \mathbf{S}^{D}_{\alpha}\,,
\qquad
[\mathbf{T}^{- A}_{\ \, B}, \mathbf{S}^{C}_{\alpha} ] = 0\,,
\end{equation}
\begin{equation}\label{CA-N-AQ}
[\mathbf{A}, \mathbf{Q}^{A}_{\alpha} ] = - \mathbf{S}^{A}_{\alpha}\,, \quad
[\mathbf{A}, \mathbf{S}^{A}_{\alpha} ] = 0\,.
\end{equation}

{}For what follows, it will be useful to have the generators $\mathbf{T}^{+ A}_{\ \, B}\in {usp}(4)$
in the splitting ${usp}(2){\oplus} {usp}(2)$ basis. This notation corresponds to the following coset decomposition
\begin{equation}\label{usp-dec1}
\mathrm{USp}(4) \ \simeq \ \frac{\mathrm{USp}(4)}{\mathrm{USp}(2){\otimes}\mathrm{USp}(2)}\,\cdot\,
\Big[ \mathrm{USp}(2){\otimes}\mathrm{USp}(2)\Big]\,,
\end{equation}
such that
\begin{eqnarray}\label{usp-dec2}
\Big( \mathbf{T}^{+ a}_{\ \, b}, \mathbf{T}^{+ \tilde a}_{\ \, \tilde b} \Big) & \in & {usp}(2){\oplus}{usp}(2)\,, \\
\Big( \mathbf{T}^{+ \tilde a}_{\ \, b}, \mathbf{T}^{+ a}_{\ \, \tilde b} \Big) & \in & {usp}(4) {\ominus}
[{usp}(2){\oplus} {usp}(2)] \,.\label{usp-dec3}
\end{eqnarray}

{}From (\ref{resc-o42}) it follows that the coset generators ${k}^{(3)}$ are rescaled ($h^{(3)}=\mathbf{h}{}^{(3)}$,
${k}^{(3)}=c\,\mathbf{k}{}^{(3)}$) and in the limit $c\to\infty$ we get
the inhomogeneous extension  of ${usp}(4)\cong {o}(5)$ internal algebra
\begin{equation}\label{h-k-til}
[\mathbf{h}{}^{(3)}, \mathbf{h}{}^{(3)}] \subset \mathbf{h}{}^{(3)}\,, \qquad
[\mathbf{h}{}^{(3)}, \mathbf{k}{}^{(3)}] \subset \mathbf{k}{}^{(3)}\,, \qquad
[\mathbf{k}{}^{(3)}, \mathbf{k}{}^{(3)}] =0\,.
\end{equation}
The five commuting generators $\mathbf{T}^{- A}_{\ \,B}$ of $\mathbf{k}{}^{(3)}$ describe a kind of curved internal momenta.
Thus in the contraction limit $c\to \infty$ one gets the following $\mathcal{N}=4$ Galilean internal inhomogeneous symmetry
algebra~$\mathbf{T}$
\begin{equation}\label{mosluke3.27}
{\mathbf{T}} = \mathbf{h}{}^{(3)}\,\subset\!\!\!\!\!+\, \mathbf{k}{}^{(3)}\,,
\qquad \qquad
\begin{array}{l}
\mathbf{h}{}^{(3)}= {u}(2;\mathbb{H})= {usp}(4)\,,
\\[2pt]
\mathbf{k}{}^{(3)}={\mathscr{A}}^{\,5} \quad \hbox{(Abelian)} \,.
\end{array}
\end{equation}
We will denote the corresponding inhomogeneous group by $\mathrm{IUSp}(4)\,$. The abelian generator $\mathbf{A}$ can
be added to the ideal $\mathbf{k}{}^{(3)}$, so extending it to six-dimensional one.\footnote{We recall here
that the replacement of $\mathrm{SU}(2N)$ internal symmetry by $\mathrm{IUSp}(2N)$ in the presence of
NR contraction occurred as well in the derivation of Galilean conformal superalgebra
\cite{fli6,fli7,FL11} from extended relativistic conformal superalgebra $\mathrm{SU}(2,2|2N)$.}

The action of the $\mathrm{IUSp}(4)\,$ generators in the $\mathrm{USp}(2)\otimes\mathrm{USp}(2)$ splitting basis,
$\mathbf{T}^{+ A}_{\ \, B} = (\mathbf{T}^{+ a}_{\ \, b}, \mathbf{T}^{+ \tilde{a}}_{\ \, \tilde{b}}, \mathbf{T}^{+ {a}}_{\ \, \tilde{b}},
\mathbf{T}^{+ \tilde{a}}_{\ \, {b}})$ on the supercharges in the same basis,
$\mathbf{Q}^{A}_{\alpha}=(\mathbf{Q}^{a}_{\alpha},\mathbf{Q}^{\tilde a}_{\alpha}),  \;
\mathbf{S}^{A}_{\alpha}=(\mathbf{S}^{a}_{\alpha},\mathbf{S}^{\tilde a}_{\alpha})$, as well as on the central charges,
can be easily found from the relations \p{GCA-N-QT-pl}. For instance, the commutation relations between $\mathbf{T}^{+ {a}}_{\ \, \tilde{b}}$
and central charges are given by
\bea \lb{T+X}
 [\mathbf{T}^{+}{}^d_{\tilde c}, \mathbf{X}^{a\tilde b}] = \delta^{\tilde b}_{\tilde c} \epsilon^{da}(\mathbf{X}_2 - \mathbf{X}_1)\,,
\qquad  [\mathbf{T}^{+}{}^d_{\tilde c}, \mathbf{X}_1] = - [\mathbf{T}^{+}{}^d_{\tilde c}, \mathbf{X}_2]  =
 \epsilon_{\tilde c \tilde b}\mathbf{X}^{d\tilde b}\,,
\eea
and by similar formulas for $\mathbf{Y}^{a\tilde b}\,, \,\mathbf{Y}_1, \mathbf{Y}_2\,$. It follows from these relations
that the full set of central charges splits into two  $\mathrm{USp}(4)\cong \mathrm{O}(5)$ five-vectors
($\mathbf{X}^{a\tilde b}$, $\mathbf{X}_1 - \mathbf{X}_2$) and ($\mathbf{Y}^{a\tilde b}$, $\mathbf{Y}_1 - \mathbf{Y}_2$)
and two $\mathrm{O}(5)$ singlets (isoscalars) ($\mathbf{X}_1 + \mathbf{X}_2$), ($\mathbf{Y}_1 + \mathbf{Y}_2$\,).

The first relation in (\ref{GCA-N-QT-min}) amounts to the following set of relations  in the splitting basis
\bea
&&[\mathbf{T}^{- a}_{\ \, b}, \mathbf{Q}^c_\alpha] = \frac 12 \delta^a_b \mathbf{S}^c_\alpha\,, \;\quad
[\mathbf{T}^{- \tilde a}_{\ \, \tilde b}, \mathbf{Q}^c_\alpha] = -\frac 12 \delta^{\tilde a}_{\tilde b} \mathbf{S}^c_\alpha\,, \quad
[\mathbf{T}^{- a}_{\ \, \tilde b}, \mathbf{Q}^c_\alpha] = -\epsilon^{ac}\epsilon_{\tilde b \tilde d}\mathbf{S}^{\tilde d}_\alpha \nn
&&[\mathbf{T}^{- a}_{\ \, b}, \mathbf{Q}^{\tilde c}_\alpha] = -\frac 12 \delta^a_b \mathbf{S}^{\tilde c}_\alpha\,, \;\quad
[\mathbf{T}^{- \tilde a}_{\ \, \tilde b}, \mathbf{Q}^{\tilde c}_\alpha]
= \frac 12 \delta^{\tilde a}_{\tilde b} \mathbf{S}^{\tilde c}_\alpha\,, \quad
[\mathbf{T}^{- a}_{\ \, \tilde b}, \mathbf{Q}^{\tilde c}_\alpha] = \delta^{\tilde c}_{\tilde b}\mathbf{S}^{a}_\alpha\,. \lb{T-Q}
\eea

The commutation relations between $\mathbf{T}^{- A}_{\ \, B}$ and the central charges $\mathbf{X}^{A B}$ and
$\mathbf{Y}^{A B}$ read
\bea
&& [\mathbf{T}^{- C}_{\ \, D}, \mathbf{X}^{A B}] = -i (\tau^C_D)^A_E \mathbf{Y}^{B E} +i (\tau^C_D)^B_E \mathbf{Y}^{A E}\,, \nn
&&  [\mathbf{T}^{- C}_{\ \, D}, \mathbf{Y}^{A B}]  = 2i m_0 \big(\delta^A_D \Omega^{C B} -  \delta^B_D \Omega^{C A}  - \frac12
\delta^C_D \Omega^{AB} \big),
\eea
or, in the splitting basis,
\bea
&& [\mathbf{T}^{- f}_{\ \, g}, \mathbf{X}_1] = i \delta^f_g  \mathbf{Y}_1\,, \quad
[\mathbf{T}^{- f}_{\ \, g}, \mathbf{X}_2] = -i \delta^f_g  \mathbf{Y}_2 \,, \quad [\mathbf{T}^{- f}_{\ \, g}, \mathbf{X}^{a \tilde b}] =
2i\big( \delta^a_g \mathbf{Y}^{f \tilde b} - \delta^f_g \mathbf{Y}^{a \tilde b} \big), \nn
&& [\mathbf{T}^{- \tilde f}_{\ \, \tilde g}, \mathbf{X}_1] = - i \delta^{\tilde f}_{\tilde g} \mathbf{Y}_1\,, \quad
[\mathbf{T}^{- \tilde f}_{\ \, \tilde  g}, \mathbf{X}_2] = i \delta^{\tilde f}_{\tilde g}  \mathbf{Y}_2 \,, \quad
[\mathbf{T}^{- \tilde f}_{\ \, \tilde g}, \mathbf{X}^{a \tilde b}] =
2i\big( \delta^{\tilde b}_{\tilde g} \mathbf{Y}^{a \tilde f} - \delta^{\tilde f}_{\tilde g} \mathbf{Y}^{a \tilde b} \big), \nn
&& [\mathbf{T}^{- f}_{\ \, \tilde  g}, \mathbf{X}_1] = -i \epsilon_{\tilde  g \tilde c}\mathbf{Y}^{f \tilde c}\,, \quad
[\mathbf{T}^{- f}_{\ \, \tilde  g}, \mathbf{X}_2] = -i \epsilon_{\tilde  g \tilde c}\mathbf{Y}^{f \tilde c}\,, \quad
[\mathbf{T}^{- f}_{\ \, \tilde  g}, \mathbf{X}^{a \tilde b}] =
-i \delta^{\tilde b}_{\tilde g} \epsilon^{fa}\big(\mathbf{Y}_1 + \mathbf{Y}_2\big), \lb{TX}
\eea
and
\bea
&& [\mathbf{T}^{- f}_{\ \, g}, \mathbf{Y}_1] = i m_0\,\delta^f_g\,, \qquad
[\mathbf{T}^{- f}_{\ \, g}, \mathbf{Y}_2] = -i m_0\,\delta^f_g\,,\qquad [\mathbf{T}^{- f}_{\ \, g}, \mathbf{Y}^{a \tilde b}] = 0\,, \nn
&& [\mathbf{T}^{- \tilde f}_{\ \, \tilde g}, \mathbf{Y}_1] = - i m_0\,\delta^{\tilde f}_{\tilde g}\,, \qquad
[\mathbf{T}^{- \tilde f}_{\ \, \tilde  g}, \mathbf{Y}_2] = i m_0 \,\delta^{\tilde f}_{\tilde g}\,, \qquad
[\mathbf{T}^{- \tilde f}_{\ \, \tilde g}, \mathbf{Y}^{a \tilde b}] = 0\,, \nn
&& [\mathbf{T}^{- f}_{\ \, \tilde  g}, \mathbf{Y}_1] = 0 \,, \qquad [\mathbf{T}^{- f}_{\ \, \tilde  g}, \mathbf{Y}_2] = 0\,, \qquad
[\mathbf{T}^{- f}_{\ \, \tilde  g}, \mathbf{Y}^{a \tilde b}] = 2im_0 \,\delta^{\tilde b}_{\tilde g} \epsilon^{af}. \lb{TY}
\eea
The commutation relations between the $\mathrm{U}(1)$ axial generator $\mathbf{A}$ and the central charges have a similar structure.:
\bea
&& [\mathbf{A}, \mathbf{X}_1] = -2i\, \mathbf{Y}_1\,, \quad [\mathbf{A}, \mathbf{X}_2] = -2i\, \mathbf{Y}_2\,, \quad
[\mathbf{A}, \mathbf{X}^{a\tilde b}] = -2i\, \mathbf{Y}^{a\tilde b}\,, \nn
&& [\mathbf{A}, \mathbf{Y}_1] = -2i\, m_0\,, \quad [\mathbf{A}_0, \mathbf{Y}_2] = -2i\, m_0\,, \quad
[\mathbf{A}, \mathbf{Y}^{a\tilde b}] = 0\,. \lb{A0XY}
\eea

Our last remark concerns the $\mathcal{N}{=}\,4$ Galilean algebra
with the diagonal choice (\ref{eqfli1.5}), \p{Z-Om} for the central charge matrix.
Recalling \p{contr-QQ-pm-2} - \p{contr-QQ-+--2}, we observe that in this case $\mathcal{N}{=}\,4$ Galilean algebra
(with suitable restriction of R-symmetry algebras taken into account) reduces to the sum of two $\mathcal{N}{=}\,2$ Galilean superalgebras
spanned by the supercharge pairs $(\mathbf{Q}^{a}_{\alpha}, \mathbf{S}^{a}_{\alpha}),
(\mathbf{Q}^{\tilde a}_{\alpha}, \mathbf{S}^{\tilde a}_{\alpha})$,
with common  generators $\mathbf{H}$ and $\mathbf{P}_i$.  The only way to avoid such a splitting is
to switch on the off-diagonal central charges as in  \p{VectZdop}. If we consider an extended  $\mathcal{N}{=}\,4$ Galilean algebra,
with the R-symmetry generators $T^{+ A}_{\;B}$ included, the  $\mathcal{N}{=}\,2$ subsectors in  \p{contr-QQ-pm-2} - \p{contr-QQ-+--2}
will be intertwined  by the generators $T^{+ a}_{\;\tilde b}$, e.g., $[T^{+ \tilde a}_{\;b}, \mathbf{Q}^c_\alpha] = \delta^c_b
 \mathbf{Q}^{\tilde a}_\alpha\,, \;[T^{+ \tilde a}_{\;b}, \mathbf{Q}^{\tilde c}_\alpha] =
 \epsilon^{\tilde a\,\tilde c}\mathbf{Q}_{b\,\alpha}\,$. In this case, the splitting into two  $\mathcal{N}{=}\,2$ algebras
arises only when we eliminate the generators $\mathbf{T}^{+}{}^a_{\tilde b}$
from the R-symmetry algebra.

To avoid a possible confusion, note that the R-symmetry
is described by the group of outer automorphisms of superalgebras
and its generators
do not appear in the r.h.s. of the (anti)commutators (distinctly  from
central charges). Therefore, when constructing the specific models,  we can restrict the R-symmetry group to some of its subgroup.
The maximal R-symmetry group $\mathrm{USp}(4) \sim \mathrm{O}(5)$ can be ensured in two distinct cases:
either for the choice  \p{Z-1} with ${\cal Z}$
being $\mathrm{USp}(4) \sim \mathrm{O}(5)$ invariant (the same if ${\cal Z}$ is an operator or a number), or for the generic choice \p{Z-matr}, with
$\mathbf{X}_1 - \mathbf{X}_2, \mathbf{X}^{a\tilde b}$ and $\mathbf{Y}_1 - \mathbf{Y}_2, \mathbf{Y}^{a\tilde b}$ forming two
independent $\mathrm{O}(5)$ vectors (see (\ref{T+X})), and  with two $\mathrm{O}(5)$ singlets $(\mathbf{X}_1 + \mathbf{X}_2)$, $(\mathbf{Y}_1 + \mathbf{Y}_2)$
accommodating the remaining two central charges.

In the second case one has an additional freedom to eliminate, without breaking  $\mathrm{O}(5)$  covariance,
either all $\mathbf{Y}$ central charges or
all $\mathbf{X}$ central charges, and further choose, e.g., $\mathbf{X}_2 + \mathbf{X}_1 =0$ or $\mathbf{Y}_2 +\mathbf{Y}_1 =0\,$.
As was already mentioned, with the general option \p{Z-matr} the choice of numerical central charges necessarily breaks
$\mathrm{O}(5)$ R-symmetry down to $\mathrm{O}(3)\,$.
\\

\noindent {\bf iv) Hermitian basis}
\\

One can alternatively formulate NR  $\mathcal{N}{=}\,4,$ $d=3$ superalgebra by using NR contraction of Hermitian pairs of supercharges
which are self-conjugate with respect to Hermitian conjugation (see  (\ref{Q-pm-H})).
We define the set of unconstrained independent supercharges spanning $\mathcal{N}{=}\,4$ Galilean Hermitian superalgebra as follows
\begin{equation}\label{Herm-QS}
\begin{array}{llll}
\mathbf{Q}_{\alpha}\equiv\mathbf{Q}^{1}_{\alpha}, \qquad
&\bar{\mathbf{Q}}_{\dot\alpha}\equiv\bar{\mathbf{Q}}_{\dot\alpha 1}\,,\qquad
&\tilde{\mathbf{Q}}_{\alpha}\equiv\mathbf{Q}^{\tilde 1}_{\alpha}, \qquad
&\bar{\tilde{\mathbf{Q}}}_{\dot\alpha}\equiv \bar{\mathbf{Q}}_{\dot\alpha \tilde 1}\,,\\ [6pt]
\mathbf{S}_{\alpha}\equiv\mathbf{S}^{1}_{\alpha}, \qquad
&\bar{\mathbf{S}}_{\dot\alpha}\equiv\bar{\mathbf{S}}_{\dot\alpha 1}\,,\qquad
&\tilde{\mathbf{S}}_{\alpha}\equiv\mathbf{S}^{\tilde 1}_{\alpha}, \qquad
&\bar{\tilde{\mathbf{S}}}_{\dot\alpha}\equiv \bar{\mathbf{S}}_{\dot\alpha \tilde 1}\,.
\end{array}
\end{equation}
One gets
\begin{equation}\label{contr-QQ-pm-2a}
\begin{array}{rcl}
\{ \mathbf{Q}_{\alpha} , \bar{\mathbf{Q}}_{\dot\beta} \}&=&2\,
\delta_{{\alpha} \dot{\beta}}\, \Big( \mathbf{H} + \mathbf{X}_{1}\Big)\,, \\  [6pt]
\{ \tilde{\mathbf{Q}}_{\alpha}, \bar{\tilde{\mathbf{Q}}}_{\dot\beta} \}&=&2\,
\delta_{{\alpha} \dot{\beta}}\, \Big( \mathbf{H} + \mathbf{X}_{2}\Big)\,,\\  [6pt]
\{ \mathbf{Q}_{\alpha} , \tilde{\mathbf{Q}}_{\beta} \} &=& 2 \epsilon_{\alpha\beta} \mathbf{X}^{1 \tilde 1}\,, \qquad
\{ \bar{\mathbf{Q}}_{\dot\alpha} , \bar{\tilde{\mathbf{Q}}}_{\dot\beta} \} \ = \ 2 \epsilon_{\dot\alpha\dot\beta} \mathbf{X}^{2\tilde 2}\,,\\  [6pt]
\{ \mathbf{Q}_{\alpha} , \bar{\tilde{\mathbf{Q}}}_{\dot\beta} \} &=& - 2 \delta_{\alpha\dot\beta} \mathbf{X}^{1 \tilde 2}\,, \qquad
\{\tilde{\mathbf{Q}}_{\alpha} , \bar{\mathbf{Q}}_{\dot\beta} \} \ = \ 2 \delta_{\alpha\dot\beta} \mathbf{X}^{2\tilde 1}\,,
\end{array}
\end{equation}
\begin{equation}\label{contr-SS-pm-2a}
\begin{array}{rcl}
\{ \mathbf{S}_{\alpha} , \bar{\mathbf{S}}_{\dot\beta} \}&=&4\,
\delta_{{\alpha} \dot{\beta}}\, m_0\,, \\  [6pt]
\{ \tilde{\mathbf{S}}_{\alpha}, \bar{\tilde{\mathbf{S}}}_{\dot\beta} \}&=&4\,
\delta_{{\alpha} \dot{\beta}}\, m_0\,,\\  [6pt]
\{ \mathbf{S}_{\alpha}, \bar{\tilde{\mathbf{S}}}_{\dot\beta} \}&=& \{ \tilde{\mathbf{S}}_{\alpha}, \bar{\mathbf{S}}_{\dot\beta} \}\ =\ 0\,,
\end{array}
\end{equation}
\begin{equation}\label{contr-QQ-+--2a}
\begin{array}{rcl}
\{ \mathbf{Q}_{\alpha}, \bar{\mathbf{S}}_{\dot\beta} \}&=&2\,
\Big( (\sigma_i )_{{\alpha} \dot{\beta}}\mathbf{P}_i - i\delta_{{\alpha} {\beta}}\,\mathbf{Y}_{1}\Big)\,, \\  [6pt]
\{ \tilde{\mathbf{S}}_{\alpha}, \bar{\tilde{\mathbf{Q}}}_{\dot\beta} \}&=&-2\,
\Big( (\sigma_i )_{{\alpha} {\dot\beta}}\mathbf{P}_i -
i\delta_{{\alpha} {\dot\beta}}\,\mathbf{Y}_{2}\Big)\,, \\  [6pt]
\{ \mathbf{Q}_{\alpha} , \tilde{\mathbf{S}}_{\beta} \} &=& 2 i\epsilon_{\alpha\beta} \mathbf{Y}^{1 \tilde 1}\,, \qquad
\{ \bar{\mathbf{Q}}_{\dot\alpha} , \bar{\tilde{\mathbf{S}}}_{\dot\beta} \} \ = \ -2 i\epsilon_{\dot\alpha\dot\beta}
\mathbf{Y}^{2\tilde 2}\,, \\  [6pt]
\{ \mathbf{Q}_{\alpha} , \bar{\tilde{\mathbf{S}}}_{\dot\beta} \} &=& 2 i\delta_{\alpha\dot\beta} \mathbf{Y}^{1 \tilde 2}\,, \qquad
\{\tilde{\mathbf{S}}_{\alpha} , \bar{\mathbf{Q}}_{\dot\beta} \} \ = \ 2 i \delta_{\alpha\dot\beta} \mathbf{Y}^{2\tilde 1}\,.
\end{array}
\end{equation}

The Hermitian form of  $\mathcal{N}{=}\,4,$ $d{=}\,3$ Galilean superalgebra (\ref{contr-QQ-pm-2a})-(\ref{contr-QQ-+--2a}) permits
to obtain
the generalized positivity conditions for the Hamiltonian $\mathbf{H}$. From first two formulas in (\ref{contr-QQ-pm-2a})
one derives that for any normalized state $|\Psi\rangle$ belonging to the Hilbert space of physical states of the models the following conditions hold
\begin{equation}\label{pos-HX}
\begin{array}{lllll}
{\displaystyle \frac14\,\sum_\alpha}\langle\Psi|\,\{ \mathbf{Q}_{\alpha} , \bar{\mathbf{Q}}_{\dot\alpha} \}\,|\Psi\rangle&=&
{\displaystyle \frac14\,\sum_{|\,\Phi\rangle}\sum_\alpha}\,\Big|\langle\Psi|\,\mathbf{Q}_{\alpha} \,|\,\Phi\rangle\Big|^2&=&
\langle\Psi| \left( \mathbf{H} + \mathbf{X}_{1}\right)|\Psi\rangle \ \ \geq \ 0\,, \\  [6pt]
{\displaystyle \frac14\,\sum_\alpha}\langle\Psi|\,\{ \tilde{\mathbf{Q}}_{\alpha} , \bar{\tilde{\mathbf{Q}}}_{\dot\alpha} \}\,|\Psi\rangle&=&
{\displaystyle \frac14\,\sum_{|\,\Phi\rangle}\sum_\alpha}\,\Big|\langle\Psi|\,\tilde{\mathbf{Q}}_{\alpha} \,|\,\Phi\rangle\Big|^2&=&
\langle\Psi| \left( \mathbf{H} + \mathbf{X}_{2}\right)|\Psi\rangle \ \ \geq \ 0\,.
\end{array}
\end{equation}
In dynamical models (like those of Sect.\,4) the central charges $\mathbf{X}_{1}$, $\mathbf{X}_{2}$
 are represented on the normalized states ${|\Psi\rangle}$ by the mass-like parameters $m_{1}$, $m_{2}$, so from \p{pos-HX}
one gets the lower bound on the energy values $E_\Psi={\langle\Psi|} \, \mathbf{H} \,{|\Psi\rangle}\geq -{\mathrm{min}}(m_{1},m_{2})$.

\setcounter{equation}{0}
\section{Nonlinear realizations of  $\mathcal{N}{=}\,4,$ $d{=}\,3$ Galilean supersymmetries}

\quad\, In the nonlinear realization of $\mathcal{N}{=}\,4,$ $d{=}\,3$ Galilean supersymmetries we will assume that
the linearization subgroup  $H$ involves the 3-dimensional space rotations
generators $\mathbf{J}^{ik}$, the internal symmetry $\mathrm{USp}(4)$ generators $\mathbf{T}^{+ A}_{\ \, B}$ and the abelian generator
$\mathbf{A}_0\,$. All other generators
are placed in the coset $G=SG(3;4|12)/H$. Some of the parameters belonging to $G$ can be relocated into the linearization subgroup $H$
just by nullifying the respective coset parameters. The coset element $G$ can be written explicitly as
\bea
G = G_{(1)}\, G_{(2)} \,G_{(3)} \,G_{(4)} \,G_{(5)} \,G_{(6)} \equiv \hat{G} \,G_{(6)} \,,
\label{factor-G}
\eea
where
\bea
&&G_{(1)} = \exp i\{t \mathbf{H} + x^i \mathbf{P}^i\}\,, \quad G_{(2)}  = \exp i\{\xi_a^\alpha \mathbf{Q}^a_\alpha
+  \xi_{\tilde a}^\alpha \mathbf{Q}^{\tilde a}_\alpha\}\,,
\quad G_{(3)} =  \exp i\{\theta^\alpha_{a}\mathbf{S}^{a}_\alpha + \theta^\alpha_{\tilde a}\mathbf{S}^{\tilde a}_\alpha \}\,, \nn
&& G_{(4)} =  \exp i\{k^i \mathbf{B}^i\}\,, \quad G_{(5)} = \exp i\{ s \mathbf{M} +  h_{1}\mathbf{X}_1 + h_{2}\mathbf{X}_2 +   h_{a\tilde b}\mathbf{X}^{a \tilde b}
+ f_{1}\mathbf{Y}_1 + f_{2}\mathbf{Y}_2 +   f_{a\tilde b}\mathbf{Y}^{a \tilde b}\}\,, \nn
&& G_{(6)} = \exp i\{ u_{a}^b \mathbf{T}^{- a}_{\ \, b}
+  u_{\tilde a}^{\tilde b} \mathbf{T}^{- \tilde a}_{\ \, \tilde b} + u_{a}^{\tilde b} \mathbf{T}^{- a}_{\ \, \tilde b}\}.
\label{factor}
\eea
The  factors $G_{(1)}$, $G_{(4)}$ are parametrized by $d{=}\,3$ Galilei group parameters (see (\ref{eqfli1.1})),
$G_{(5)}$ by the central charge parameters dual to the Galilean central charges,
$G_{(6)}$ represents  the abelian 5-dimensional coset $\mathrm{IUSp}(4)/\mathrm{USp}(4)$ and
$G_{(2)}$, $G_{(3)}$ collect the parameters of the fermionic (odd) sector of $\mathrm{SG}(3;4|12)$.

The odd generators satisfy the symplectic-Majorana conditions (see also (\ref{MW-cond}))
\begin{equation}\label{MW-cond-G}
(\mathbf{Q}^{a}_{{\alpha}})^\dagger = \epsilon^{\alpha\beta}\epsilon_{ab}\, \mathbf{Q}^{b}_{\beta} \,,\quad
(\mathbf{Q}^{\tilde a}_{{\alpha}})^\dagger = \epsilon^{\alpha\beta}\epsilon_{\tilde a\tilde b}\, \mathbf{Q}^{\tilde b}_{\beta} \,,\qquad
(\mathbf{S}^{a}_{{\alpha}})^\dagger = -\epsilon^{\alpha\beta}\epsilon_{ab}\, \mathbf{S}^{b}_{\beta} \,,\quad
(\mathbf{S}^{\tilde a}_{{\alpha}})^\dagger = -\epsilon^{\alpha\beta}\epsilon_{\tilde a\tilde b}\, \mathbf{S}^{\tilde b}_{\beta} \,.
\end{equation}
The Grassmann coordinates dual to these odd generators satisfy similar pseudo-reality conditions
\begin{equation}\label{MW-cond-G-c}
(\xi_{a}^{{\alpha}})^* = \epsilon_{\alpha\beta}\epsilon^{ab}\, \xi_{b}^{\beta} \,,\quad
(\xi_{\tilde a}^{{\alpha}})^* = \epsilon_{\alpha\beta}\epsilon^{\tilde a\tilde b}\, \xi_{\tilde b}^{\beta} \,,\qquad
(\theta_{a}^{{\alpha}})^* = -\epsilon_{\alpha\beta}\epsilon^{ab}\, \theta_{b}^{\beta} \,,\quad
(\theta_{\tilde a}^{{\alpha}})^* = -\epsilon_{\alpha\beta}\epsilon^{\tilde a\tilde b}\, \theta_{\tilde b}^{\beta} \,.
\end{equation}
Being dual to the relations (\ref{XY-pm-2}), the reality conditions for the tensorial central charge coordinates read
\begin{equation}\label{hf-pm-2}
\left(h_{a \tilde b}\right)^*   =
\epsilon^{ac} \epsilon^{\tilde b\tilde d}\,h_{c \tilde d}\,, \qquad
\left(f_{a \tilde b}\right)^*  =
\epsilon^{ac} \epsilon^{\tilde b\tilde d}\,f_{c \tilde d}\,.
\end{equation}

The full set of the left-covariant MC one-forms is given by
\be
G^{-1} d G = G_{(6)}^{-1}( \hat{G}^{-1}d\hat{G}) G_{(6)} + G_{(6)}^{-1} d G_{(6)}, \lb{hatnehat}
\ee
where
\bea
\hat{G}^{-1}d\hat{G} &=&  G_{(4)}^{-1}\, G_{(3)}^{-1}\, G_{(2)}^{-1}\,\Big(G_{(1)}^{-1} d  G_{(1)} \Big)\, G_{(2)} \, G_{(3)}\, G_{(4)}
+  G_{(4)}^{-1}\, G_{(3)}^{-1} \,\Big( G_{(2)}^{-1} d  G_{(2)} \Big)\, G_{(3)}\, G_{(4)}  \nn
&&
+\, G_{(4)}^{-1}\, \Big(G_{(3)}^{-1} d  G_{(3)}\Big)\,G_{(4)} + G_{(4)}^{-1} d  G_{(4)} + G_{(5)}^{-1} d  G_{(5)}   \nn
&:=& (I) + (II) + (III) + (IV) + (V)\,.\lb{HaT}
\eea
A straightforward calculation yields
\bea
G_{(1)}^{-1} d  G_{(1)}  \!&\!=\!&\! i\, \Big( dt \mathbf{H} + dx_i\mathbf{P}_i \Big)\,, \nn
G_{(2)}^{-1} d  G_{(2)} \!&\!=\!&\! i\,\Big(d\xi^\alpha_a \mathbf{Q}^a_\alpha
 + d\xi^\alpha_{\tilde a} \mathbf{Q}^{\tilde a}_\alpha \Big) - \Big(\xi^\alpha_a  d\xi^a_\alpha +
 \xi^\alpha_{\tilde a}  d\xi^{\tilde a}_\alpha \Big)\, \mathbf{H}  \,,\nn
&& -\,  \Big(\xi^\alpha_a  d\xi^a_\alpha\Big)\,\mathbf{X}_1 -
 \Big(\xi^\alpha_{\tilde a}  d\xi^{\tilde a}_\alpha\Big)\, \mathbf{X}_2 - \Big(\xi^\alpha_a  d\xi_{\alpha \tilde b}  - \xi^\alpha_{\tilde b}
 d\xi_{\alpha a}\Big)\, \mathbf{X}^{a \tilde b}\,, \nn
(III) \ = \ G_{(3)}^{-1} d  G_{(3)}
\!&\!=\!&\! i\,\Big(d\theta^\alpha_a \mathbf{S}^a_\alpha + d\theta^\alpha_{\tilde a} \mathbf{S}^{\tilde a}_\alpha \Big) +
2\, \Big(\theta^\alpha_a  d\theta^a_\alpha +
 \theta^\alpha_{\tilde a}  d\theta^{\tilde a}_\alpha \Big)\, \mathbf{M} \,, \nn
(IV) \ = \ G_{(4)}^{-1} d  G_{(4)} \!&\!=\!&\! i\, dk^i \mathbf{B}^i\,, \nn
(V) \ = \  G_{(5)}^{-1} d  G_{(5)} \!&\!=\!&\! i\left(ds \mathbf{M} +  dh_{1}\mathbf{X}_1 + dh_{2}\mathbf{X}_2 +   dh_{a\tilde b}\mathbf{X}^{a \tilde b}
+ df_{1}\mathbf{Y}_1 + df_{2}\mathbf{Y}_2 +   df_{a\tilde b}\mathbf{Y}^{a \tilde b} \right).
\eea
The remaining part of \p{HaT} is as follows
\bea
(I) &=& i \left[ dt  \mathbf{H} + \big( dx_i + k_i dt\big)  \mathbf{P}_i +
\big(k_i dx_i + \frac12\, k^2 dt \big) \mathbf{M}\right] , \lb{IIandIV}
\eea
\bea
(II) &=&  G_{(2)}^{-1} d  G_{(2)} - 2 \left[ (\sigma_i)_{\alpha\beta}\big( \theta^{b \alpha} d\xi^\beta_b
+\theta^{\tilde b \alpha} d\xi^\beta_{\tilde b}\big) +\frac12\, k_i  \big(\xi^{\alpha}_a d\xi^a_\alpha +
\xi^\alpha_{\tilde a} d\xi^{\tilde a}_\alpha\big) \right]  \mathbf{P}_i \nn
&& +\,\frac{i}{2}\, k_i(\sigma_i)_{\alpha}{}^{\beta}\big(d\xi^\alpha_a \mathbf{S}^a_\beta
+ d\xi^\alpha_{\tilde a} \mathbf{S}^{\tilde a}_\beta \big) \nn
&& -\, \left[ 2 k_i(\sigma_i)_{\alpha \beta}\big( \theta^{b \alpha} d\xi^\beta_b
+\theta^{\tilde b}_\alpha d\xi^\beta_{\tilde b}\big) + \frac{1}{2}\, k^2 \big(\xi^{\alpha}_a d\xi^a_\alpha +
\xi^\alpha_{\tilde a} d\xi^{\tilde a}_\alpha\big) \right] \mathbf{M} \nn
&& + \,2i (\theta^{\alpha b}d\xi_{\alpha b})\mathbf{Y}_1 + 2i (\theta^{\alpha \tilde b}d\xi_{\alpha \tilde b})\mathbf{Y}_2
- 2i \big( \theta^\alpha_a d\xi_{\alpha \tilde b}-  \theta^\alpha_{\tilde b} d\xi_{\alpha a}\big) \mathbf{Y}^{a\tilde b}\,.
\lb{III}
\eea
We can write the formula \p{HaT} in the following way
\bea
\hat{G}^{-1}d \hat{G} := i \sum_K \hat{\omega}_{(K)} T_{(K)}\,,
\eea
where $T_{(K)}$ stand for all coset $G$ generators,
and $\hat{\omega}_{(K)}$ denote the corresponding MC one-forms. We obtain
\bea
&&\hat{\omega}_{(Q) \;a}^\alpha = d\xi^\alpha_a\,, \quad \hat{\omega}_{(Q) \;{\tilde a}}^\alpha =  d\xi^\alpha_{\tilde a}\,, \quad
\hat{\omega}_{(S) \;a}^\alpha = d\theta^\alpha_a + \frac12\,  k_i(\sigma_i)_{\beta}{}^{\alpha} d\xi^\beta_a\,, \quad
\hat{\omega}_{(S) \;\tilde a}^\alpha = d\theta^\alpha_{\tilde a} + \frac12\,  k_i(\sigma_i)_{\beta}{}^{\alpha} d\xi^\beta_{\tilde a}, \nn
&& \hat\omega_{(H)} = dt + i\big(\xi^\alpha_a  d\xi^a_\alpha +
 \xi^\alpha_{\tilde a}  d\xi^{\tilde a}_\alpha\big)\,, \qquad \hat\omega_{(B)\;i} = d k_i\,, \nn
&&\hat\omega_{(P)\;i} = \left[ dx_i  +2i (\sigma_i)_{\alpha\beta}
 \big(\theta^{b \alpha} d\xi^\beta_b
+\theta^{\tilde b \alpha} d\xi^\beta_{\tilde b}\big)\right] + k_i\, \hat\omega_{(H)}\,, \nn
&& \hat\omega_{(M)} = ds + k_i \, \hat\omega_{(P)\;i} - \frac12\, k^2 \hat\omega_{(H)}  - 2 i\big(\theta^\alpha_a  d\theta^a_\alpha +
 \theta^\alpha_{\tilde a}  d\theta^{\tilde a}_\alpha \big) \,,  \nn
&& \hat\omega_{(X)\,1} = dh_1 + i \xi^\alpha_a  d\xi^a_\alpha\,, \quad  \hat\omega_{(X)\,2} = dh_2 + i \xi^\alpha_{\tilde a}
d\xi^{\tilde a}_\alpha\,, \quad  \hat\omega_{(X)\,a\tilde b} = d h_{a \tilde b} + i
\big( \xi^\alpha_a  d\xi_{\alpha \tilde b}  - \xi^\alpha_{\tilde b}
 d\xi_{\alpha a}\big)\,, \nn
&&  \hat\omega_{(Y)\,1} = df_1 + 2\theta^{\alpha a}  d\xi_{a \alpha}\,, \; \hat\omega_{(Y)\,2} = df_2 + 2 \theta^{\alpha \tilde a}
d\xi_{\alpha \tilde a}\,, \;  \hat\omega_{(Y)\,a\tilde b} = d f_{a \tilde b} -2
\big( \theta^\alpha_a  d\xi_{\alpha \tilde b}  - \theta^\alpha_{\tilde b}
 d\xi_{\alpha a}\big)\,,
\lb{Car-f-hat}
\eea
where $k^2:=k_i k_i$.
The MC one-forms describing the whole coset $G$ are defined as follows
\bea
{G}^{-1}d {G} := i \sum_K {\omega}_{(K)} T_{(K)}
\eea
and can be calculated by the formulas \p{factor} and \p{hatnehat}. We observe that
\be
G_{(6)}^{-1} d G_{(6)} = i\Big(  du_{a}^b \mathbf{T}^{- a}_{\ \, b}
+  du_{\tilde a}^{\tilde b} \mathbf{T}^{- \tilde a}_{\ \, \tilde b} + du_{a}^{\tilde b} \mathbf{T}^{- a}_{\ \, \tilde b} \Big)
\ee
and, further,
\bea
G_{(6)}^{-1} \mathbf{Q}^a_\alpha G_{(6)} &=& \mathbf{Q}^a_\alpha+\frac{i}{2} \big( u^{\tilde b}_{\tilde b} - u^d_d\big)  \mathbf{S}^a_\alpha
 - iu^{a\tilde b}  \mathbf{S}_{\alpha \tilde b}\,, \qquad
G_{(6)}^{-1} \mathbf{Q}^{\tilde a}_\alpha G_{(6)} \ =\ \mathbf{Q}^{\tilde a}_\alpha -\frac{i}{2} \big( u^{\tilde b}_{\tilde b} - u^d_d\big)  \mathbf{S}^{\tilde a}_\alpha
 - iu^{\tilde a}_{b}  \mathbf{S}_{\alpha}^{b}\,, \nn
G_{(6)}^{-1} \mathbf{Y}_1 G_{(6)}  &=&   \mathbf{Y}_1 + \big(u^a_a - u^{\tilde a}_{\tilde a}\big)\mathbf{M}\,, \qquad
G_{(6)}^{-1} \mathbf{Y}_2 G_{(6)} \ = \ \mathbf{Y}_2 -  \big(u^a_a - u^{\tilde a}_{\tilde a}\big)\mathbf{M}\,, \nn
G_{(6)}^{-1} \mathbf{Y}^{a\tilde b} G_{(6)} &=& \mathbf{Y}^{a\tilde b} + 2  u^{a\tilde b}\mathbf{M}\,, \nn
G_{(6)}^{-1} \mathbf{X}_1 G_{(6)}  &=&   \mathbf{X}_1 + \big(u^a_a - u^{\tilde a}_{\tilde a}\big)\mathbf{Y}_1 + u_{a\tilde c}
\mathbf{Y}^{a \tilde c} + \frac12 \left[\big(u^a_a - u^{\tilde a}_{\tilde a}\big)^2 + 2u_{a\tilde c}u^{a \tilde c}\right] \mathbf{M}\,, \nn
G_{(6)}^{-1} \mathbf{X}_2 G_{(6)}  &=&  \mathbf{X}_2 - \big(u^a_a - u^{\tilde a}_{\tilde a}\big)\mathbf{Y}_2 + u_{a\tilde c}
\mathbf{Y}^{a \tilde c}  + \frac12 \left[\big(u^a_a - u^{\tilde a}_{\tilde a}\big)^2 + 2u_{a\tilde c}u^{a \tilde c}\right]\mathbf{M} \,, \nn
G_{(6)}^{-1} \mathbf{X}^{a\tilde b} G_{(6)} &=& \mathbf{X}^{a\tilde b} + 2 u^a_d \mathbf{Y}^{d\tilde b} +
2 u^{\tilde b}_{\tilde d} \mathbf{Y}^{a\tilde d} - 2\big( u^c_c + u^{\tilde c}_{\tilde c}\big)\mathbf{Y}^{a\tilde b}  +
u^{a\tilde b}\big(\mathbf{Y}_1 + \mathbf{Y}_2 \big)  \nn
&& -\, 2 \left[\big( u^c_c + u^{\tilde c}_{\tilde c}\big) u^{a\tilde b}
-  u^{a\tilde d} u^{\tilde b}_{\tilde d} -  u^a_d u^{d \tilde b}\right]\mathbf{M} \,.\lb{G6XY}
\eea
We see, in particular, that
\begin{equation}\label{H-hH}
\omega_{(H)} =\hat\omega_{(H)} \,, \qquad  \omega_{(P)\;i} =\hat\omega_{(P)\;i} \,.
\end{equation}

Let us find supersymmetry transformations of the coset coordinates.
For this purpose we will use the well known formula
$iG^{-1}(\varepsilon\cdot T)G=
G^{-1}\delta G +\delta h$, where
$T$ denotes the collection of coset generators and
$\delta h$ defines induced transformations of the
stability subgroup $h_{ind}=1+\delta h$ (see, e.g., \cite{Zum}).

Supersymmetry transformations generated by the left action of generators $\mathbf{Q}^a_\alpha$, $\mathbf{Q}^{\tilde a}_\alpha\,$
\begin{equation}\label{Q-tr-gr}
\exp i\{\varepsilon_a^\alpha \mathbf{Q}^a_\alpha
+  \varepsilon_{\tilde a}^\alpha \mathbf{Q}^{\tilde a}_\alpha\}\,G=G^\prime h\,,
\end{equation}
where $\varepsilon_a^\alpha$, $\varepsilon_{\tilde a}^\alpha$ are odd constant parameters, lead to the
following transformations of the coset coordinates:
\bea
&& \delta_\varepsilon\xi_a^\alpha=\varepsilon_a^\alpha \,, \qquad
\delta_\varepsilon\xi_{\tilde a}^\alpha=\varepsilon_{\tilde a}^\alpha \,, \qquad
\delta_\varepsilon t=-i\left(\varepsilon_a^\alpha \xi^a_\alpha + \varepsilon_{\tilde a}^\alpha \xi^{\tilde a}_\alpha \right),\nn
&& \delta_\varepsilon h_1=-i \varepsilon_a^\alpha \xi^a_\alpha \,, \qquad
\delta_\varepsilon h_2=-i \varepsilon_{\tilde a}^\alpha \xi^{\tilde a}_\alpha  \,, \qquad
\delta_\varepsilon h_{a\tilde b}=-i\left(\varepsilon_a^\alpha \xi_{\alpha\tilde b} - \varepsilon_{\tilde b}^\alpha \xi_{\alpha a} \right)\,,\nn
&& \delta_\varepsilon x_i=0\,, \qquad \delta_\varepsilon k_i=0\,,  \nn
&& \delta_\varepsilon\theta_a^\alpha=0 \,, \qquad
\delta_\varepsilon\theta_{\tilde a}^\alpha=0 \,, \qquad
\delta_\varepsilon s=0 \,,  \nn
&& \delta_\varepsilon f_1=0 \,, \qquad
\delta_\varepsilon f_2=0  \,, \qquad
\delta_\varepsilon f_{a\tilde b}=0\,, \nn
&& \delta_\varepsilon u_{a}^{b}=0 \,, \qquad
\delta_\varepsilon u_{\tilde a}^{\tilde b}=0  \,, \qquad
\delta_\varepsilon u_{a}^{\tilde b}=0\,.\lb{Q-tr-comp}
\eea
The second half of the odd left shifts, those generated by $\mathbf{S}^a_\alpha$, $\mathbf{S}^{\tilde a}_\alpha$,
\begin{equation}\label{S-tr-gr}
\exp i\{\eta^\alpha_{a}\mathbf{S}^{a}_\alpha + \eta^\alpha_{\tilde a}\mathbf{S}^{\tilde a}_\alpha \}
\,G=G^\prime h\,,
\end{equation}
lead to the transformations
\bea
&& \delta_\eta\xi_a^\alpha=0 \,, \qquad
\delta_\eta\xi_{\tilde a}^\alpha=0 \,, \qquad
\delta_\eta t=0,\nn
&& \delta_\eta h_1=0 \,, \qquad
\delta_\eta h_2=0  \,, \qquad
\delta_\eta h_{a\tilde b}=0\,,\nn
&& \delta_\eta x_i=2i\left(\eta_a^\alpha \xi^{\beta a} + \eta_{\tilde a}^\alpha \xi^{\beta \tilde a} \right)(\sigma_i)_{\alpha\beta}\,, \qquad
\delta_\eta k_i=0\,,  \nn
&& \delta_\eta\theta_a^\alpha=\eta_a^\alpha \,, \qquad
\delta_\eta\theta_{\tilde a}^\alpha=\eta_{\tilde a}^\alpha \,, \qquad
\delta_\eta s=2i\left(\eta_a^\alpha \theta^a_\alpha + \eta_{\tilde a}^\alpha \theta^{\tilde a}_\alpha \right) \,,  \nn
&& \delta_\eta f_1= 2 \eta_a^\alpha \xi^a_\alpha \,, \qquad
\delta_\eta f_2= 2 \eta_{\tilde a}^\alpha \xi^{\tilde a}_\alpha  \,, \qquad
\delta_\eta f_{a\tilde b}=2\left(\eta_a^\alpha \xi_{\alpha\tilde b} - \eta_{\tilde b}^\alpha \xi_{\alpha a} \right), \nn
&& \delta_\eta u_{a}^{b}=0 \,, \qquad
\delta_\eta u_{\tilde a}^{\tilde b}=0  \,, \qquad
\delta_\eta u_{a}^{\tilde b}=0\,,
\lb{S-tr-comp}
\eea
where $\eta_a^\alpha$, $\eta_{\tilde a}^\alpha$ are the appropriate odd parameters.  It follows
from (\ref{Q-tr-comp}) and (\ref{S-tr-comp}) that the three-vector $k_i$ and ``harmonic variables'' $u$
are inert under all supersymmetry transformations.

{}From the form of the supersymmetry transformations (\ref{Q-tr-comp}) follows that the set of coordinates
($t$; $\xi_{a}^\alpha$, $\xi_{\tilde a}^\alpha$; $h_{1}$, $h_{2}$, $h_{a\tilde b}$) is closed under the action of
$\mathbf{Q}$ - supersymmetry,  while this supersymmetry does not act on the remaining coordinates ($x_i$; $\theta_{a}^\alpha$, $\theta_{\tilde a}^\alpha$; $s$, $f_{1}$, $f_{2}$, $f_{a\tilde b}$).
Alternatively, $\mathbf{S}$ - supersymmetry (\ref{S-tr-comp})
leaves inert the subset ($t$; $\xi_{a}^\alpha$, $\xi_{\tilde a}^\alpha$; $h_{1}$, $h_{2}$, $h_{a\tilde b}$)
and transforms the remaining coordinates ($x_i$; $\theta_{a}^\alpha$, $\theta_{\tilde a}^\alpha$; $s$, $f_{1}$, $f_{2}$, $f_{a\tilde b}$)
(however with dependence of $\delta_\eta x_i$, $\delta_\eta f_1$, $\delta_\eta f_2$,
$\delta_\eta f_{a\tilde b}$ on $\xi$-variables). This split of the full set of coset parameters into
two subsets, each closed under the action of one half of the supersymmetries and inert under another half, is due to the choice of coset parametrization
(\ref{factor-G}), (\ref{factor}) with the particular order of the factors $G_2$ and $G_3$, $G=\ldots G_2 G_3 \ldots\,$.
In \cite{fli17}, there was used another parametrization, $G=\ldots G_3 G_2 \ldots$, and the
separation of $\mathbf{Q}$ - and $\mathbf{S}$ - transformations into two sectors could not be seen.

The closure of the transformations  (\ref{Q-tr-comp}) and  (\ref{S-tr-comp}) generates all the bosonic transformations of $G$
which do not belong to the stability subgroup $H$. The transformations of subgroup $H$ are realized as some linear homogeneous maps of
the coset fields and MC 1-forms. The abelian generators $\mathbf{T}^-{}^A_B$ do not appear in the closure of fermionic generators,
so the left shifts by these generators should be considered separately. The corresponding transformations of the coset parameters
can be found explicitly, using the formulas  \p{G6XY}. The coset parameters $u^A_B$ are changed only by the pure shifts. Actually, in this
paper we will not make use of these $\mathbf{T}^-{}^A_B$ transformations.

We also observe that all MC forms $\hat{\omega}_{(K)}$ (see (\ref{Car-f-hat})) transform linearly
under $H$ transformations and are inert with respect to the odd transformations (\ref{Q-tr-comp}) and  (\ref{S-tr-comp}).

\setcounter{equation}{0}
\section{The phase-space formulation of $\mathcal{N}=4,$ $d=3$ Galilean superparticle model and $\kappa$-gauge freedom}

\quad\, Let us describe the mechanical system on the coset $G$ with evolution parameter $\tau$ and with
all parameters of $G$ promoted to the $d{=}\,1$ fields: $t=t(\tau)$, $x_i=x_i(\tau)$,
$k_i=k_i(\tau)$, $\xi^\alpha_{a}=\xi^\alpha_{a}(\tau)$, $\xi^\alpha_{\tilde a}=\xi^\alpha_{\tilde a}(\tau)$,
$\theta^\alpha_{a}=\theta^\alpha_{a}(\tau)$, $\theta^\alpha_{\tilde a}=\theta^\alpha_{\tilde a}(\tau)$, {\it etc}.
We shall deal with the simplified situation, with all internal coordinates $u^A_B$ being suppressed, which means that we transfer
the generators $\mathbf{T}^-{}^A_B$ into the stability subgroup and use the ``truncated'' MC one-forms $\hat\omega_{(K)}\,$.
We will not employ the strict invariance of the superparticle actions under these abelian outer automorphisms, as well under
the full compact R-symmetry $\mathrm{USp}(4)$. Only the symmetries under  some particular subgroups  of the latter, as well as
the $\mathrm{O}(3)$ space symmetry generated by $\mathbf{J}_{ij}$, will be respected.

We can covariantly eliminate the fields $k_i(t)$ by imposing the following algebraic inverse Higgs constraints \cite{IO}
\begin{equation}\label{inH-G}
\omega_{(P)\;i} =\hat\omega_{(P)\;i} =0 \,.
\end{equation}
Taking into account that
\begin{equation}\label{om-P}
\omega_{(P)\;i} =d\tau\left(\pi_i + k_i \pi_0\right) \,,
\end{equation}
where
\begin{equation}\label{pi-i0}
\pi_i :=\dot x_i  +2i (\sigma_i)_{\alpha\beta} \left(\theta^{b \alpha} \dot \xi^\beta_b
+\theta^{\tilde b \alpha} \dot \xi^\beta_{\tilde b}\right) ,\qquad
\pi_0 :=\dot t+ i\left(\xi^\alpha_a  \dot\xi^a_\alpha + \xi^\alpha_{\tilde a}  \dot\xi^{\tilde a}_\alpha\right),
\end{equation}
the constraints (\ref{inH-G}) are solved by
\begin{equation}\label{k-i}
k_i \ =\  -\,\frac{\pi_i}{\pi_0} \,,
\end{equation}
where $\dot x_i=d x_i/d\tau$, $\dot t=d t/d\tau$, etc.

Using  (\ref{inH-G}), one obtains that
\begin{equation}\label{oM-iH}
\hat\omega_{(M)} = d\tau\left[\dot s - \frac{\pi_i \pi_i}{2\pi_0}  -
2 i\big(\theta^\alpha_a  \dot\theta^a_\alpha +  \theta^\alpha_{\tilde a}  \dot\theta^{\tilde a}_\alpha \big) \right] \,.
\end{equation}

\subsection{Simplest bosonic case: Schr\"{o}dinger NR particle}

\quad\, As the instructive step we consider the standard bosonic Schr\"{o}dinger particle.
We recall how to  derive the action of non-relativistic massive particle,
which, after quantization, leads to the non-relativistic Schr\"{o}dinger equation.

Such an action is obtained from the MC one-form (\ref{oM-iH}), which in the pure bosonic case  is given by
\begin{equation}\label{oM-iH-b}
\hat\omega_M = d\tau\left(\dot s - \frac{\dot x_i \dot x_i}{2\dot t} \right) \,.
\end{equation}
Selecting the rest mass as the normalization factor and omitting a total $\tau$ derivative, we obtain
\begin{equation}\label{act-b}
S_0=-m_0\int\hat\omega_{(M)} = \int d\tau\, L_0 = m_0\int d\tau\, \frac{\dot x_i \dot x_i}{2\dot t} \,.
\end{equation}
It leads to NR particle model studied in \cite{fli4,fli17}.

The action (\ref{act-b}) provides the canonical momenta
\begin{equation}\label{mom-b}
p_{\,t}=-\,m_0\,\frac{\dot x_i \dot x_i}{2{\dot t}^2}\,, \qquad p_{\,x}{}_i=m_0\,\frac{\dot x_i}{{\dot t}}
\end{equation}
and the vanishing canonical Hamiltonian:
\begin{equation}\label{H-b}
H_0=p_{\,t}\dot t+ p_{\,x}{}_i \dot x_i-L_0=0\,.
\end{equation}
The expressions (\ref{mom-b}) imply the first-class constraint
defining free NR energy-momentum dispersion relation
called free Schr\"{o}dinger constraint
\begin{equation}\label{constr-b}
p_{\,t}+ \frac{p_{\,x}{}_ip_{\,x}{}_i}{2m_0} \ \approx \  0\,,
\end{equation}
which, after quantization  in the Schr\"{o}dinger realization,
\begin{equation}\label{p-t-x}
p_{\,t}=-\,i\hbar\,\frac{\partial}{\partial t}\,,\qquad p_{\,x}{}_i=-\,i\hbar\,\frac{\partial}{\partial x_i}\,,
\end{equation}
gives the non-relativistic Schr\"{o}dinger equation for a free NR particle of mass $m_0$.

\subsection{The superparticle model with vanishing off-diagonal central charges }

\quad\, As the next step,  we consider the action with the Lagrangian density taken as a linear combination of the
MC one-forms associated with central charges described by Jordanian quasi-diagonal form of the central charge matrix
\begin{equation}\label{act-1}
S_1\ =\ -\ m_0\int\hat\omega_{(M)} \ + \ \int \Big( a\,\hat\omega_{(H)}
+m_1\,\hat\omega_{(X)\,1} + m_2\,\hat\omega_{(X)\,2}
+ \mu_1\,\hat\omega_{(Y)\,1} + \mu_2\,\hat\omega_{(Y)\,2}\Big)\,,
\end{equation}
where $a$, $m_1$, $m_2$, $\mu_1$, $\mu_2$ are real constants. The choice of these parameters specifies
the explicit form of odd constraints, including the first class ones generating local $\kappa$-symmetries.

Using the expressions of the MC forms  (\ref{Car-f-hat}), (\ref{oM-iH}) and omitting total derivative terms,
we get from (\ref{act-1}) the Lagrangian
\bea
L_1&=&m_0\,\frac{\pi_i \pi_i}{2\pi_0} \ + \
2i\,m_0\,\big(\theta^\alpha_a  \dot\theta^a_\alpha \ + \  \theta^\alpha_{\tilde a}  \dot\theta^{\tilde a}_\alpha \big)
\ + \  i\,a\big(\xi^\alpha_a  \dot\xi^a_\alpha \ + \  \xi^\alpha_{\tilde a}  \dot\xi^{\tilde a}_\alpha\big) \nn
&&  \ + \ \ i\,m_1 \, \xi^\alpha_a  \dot\xi^a_\alpha \ + i\, m_2 \, \xi^\alpha_{\tilde a} \dot\xi^{\tilde a}_\alpha
 \ + \ 2 \mu_1\, \theta^{\alpha a}  \dot\xi_{a \alpha} \ + \ 2 \mu_2\, \theta^{\alpha \tilde a} \dot\xi_{\alpha \tilde a}\,,
\label{Lagr-1}
\eea
where $\pi_i=\dot x_i  +2i (\sigma_i)_{\alpha\beta} \left(\theta^{b \alpha} \dot \xi{}^\beta_b
+\theta^{\tilde b \alpha} \dot \xi{}^\beta_{\tilde b}\right)$ and
$\pi_0=\dot t+ i\left(\xi^\alpha_a  \dot\xi^a_\alpha + \xi^\alpha_{\tilde a}  \dot\xi^{\tilde a}_\alpha\right)$
were defined in (\ref{pi-i0}).

Without the loss of generality, the terms  proportional to $a$ in  (\ref{Lagr-1})  can be omitted because they can be re-absorbed by
the redefinition of $m_1$ and  $m_2$. Therefore we will put $a=0$ (see the same condition in \cite{fli17}, assumed, however, for another reason).
Then the Lagrangian  (\ref{Lagr-1}) produces the following bosonic momenta
\begin{equation}\label{mom-b-1}
p_{\,t}=-\,m_0\,\frac{\pi_i \pi_i}{2({\pi_0})^2}\,, \qquad p_{\,x}{}_i=m_0\,\frac{\pi_i}{\pi_0}\,,
\end{equation}
and the fermionic ones
\bea
p_\xi{}_\alpha^{a}&=& 2m_0i\,\frac{\pi_i}{\pi_0}\,(\sigma_i)_{\alpha\beta}\theta^{\beta b}
+i\left(m_1- m_0\,\frac{\pi_i \pi_i}{2(\pi_0)^2}\right)\xi_\alpha^{a}-2\mu_1\theta_\alpha^{a}\,, \nn
p_\xi{}_\alpha^{\tilde a}&=& 2m_0i\,\frac{\pi_i}{\pi_0}\,(\sigma_i)_{\alpha\beta}\theta^{\beta\tilde  b}
+i\left(m_2- m_0\,\frac{\pi_i \pi_i}{2(\pi_0)^2}\right)\xi_\alpha^{\tilde a}-2\mu_2\theta_\alpha^{\tilde a}\,, \nn
p_\theta{}_\alpha^{a}&=& 2m_0 i\,\theta_\alpha^{a}\,, \nn
p_\theta{}_\alpha^{\tilde a}&=& 2m_0 i\,\theta_\alpha^{\tilde a}\,.
\label{mom-f-1}
\eea
In accordance with  (\ref{MW-cond-G}) and (\ref{MW-cond-G-c}), the odd momenta satisfy the symplectic-Majorana conditions
\begin{equation}\label{MW-cond-G-p}
(p_\xi{}^{a}_{{\alpha}})^* = -\epsilon^{\alpha\beta}\epsilon_{ab}\, p_\xi{}^{b}_{\beta} \,,\quad
(p_\xi{}^{\tilde a}_{{\alpha}})^* = -\epsilon^{\alpha\beta}\epsilon_{\tilde a\tilde b}\, p_\xi{}^{\tilde b}_{\beta} \,,\quad
(p_\theta{}^{a}_{{\alpha}})^* = \epsilon^{\alpha\beta}\epsilon_{ab}\, p_\theta{}^{b}_{\beta} \,,\quad
(p_\theta{}^{\tilde a}_{{\alpha}})^* = \epsilon^{\alpha\beta}\epsilon_{\tilde a\tilde b}\, p_\theta{}^{\tilde b}_{\beta} \,.
\end{equation}

The model with the Lagrangian (\ref{Lagr-1}) is invariant under the NR supersymmetry transformations
(\ref{Q-tr-comp}) and (\ref{S-tr-comp}). Note that all terms in (\ref{Lagr-1}), except the first one,
are of the Wess-Zumino type. Noether charges which generate  the supersymmetry transformations  (\ref{Q-tr-comp}) and (\ref{S-tr-comp})
are as follows
\bea
\mathbf{Q}_\alpha^{a}&=& p_\xi{}_\alpha^{a}
+i\left(p_{\,t}+m_1\right)\xi_\alpha^{a} \,, \nn
\mathbf{Q}_\alpha^{\tilde a}&=& p_\xi{}_\alpha^{\tilde a}
+i\left(p_{\,t}+m_2\right)\xi_\alpha^{\tilde a}  \,, \nn
\mathbf{S}_\alpha^{a}&=& p_\theta{}_\alpha^{a}+2m_0 i\,\theta_\alpha^{a} -2i\,p_{x\,}{}_{\alpha\beta}\xi^{\beta a}-2\mu_1\xi_\alpha^{a}\,, \nn
\mathbf{S}_\alpha^{\tilde a}&=& p_\theta{}_\alpha^{\tilde a}+2m_0 i\,\theta_\alpha^{\tilde a}-2i\,p_{x\,}{}_{\alpha\beta}\xi^{\beta\tilde  a}
-2\mu_2\xi_\alpha^{\tilde a} \,,
\label{gen-f-1}
\eea
where $p_{x\,}{}_{\alpha\beta}:=p_{x}{}_i\,(\sigma_i)_{\alpha\beta}$.
Using the canonical Poisson brackets,
\bea
&&\{t, p_{\,t}\}=1\,,\qquad \{ x_i, p_{\,x}{}_j \}=\delta_{ij} \,, \nn
&&\{ \xi^\alpha_{a}, p_\xi{}_\beta^{b} \}=\delta^\alpha_\beta \delta_{a}^{b}\,,\quad
\{ \xi^\alpha_{\tilde a}, p_\xi{}_\beta^{\tilde b} \}=\delta^\alpha_\beta \delta_{\tilde a}^{\tilde b}\,,\qquad
\{ \theta^\alpha_{a}, p_\theta{}_\beta^{b} \}=\delta^\alpha_\beta \delta_{a}^{b}\,,\quad
\{ \theta^\alpha_{\tilde a}, p_\theta{}_\beta^{\tilde b} \}=\delta^\alpha_\beta \delta_{\tilde a}^{\tilde b}\,,
\label{PB-1}
\eea
we find  the non-vanishing Poisson brackets of the classical NR supersymmetry generators  (\ref{gen-f-1}):
\begin{equation}\label{PB-QQ-pm-2}
\begin{array}{lll}
&\{ \mathbf{Q}^{a}_{\alpha} , \mathbf{Q}^{b}_{\beta} \}=2i\,
\epsilon^{ab} \, \epsilon_{{\alpha} {\beta}}\, ( p_t + m_{1})\,, \qquad &
\{ \mathbf{Q}^{\tilde a}_{\alpha}, \mathbf{Q}^{\tilde b}_{\beta} \}=2i\,
\epsilon^{\tilde a\tilde b} \, \epsilon_{{\alpha} {\beta}}\, ( p_t + m_{2})\,,
\\ [6pt]
&\{ \mathbf{S}^{a}_{\alpha} , \mathbf{S}^{b}_{\beta} \}=4i\,
\epsilon^{ab} \, \epsilon_{{\alpha} {\beta}}\, m_0\,, \qquad &
\{ \mathbf{S}^{\tilde a}_{\alpha}, \mathbf{S}^{\tilde b}_{\beta} \}=4i\,
\epsilon^{\tilde a\tilde b} \, \epsilon_{{\alpha} {\beta}}\, m_0\,,
\\ [6pt]
&\{ \mathbf{Q}^{a}_{\alpha}, \mathbf{S}^{b}_{\beta} \}=2i\,
\epsilon^{ab} \,  ( p_{x\,}{}_{\alpha\beta} + i\mu_1\epsilon_{{\alpha} {\beta}})\,, \qquad &
\{ \mathbf{Q}^{\tilde a}_{\alpha}, \mathbf{S}^{\tilde b}_{\beta} \}=2i\,
\epsilon^{\tilde a\tilde b} \,  ( p_{x\,}{}_{\alpha\beta} +
i\mu_2\epsilon_{{\alpha} {\beta}})\,.
\end{array}
\end{equation}
The Poisson brackets (\ref{PB-QQ-pm-2}) are the classical counterparts of the anticommutators (\ref{contr-QQ-pm-2})-(\ref{contr-QQ-+--2}).
We see that in the model (\ref{Lagr-1}) the parameters $m_1$, $m_2$ and $\mu_1$, $\mu_2$ generate the constant central charges
$\mathbf{X}_1$, $\mathbf{X}_2$ and $\mathbf{Y}_1$, $\mathbf{Y}_2\,$.\footnote{Note that
the value of central charge $\mathbf{M}$ in the model (\ref{Lagr-1}) is equal to $-m_0$ as we see from
the comparison of (\ref{PB-QQ-pm-2}) and (\ref{contr-QQ-pm-2})-(\ref{contr-QQ-+--2}). The appearance of
the sign minus here is the consequence of the choice of the sign minus in front of the one-form $\omega_{(M)}$ in action (\ref{act-1}),
what leads to the Lagrangian (\ref{Lagr-1}) with standard Lagrangian in the bosonic limit.}

Canonical Hamiltonian of the model (\ref{Lagr-1}) is vanishing as in the bosonic case
\begin{equation}\label{H-1}
H_1=p_{\,t}\dot t+ p_{\,x}{}_i \dot x_i + p_\xi{}^{a}_{{\alpha}} \dot\xi_{a}^{{\alpha}},
+ p_\xi{}^{\tilde a}_{{\alpha}} \dot\xi_{\tilde a}^{{\alpha}}
+ p_\theta{}^{a}_{{\alpha}}\dot\theta_{a}^{{\alpha}} + p_\theta{}^{\tilde a}_{{\alpha}}\dot\theta_{\tilde a}^{{\alpha}} -L_1=0\,,
\end{equation}
which indicates the reparametrization invariance of the model.

The expressions for the bosonic momenta (\ref{mom-b-1}) lead again to the free Schr\"{o}dinger constraint (\ref{constr-b}):
\begin{equation}\label{constr-b-1}
T \ := \ p_{\,t}+ \frac{p_{\,x}{}_i p_{\,x}{}_i}{2m_0} \ \approx \  0\,.
\end{equation}
The definitions (\ref{mom-f-1}) of fermionic momenta  lead to the constraints:
\bea
D_\xi{}_\alpha^{a}&:=& p_\xi{}_\alpha^{a}-i\left(p_{\,t}+m_1\right)\xi_\alpha^{a}-2i\,p_{x\,}{}_{\alpha\beta}\theta^{\beta a}
+2\mu_1\theta_\alpha^{a} \ \approx \  0 \,, \nn
D_\xi{}_\alpha^{\tilde a}&:=& p_\xi{}_\alpha^{\tilde a}-i\left(p_{\,t}+m_2\right)\xi_\alpha^{\tilde a}
-2i\,p_{x\,}{}_{\alpha\beta}\theta^{\beta\tilde  a}
+2\mu_2\theta_\alpha^{\tilde a} \ \approx \  0 \,, \nn
D_\theta{}_\alpha^{a}&:=& p_\theta{}_\alpha^{a}-2m_0 i\,\theta_\alpha^{a} \ \approx \  0 \,, \nn
D_\theta{}_\alpha^{\tilde a}&:=& p_\theta{}_\alpha^{\tilde a}-2m_0 i\,\theta_\alpha^{\tilde a} \ \approx \  0 \,.
\label{constr-f-1}
\eea

Using (\ref{PB-1}), we obtain the non-vanishing Poisson brackets for the system of constraints  (\ref{constr-b-1}) and (\ref{constr-f-1})
\bea
&&\{D_\xi{}_\alpha^{a}, D_\xi{}_\beta^{b}\}=-2i\left(p_{\,t}+m_1\right)\epsilon_{\alpha\beta}\epsilon^{ab}\,,\qquad
\{ D_\xi{}_\alpha^{\tilde a}, D_\xi{}_\beta^{\tilde b} \}=-2i\left(p_{\,t}+m_2\right)\epsilon_{\alpha\beta}\epsilon^{\tilde a\tilde b} \,, \nn
&&\{D_\theta{}_\alpha^{a}, D_\theta{}_\beta^{b}\}=-4m_0i\,\epsilon_{\alpha\beta}\epsilon^{ab}\,,\qquad
\{ D_\theta{}_\alpha^{\tilde a}, D_\theta{}_\beta^{\tilde b} \}=-4m_0i\,\epsilon_{\alpha\beta}\epsilon^{\tilde a\tilde b} \,, \nn
&&\{D_\xi{}_\alpha^{a}, D_\theta{}_\beta^{b}\}=
-2i\Big(p_{x\,}{}_{\alpha\beta}+\mu_1i\epsilon_{\alpha\beta}\Big)\epsilon^{ab}\,, \nn
&&\{ D_\xi{}_\alpha^{\tilde a}, D_\theta{}_\beta^{\tilde b} \}=
-2i\Big(p_{x\,}{}_{\alpha\beta}+\mu_2i\epsilon_{\alpha\beta}\Big)\epsilon^{\tilde a\tilde b}\,.
\label{PB-f-constr}
\eea

We will be interested in the NR superparticle models possessing local fermionic $\kappa$ symmetry \cite{AzLuk82, Siegel},
after imposing suitable relations between the parameters of the model (see, e.g., \p{cond-inv-1}, \p{cond-inv-2} below).
In the ${\cal N}=2\,, d=2$ case this kind of NR superparticles was considered in  \cite{fli17}.
We recall that in the phase space formulation, $\kappa$ symmetry is generated by the first class odd constraints.

Let us determine
the values of central charges in the model which imply the first class odd constraints.
For that purpose we should calculate the determinant of
the 16-dimensional matrix of the Poisson brackets of fermionic constraints (\ref{PB-f-constr}),
in the presence of the bosonic constraint (\ref{constr-b-1}), and assume that this determinant becomes zero.

Defining $D_{\mathcal{A}}$ $:=$ ($D_\xi{}_\alpha^{a}$, $D_\xi{}_\alpha^{\tilde a}$, $D_\theta{}_\alpha^{a}$,
$D_\theta{}_\alpha^{\tilde a})$
and
\begin{equation}\label{DD-P}
\{ D_{\mathcal{A}}, D^{\mathcal{B}} \} \ = \ 2i\, \mathcal{P}_{\mathcal{A}}{}^{\mathcal{B}}\,,
\end{equation}
we find that the determinant of the matrix $\mathcal{P}$ of the fermionic
constraints (\ref{PB-f-constr}) is given, modulo a multiplicative constant, by the expression
\begin{equation}\label{det-PB-1}
(m_0)^8\left[ p_{\,t}+ \frac{p_{\,x}{}_i p_{\,x}{}_i}{2m_0} + m_1 +\frac{(\mu_1)^2}{2m_0}\right]^{\,4}
\left[ p_{\,t}+ \frac{p_{\,x}{}_i p_{\,x}{}_i}{2m_0} + m_2 +\frac{(\mu_2)^2}{2m_0}\right]^{\,4}\,.
\end{equation}
Thus, first class odd constraints are present  at least under one of the following two conditions
\begin{equation}\label{cond-inv-1}
m_1 =-\frac{(\mu_1)^2}{2m_0}\,,
\end{equation}
\begin{equation}\label{cond-inv-2}
m_2 =-\frac{(\mu_2)^2}{2m_0}\,.
\end{equation}

If the condition (\ref{cond-inv-1}) is valid, half of the odd constraints linear in $D_\xi{}_\alpha^{a}$, $D_\theta{}_\alpha^{a}$
are first class. Explicitly, these constraints are
\begin{equation}\label{constr-f-f1-1}
F_\xi{}_\alpha^{a} \ := \ D_\xi{}_\alpha^{a} +
\frac{1}{2m_0}\Big( p_{x\,}{}_{\alpha}{}^{\beta}  -i\mu_1\delta_\alpha^\beta\Big) D_\theta{}_\beta^{a} \ \approx 0
\end{equation}
The Poisson brackets involving the constraints (\ref{constr-f-f1-1}) form the following set
\bea
\{F_\xi{}_\alpha^{a}, D_\xi{}_\beta^{b}\}&=&-2i\epsilon_{\alpha\beta}\epsilon^{ab}\,T
-2i\epsilon_{\alpha\beta}\epsilon^{ab}\left( m_1 +\frac{(\mu_1)^2}{2m_0}\right)\,,\qquad
\{ F_\xi{}_\alpha^{a}, D_\theta{}_\beta^{b} \} \ = \ 0 \,, \nn
\{F_\xi{}_\alpha^{a}, F_\xi{}_\beta^{b}\}&=&-2i\epsilon_{\alpha\beta}\epsilon^{ab}\,T
-2i\epsilon_{\alpha\beta}\epsilon^{ab}\left( m_1 +\frac{(\mu_1)^2}{2m_0}\right)\,.
\label{PB-1f-constr}
\eea
We see that the full set of the original constraints ($D_\xi{}_\alpha^{a}$, $D_\theta{}_\alpha^{a}$)
is equivalent to the set ($F_\xi{}_\alpha^{a}$, $D_\theta{}_\alpha^{a}$), where $F_\xi{}_\alpha^{a}$
are first class, and $D_\theta{}_\alpha^{a}$ are second class.

The analysis of the second half of the odd constraints (with tilded indices) is performed quite analogously.
If the condition (\ref{cond-inv-2}) holds, the constraints
\begin{equation}\label{constr-f-f2-1}
F_\xi{}_\alpha^{\tilde a} \ := \ D_\xi{}_\alpha^{\tilde a} +
\frac{1}{2m_0}\Big( p_{x\,}{}_{\alpha}{}^{\beta} -i\mu_2\delta_\alpha^\beta\Big) D_\theta{}_\beta^{\tilde a} \ \approx 0\,.
\end{equation}
are first class, with the following Poisson brackets  with all odd constraints
\bea
\{F_\xi{}_\alpha^{\tilde a}, D_\xi{}_\beta^{\tilde b}\}&=&-2i\epsilon_{\alpha\beta}\epsilon^{\tilde a\tilde b}\,T
-2i\epsilon_{\alpha\beta}\epsilon^{\tilde a\tilde b}\left( m_2 +\frac{(\mu_2)^2}{2m_0}\right)\,,\qquad
\{ F_\xi{}_\alpha^{\tilde a}, D_\theta{}_\beta^{\tilde b} \} \ = \ 0 \,, \nn
\{F_\xi{}_\alpha^{\tilde a}, F_\xi{}_\beta^{\tilde b}\}&=&-2i\epsilon_{\alpha\beta}\epsilon^{\tilde a\tilde b}\,T
-2i\epsilon_{\alpha\beta}\epsilon^{\tilde a\tilde b}\left( m_2 +\frac{(\mu_2)^2}{2m_0}\right)\,.
\label{PB-2f-constr}
\eea
The set of constraints ($D_\xi{}_\alpha^{\tilde a}$, $D_\theta{}_\alpha^{\tilde a}$)
is therefore equivalent to the set ($F_\xi{}_\alpha^{\tilde a}$, $D_\theta{}_\alpha^{\tilde a}$), with $F_\xi{}_\alpha^{\tilde a}$
being first class and $D_\theta{}_\alpha^{\tilde a}$ second class.

In Section\,5 we will study the quantization of the superparticle model  defined by the action  (\ref{act-1}),  (\ref{Lagr-1})
possessing $\kappa$-symmetries due to the presence of conditions (\ref{cond-inv-1}) and
(\ref{cond-inv-2}).
Using (\ref{constr-f-1}),
we obtain the following explicit form of the first class constraints (\ref{constr-f-f1-1}) and  (\ref{constr-f-f2-1})
generating $\kappa$-symmetries
\bea
F_\xi{}_\alpha^{a}&=& p_\xi{}_\alpha^{a}
-i\left(p_{\,t}+m_1\right)\xi_\alpha^{a} +
\frac{1}{2m_0}\Big( p_{x\,}{}_{\alpha}{}^{\beta} -i\mu_1\delta_\alpha^\beta\Big)
\Big( p_\theta{}_\beta^{a}+2m_0 i\,\theta_\beta^{a} \Big) \ \approx \ 0 \,, \nn
F_\xi{}_\alpha^{\tilde a}&=& p_\xi{}_\alpha^{\tilde a}
-i\left(p_{\,t}+m_2\right)\xi_\alpha^{a} +
\frac{1}{2m_0}\Big( p_{x\,}{}_{\alpha}{}^{\beta} -i\mu_2\delta_\alpha^\beta\Big)
\Big( p_\theta{}_\beta^{\tilde a}+2m_0 i\,\theta_\beta^{\tilde a} \Big) \ \approx \ 0 \,.
\label{F-constr}
\eea

We notice that, if we specialize our discussion of constraints to one sector only, with either index $a$ or index $\tilde a$, we recover
the model with smaller $\mathcal{N}{=}\,2$ Galilean supersymmetry.
Such models in $d\,{=}\,2$ case
have been studied in \cite{fli17} for the case of   $\mathcal{N}{=}\,2,$ $d\,{=}\,2$ Galilean supersymmetry
with only one central charge. The $d\,{=}\,2$ models of \cite{fli17} can be obtained as a special case of our model
with only one of the $\mathcal{N}{=}\,2$ sectors retained.

The constraints (\ref{F-constr}) generate local $\kappa$-transformations of an arbitrary phase space function $X$
by the following Poisson bracket
\begin{equation}\label{kappa-X}
\delta X =\left\{\kappa^\alpha_aF_\xi{}_\alpha^{a}+ \kappa^\alpha_{\tilde a}F_\xi{}_\alpha^{\tilde a},X\right\},
\end{equation}
where $\kappa^\alpha_{a}(\tau)$ and $\kappa^\alpha_{\tilde a}(\tau)$ are local Grassmann parameters.
The $\kappa$-transformations (\ref{kappa-X})
of the variables in the Lagrangian (\ref{Lagr-1}) with $a=0$
are as follows
\bea
&& \delta t= i\,\kappa^\alpha_a \xi_\alpha^a \ + \ i\,\kappa^\alpha_{\tilde a}\xi_\alpha^{\tilde a}\,,\qquad
\delta x_i=-\,2i\,\kappa^\alpha_a (\sigma_i)_\alpha{}^\beta\xi_\beta^a \ - \
2i\,\kappa^\alpha_{\tilde a}(\sigma_i)_\alpha{}^\beta\xi_\beta^{\tilde a} \,, \nn
&&  \delta \xi^\alpha_a= \kappa^\alpha_a \,,\qquad \delta \xi^\alpha_{\tilde a}=\kappa^\alpha_{\tilde a} \,, \nn
&&  \delta \theta^\alpha_a=\frac{\pi_i}{2\pi_0}\,\kappa^\beta_a (\sigma_i)_\beta{}^\alpha -\frac{i\mu_1}{2m_0}\kappa^\alpha_a\,,\qquad
\delta \theta^\alpha_{\tilde a}=\frac{\pi_i}{2\pi_0}\,\kappa^\beta_{\tilde a} (\sigma_i)_\beta{}^\alpha -\frac{i\mu_2}{2m_0}\kappa^\alpha_{\tilde a}\,.
\label{kappa-tr1}
\eea
Under these  transformations, the variation of Lagrangian (\ref{Lagr-1})  (with $a=0$) is
\bea
\delta L_1&=& 2i\left(m_1+\frac{(\mu_1)^2}{2m_0} \right)\kappa^\alpha_a \dot\xi_\alpha^a \ + \
2i\left(m_2+\frac{(\mu_2)^2}{2m_0} \right)\kappa^\alpha_{\tilde a}\dot\xi_\alpha^{\tilde a}  \nn
&&  -\,\frac{\partial}{\partial\tau}\left[i m_0\frac{\pi_i}{\pi_0}\,\kappa^\alpha_a (\sigma_i)_\alpha{}^\beta \theta^a_\beta \ + \
i m_1 \kappa^\alpha_a \xi_\alpha^a \ + \ \mu_1 \kappa^\alpha_a \theta_\alpha^a\right] \nn
&&  -\,\frac{\partial}{\partial\tau}\left[i m_0\frac{\pi_i}{\pi_0}\,\kappa^\alpha_{\tilde a} (\sigma_i)_\alpha{}^\beta \theta^{\tilde a}_\beta \ + \
i m_2 \kappa^\alpha_{\tilde a} \xi_\alpha^{\tilde a} \ + \ \mu_2 \kappa^\alpha_{\tilde a} \theta_\alpha^{\tilde a}\right]\,.
\label{kappa-L1}
\eea
It should be pointed out that  the variation (\ref{kappa-L1}) is a total derivative only provided
the conditions (\ref{cond-inv-1}) and (\ref{cond-inv-2}) are taken into account.
Thus we have shown explicitly that the superparticle model  (\ref{Lagr-1})
is $\kappa$-invariant only when the central charge parameters satisfy the conditions (\ref{cond-inv-1}) and (\ref{cond-inv-2}).

Using local transformations (\ref{kappa-tr1}) one can choose the gauge $\xi_\alpha^a=0$, $\xi_\alpha^{\tilde a}=0\,$.
In such a gauge, the rigid $\mathbf{Q}$ - transformations (\ref{Q-tr-comp}) should be accompanied
by the appropriate compensating  gauge transformations (see, e.g., \cite{fli17}).
In this case,  as well as in other cases considered below, we will not impose such gauges,
reserving it for more complicated ${\cal N}=4$ NR superparticle models still to be constructed, e.g.,
those formulated on external electromagnetic background.

\subsection{The superparticle model with all central charges incorporated}

\quad\, We can add to the action (\ref{act-1}) at $a=0$ the additional terms associated with off-diagonal central charges
\begin{equation}\label{act-2}
S_2\ =\  \int L_2 \,d\tau \ = \ \int \Big( \ n^{a\tilde b}\,\hat\omega_{(X)\,a\tilde b} + \nu^{a\tilde b}\,
\hat\omega_{(Y)\,a\tilde b} \ \Big)\,,
\end{equation}
where $n^{a\tilde b}$ and $\nu^{a\tilde b}$
are constants with the reality conditions as for $\mathrm{USp}(2)\otimes \mathrm{USp}(2)\simeq\mathrm{O}(4)$ bispinors
(see (\ref{XY-pm-2}), (\ref{hf-pm-2}))
\begin{equation}\label{nnu-pm-2}
\left(n^{a \tilde b}\right)^*   =
\epsilon_{ac} \epsilon_{\tilde b\tilde d}\,n^{c \tilde d}\,, \qquad
\left(\nu^{a \tilde b}\right)^*   =
\epsilon_{ac} \epsilon_{\tilde b\tilde d}\,\nu^{c \tilde d}\,.
\end{equation}
These bi-spinorial constants can be represented as internal four-vectors (isovectors)
\begin{equation}\label{nnu-sigma}
n^{a\tilde b}=i(\tilde\sigma_M)^{a\tilde b}n_M\,,\qquad \nu^{a\tilde b}=i(\tilde\sigma_M)^{a\tilde b}\nu_M\,,
\end{equation}
where $(\tilde\sigma_M)^{a\tilde b}$, $M=1,2,3,4$, are $\mathrm{O}(4)$ Euclidean $\sigma$-matrices and
$n_M=(n_M)^*$, $\nu_M=(\nu_M)^*$ are the real internal $\mathrm{O}(4)$ vectors.
Below we use the matrix realization  $(\tilde\sigma_M)^{a\tilde b}=(\sigma_i,-i1_2)$, where $\sigma_i$, $i=1,2,3$,
are the standard Pauli matrices.

The total action can be written as follows
\begin{equation}\label{act-t}
S\ =\  S_1\ + \ S_2 \,,
\end{equation}
which includes thirteen coupling constants $m_0, m_1, m_2, \mu_1, \mu_2,n_M, \nu_M\,$.
Together with the $\mathrm{O}(4)$ singlets $(m_1 - m_2)$ and $(\mu_1 - \mu_2)$, respectively,
the $\mathrm{O}(4)$ vectors $n_M$ and  $\nu_M$ can be combined in two independent constant $\mathrm{USp}(4) \sim \mathrm{O}(5)$ vectors.
The remaining  constants $(m_1 + m_2)$ and $(\mu_1 + \mu_2)$  are $\mathrm{O}(5)$ singlets.
Constant Euclidean four-vectors $n_M$ and $\nu_M$ break both $\mathrm{O}(5)$ and $\mathrm{USp}(2)\otimes \mathrm{USp}(2)\simeq \mathrm{O}(4) \subset
\mathrm{O}(5)$ internal symmetries to some their subgroups, depending on the concrete choice of the four-vector components. As we will note later, the broken $\mathrm{O}(5)$ and $\mathrm{O}(4)$ symmetries can be used to bring the constant parameters entering \p{act-t} to the set with
the diagonal unbroken R-symmetry $\mathrm{USp}(2)\simeq \mathrm{O}(3)  \subset \mathrm{USp}(2)\otimes \mathrm{USp}(2)\simeq \mathrm{O}(4)$.
With some particular choices of the parameters only $\mathrm{O}(2)$ remains as the unbroken internal symmetry.

In the presence of arbitrary central charges the preservation of the total $\mathrm{O}(5)$ or $\mathrm{O}(4)\subset \mathrm{O}(5)$
invariances as Noether symmetries can be restored either  by introducing additional harmonic variables or
by choosing a different kind of dynamics. Such dynamical model respecting $\mathrm{O}(4)$ invariance in the sector described
by off-diagonal central charges, will be discussed in Sect.\,6.

The Lagrangian in \p{act-t} is given by the following expression
\bea
L\ =&L_1+L_2\ =&m_0\,\frac{\pi_i \pi_i}{2\pi_0} \ + \
2m_0 i\,\big(\theta^\alpha_a  \dot\theta^a_\alpha \ + \  \theta^\alpha_{\tilde a}  \dot\theta^{\tilde a}_\alpha \big) \nn
&&  + \ i\,m_1 \, \xi_\alpha^a  \dot\xi_a^\alpha \ + \ i\,m_2 \, \xi_\alpha^{\tilde a} \dot\xi_{\tilde a}^\alpha
 \ + \ 2 \mu_1\, \theta^{\alpha a}  \dot\xi_{a \alpha} \ + \ 2 \mu_2\, \theta^{\alpha \tilde a} \dot\xi_{\alpha \tilde a}\nn
&&  + \ i\, n^{a\tilde b}\, \Big( \xi_{\alpha\tilde b}  \dot\xi_a^\alpha \ - \
\xi_{\alpha a}  \dot\xi_{\tilde b}^\alpha \Big)
 \ - \ 2 \, \nu^{a\tilde b}\, \Big( \theta_{\alpha\tilde b}  \dot\xi_a^\alpha \ - \
\theta_{\alpha a}  \dot\xi_{\tilde b}^\alpha \Big)\,.
\label{Lagr-t}
\eea

In comparison with (\ref{Lagr-1}), the Lagrangian (\ref{Lagr-t}) contains
two additional terms,
which involve only derivatives of $\xi$'s.
Therefore, the momenta $p_{\,t}$, $p_{\,x}{}_i$, $p_\theta{}_\alpha^{a}$, $p_\theta{}_\alpha^{\tilde a}$
are the same as in (\ref{mom-b-1}), (\ref{mom-f-1}), whereas the momenta $p_\xi{}_\alpha^{a}$, $p_\xi{}_\alpha^{\tilde a}$
acquire additional terms as compared with (\ref{mom-f-1}).
Then  it follows that the bosonic constraint $T \approx 0$ (see (\ref{constr-b-1})) and the fermionic constraints
$D_\theta{}_\alpha^{a} \approx 0$, $D_\theta{}_\alpha^{\tilde a} \approx 0$, defined by (\ref{constr-f-1}),
remain the same, whereas fermionic constraints
$D_\xi{}_\alpha^{a} \approx 0$, $D_\xi{}_\alpha^{\tilde a} \approx 0$ will acquire  additional terms in comparison with (\ref{constr-f-1}):
\bea
D_\xi{}_\alpha^{a}&=& p_\xi{}_\alpha^{a}-2i\,p_{x\,}{}_{\alpha\beta}\theta^{\beta a}
-i\left(p_{\,t}+m_1\right)\xi_\alpha^{a}+2\mu_1\theta_\alpha^{a} -i n^{a\tilde b}\xi_{\alpha\tilde b}
+2\nu^{a\tilde b}\theta_{\alpha\tilde b}\ \approx \  0 \,, \nn
D_\xi{}_\alpha^{\tilde a}&=& p_\xi{}_\alpha^{\tilde a}-2i\,p_{x\,}{}_{\alpha\beta}\theta^{\beta\tilde  a}
-i\left(p_{\,t}+m_2\right)\xi_\alpha^{\tilde a}+2\mu_2\theta_\alpha^{\tilde a}
+i n^{b\tilde a}\xi_{\alpha b}
-2\nu^{b\tilde a}\theta_{\alpha b} \ \approx \  0 \,.
\label{constr-f-2}
\eea

One can check again that the canonical Hamiltonian is vanishing.

Besides the Poisson brackets (\ref{PB-f-constr}), one gets the following additional
non-vanishing Poisson brackets of fermionic constraints
$D_\theta{}_\alpha^{a} \approx 0$, $D_\theta{}_\alpha^{\tilde a} \approx 0$ (\ref{constr-f-1})
and $D_\xi{}_\alpha^{a} \approx 0$, $D_\xi{}_\alpha^{\tilde a} \approx 0$ (\ref{constr-f-2}):
\bea
&&\{D_\xi{}_\alpha^{a}, D_\xi{}_\beta^{\tilde b}\}=-2i\,\epsilon_{\alpha\beta}\,n^{a\tilde b}\,, \nn
&&\{D_\xi{}_\alpha^{a}, D_\theta{}_\beta^{\tilde b}\}=
2\,\epsilon_{\alpha\beta}\,\nu^{a\tilde b}\,,
\qquad \{ D_\xi{}_\alpha^{\tilde a}, D_\theta{}_\beta^{b} \}=
-2\,\epsilon_{\alpha\beta}\,\nu^{b\tilde a }\,.
\label{PB-f-constr-2}
\eea

For the model with Lagrangian (\ref{Lagr-t})
the Noether charges generating the NR supersymmetry transformations  (\ref{Q-tr-comp}) and (\ref{S-tr-comp}) contain,
in comparison with the expressions  (\ref{gen-f-1}), some additional terms and take the form
\bea
\mathbf{Q}_\alpha^{a}&=& p_\xi{}_\alpha^{a}
+i\left(p_{\,t}+m_1\right)\xi_\alpha^{a} +in^{a\tilde b}\xi_{\alpha\tilde b}\,, \nn
\mathbf{Q}_\alpha^{\tilde a}&=& p_\xi{}_\alpha^{\tilde a}
+i\left(p_{\,t}+m_2\right)\xi_\alpha^{\tilde a}  -in^{b\tilde a}\xi_{\alpha b}\,, \nn
\mathbf{S}_\alpha^{a}&=& p_\theta{}_\alpha^{a}+2m_0 i\,\theta_\alpha^{a} -2i\,p_{x\,}{}_{\alpha\beta}\xi^{\beta a}-2\mu_1\xi_\alpha^{a}
-2\nu^{a\tilde b}\xi_{\alpha\tilde b}\,, \nn
\mathbf{S}_\alpha^{\tilde a}&=& p_\theta{}_\alpha^{\tilde a}+2m_0 i\,\theta_\alpha^{\tilde a}-2i\,p_{x\,}{}_{\alpha\beta}\xi^{\beta\tilde  a}
-2\mu_2\xi_\alpha^{\tilde a} +2\nu^{b\tilde a}\xi_{\alpha b}\,.
\label{gen-f-t}
\eea
The set of non-vanishing Poisson brackets between the classical supersymmetry generators  (\ref{gen-f-t}) involves
the relations (\ref{PB-QQ-pm-2}) and, in addition, the following Poisson brackets
\begin{equation}\label{PB-QQ-pm-3}
\begin{array}{lll}
&& \{ \mathbf{Q}^{a}_{\alpha}, \mathbf{Q}^{\tilde b}_{\beta} \} \ = \ 2i\,\epsilon_{\alpha\beta} n^{a \tilde b}\,,
\\ [6pt]
&& \{ \mathbf{Q}^{a}_{\alpha}, \mathbf{S}^{\tilde b}_{\beta} \} \ = \ \{ \mathbf{Q}^{\tilde b}_{\beta},
\mathbf{S}^{a}_{\alpha} \} \ = \ -2\, \epsilon_{\alpha \beta} \nu^{a \tilde b}\,.
\end{array}
\end{equation}
The Poisson brackets (\ref{PB-QQ-pm-3}) are  classical counterparts of the anticommutators (\ref{VectZdop}).
We see that the constants $n^{a\tilde b}$ and $\nu^{a\tilde b}$ of the general model (\ref{Lagr-t}) reappear at the level
of Poisson brackets in place of the central charges $\mathbf{X}^{a\tilde b}$ and $\mathbf{Y}^{a\tilde b}\,$.

\subsubsection{Analysis of the constrains}

The determinant of the Poisson brackets matrix $\mathcal{P}$ of the fermionic
constraints (\ref{constr-f-2}), (\ref{constr-f-1}) defined in \p{DD-P} in the case under consideration looks more complicated than
\p{det-PB-1}.
It is equal, up to a multiplicative constant, to the following expression:
\begin{equation}\label{det-PB-2}
\Delta \ {:=} \ (m_0)^8\left[\left( T + m_1 +\frac{(\mu_1)^2+\hat \nu}{2m_0}\right)
\left( T + m_2 +\frac{(\mu_2)^2+\hat \nu}{2m_0}\right) -\hat w \right]^{\,4}\,,
\end{equation}
where
\begin{equation}\label{def-scalars}
\hat \nu := \frac12\,\nu^{a\tilde b}\nu_{a\tilde b}=\nu_M \nu_M
\,,\qquad \hat w := \frac12\,w^{a\tilde b}w_{a\tilde b}=w_M w_M\,,
\end{equation}
$T$ is the first class constraint (\ref{constr-b-1}),
and the 4-vector $w^{a\tilde b}=i(\tilde\sigma_M)^{a\tilde b}w_M$ is defined by
\begin{equation}\label{def-w}
w^{a\tilde b} \ := \ n^{a\tilde b}+ \frac{\mu_1+\mu_2}{2m_0}\, \nu^{a\tilde b}\qquad\Leftrightarrow\qquad
w_M \ = \ n_M+ \frac{\mu_1+\mu_2}{2m_0}\,\nu_M\,.
\end{equation}

It should be added that $\hat n=\frac12\,n^{a\tilde b}\nu_{a\tilde b}=n_M n_M$ and $\hat \nu$, $\hat w$
are the length squares of $\mathrm{O}(4)$ internal symmetry vectors, i.e. one gets that
$\hat n \geq 0$, $\hat \nu \geq 0$ and $\hat w \geq 0$.
Moreover, $\hat n = 0$ leads to $n^{a\tilde b}=0$ as well as $\hat \nu = 0$ leads to $\nu^{a\tilde b}=0$;
similarly, $\hat w=0$ implies $w^{a\tilde b}=0\,$.

The odd first class constraints generating $\kappa$-symmetry, are present provided that
\begin{equation}\label{det-eq-0}
\mathrm{det}\,\mathcal{P} \ \simeq \ \Delta \ = \ 0\,.
\end{equation}
The constants $n^{a\tilde b}$, $\nu^{a\tilde b}$ enter the expression (\ref{det-PB-2}) and the equation  (\ref{det-eq-0})
only through two quantities, $\hat \nu$ and $\hat w$ defined by (\ref{def-scalars}). Since \p{det-PB-2} is not factorized,
in contrast to  \p{det-PB-1}, resolving eq. \p{det-eq-0} is a more complicated task.

The condition (\ref{det-eq-0}) is necessary for (any number of) odd first class constraint.
The full number of such constraints is found by solving the characteristic equation
\begin{equation}\label{ch-eq}
\mathrm{det}\left( \mathcal{P}-\lambda I\right)=0\,,
\end{equation}
which determines the eigenvalues $\lambda$ of the matrix $\mathcal{P}$ (see analogous consideration, e.g., in \cite{GGHT,FedZ}).
In (\ref{ch-eq}), $\lambda$ describes the spectral parameter and $I$ is the unit matrix.
The number of first class constraints is equal to the number of solutions $\lambda=0$ of the characteristic equation (\ref{ch-eq}).
In the presence of $k$ odd first class constraints among sixteen constraints $D_{\mathcal{A}}$,
the equation (\ref{ch-eq}) has the form
\begin{equation}\label{ch-eq1}
\lambda^k\left(\lambda-\lambda_1\right)\ldots \left(\lambda-\lambda_{16-k}\right)=0\,,
\qquad \mbox{where}\quad\lambda_1\neq 0\,,\quad\ldots\,,\quad\lambda_{16-k}\neq 0\,.
\end{equation}

In the model considered here the characteristic equation (\ref{ch-eq}) coincides with the equation  (\ref{det-eq-0})
in which the substitutions  $p_t\to (p_t-\lambda)$ and $m_0\to (m_0-\lambda)$ are performed.
Using the expression  (\ref{det-PB-2}) \footnote{A possible numerical coefficient before $\Delta$ can be absorbed into a rescaling
of the matrix ${\cal P}$ and, further, into the redefinition of the parameter $\lambda$ in (\ref{ch-eq}).} the characteristic equation (\ref{ch-eq})
can be written in the following form
\begin{equation}\label{ch-eq-fin}
\Delta+(m_0)^4\lambda^4\left(A+B\lambda+C\lambda^2+\lambda^3\right)^4=0\,,
\end{equation}
where
\bea
A&=& -\left( p_t + m_1 + m_0\right)\left( T + m_2 +\frac{(\mu_2)^2+\hat \nu}{2m_0}\right)-
\left( p_t + m_2 + m_0\right)\left( T + m_1 +\frac{(\mu_1)^2+\hat \nu}{2m_0}\right)\nn
&& +\,2\,n^{a\tilde b}w_{a\tilde b} \,,
\label{def-A} \\
B&=& \frac{1}{m_0}\left( p_t + m_1 + m_0\right)\left( p_t + m_2 + m_0\right)
+2T + m_1+m_2 +\frac{(\mu_1)^2+(\mu_1)^2+2\hat \nu -\hat n}{2m_0}\,,
\label{def-B} \\
C&=& -2\left( p_t + \frac{m_1+m_2}{2} + m_0\right).
\label{def-C}
\eea

As we see from (\ref{ch-eq-fin}), the condition (\ref{det-eq-0}) implies
the presence of at least four odd first class constraints in the total set of sixteen fermionic constraints
(\ref{constr-f-2}), (\ref{constr-f-1}).

The condition $A=0$, together  with (\ref{det-eq-0}), lead to the presence of eight odd first class constraints.
The condition $A=0$ requires vanishing those terms in (\ref{def-A}) which are proportional to $p_t$:
\begin{equation}\label{12-1}
m_1 +\frac{(\mu_1)^2+\hat \nu}{2m_0} +
m_2 +\frac{(\mu_2)^2+\hat \nu}{2m_0}=0\,.
\end{equation}
An additional condition which stems from $A=0$ is the vanishing of the remaining constant term:
\begin{equation}\label{12-2}
m_2\left( m_1 +\frac{(\mu_1)^2+\hat \nu}{2m_0}\right)+
m_1\left( m_2 +\frac{(\mu_2)^2+\hat \nu}{2m_0}\right)=n^{a\tilde b}w_{a\tilde b} \,.
\end{equation}
For $m_1=m_2$ it leads to the condition $n^{a\tilde b}w_{a\tilde b}=2n_M\nu_M=0$.

Further, one can show that  $B\neq 0$ and $C\neq 0$ in (\ref{ch-eq-fin}) due to  non-vanishing constant coefficients in (\ref{def-B})
and (\ref{def-C}) in front of $(p_t)^2$ and $p_t$ .

Thus, by definite choices of central charges, we can recover the cases,
when the number of odd first class constraints is quarter or half the total number of  odd constraints.
Note that, up to a sign, the algebra of the fermionic constraints (\ref{PB-f-constr}), (\ref{PB-f-constr-2})
coincides with the NR superalgebra (\ref{PB-QQ-pm-2}), (\ref{PB-QQ-pm-3}) and the number of the first class constraints
equals the number of preserved supersymmetries in BPS configurations.
Therefore, respective models describe BPS configurations preserving $1/4$ or $1/2$ of NR supersymmetry.

In the last part of this Section, we will consider in detail two special cases.

\subsubsection{The case when half of odd constraints is  first class}

This particular example is specified by the following condition on isotensorial central charges:
\begin{equation}\label{w-0}
\hat w=0
\end{equation}
or, equivalently,
\begin{equation}\label{w-00}
w^{a\tilde b} = \ n^{a\tilde b}+ \frac{\mu_1+\mu_2}{2m_0}\, \nu^{a\tilde b}=0\,.
\end{equation}
In this case the vanishing of the quantity  (\ref{det-PB-2}) (with $T\approx0$) requires that
\begin{equation}\label{eq-0-1}
\left[ m_1 +\frac{(\mu_1)^2+\hat \nu}{2m_0}\right] \left[ m_2 +\frac{(\mu_2)^2+\hat \nu}{2m_0}\right]=0\,.
\end{equation}
The relation (\ref{eq-0-1})  is obeyed provided at least one of two conditions
\begin{equation}\label{cond-inv-1a}
m_1 =-\frac{(\mu_1)^2+\hat \nu}{2m_0}\,,
\end{equation}
\begin{equation}\label{cond-inv-2a}
m_2 =-\frac{(\mu_2)^2+\hat \nu}{2m_0},
\end{equation}
or both of them are fulfilled. The conditions (\ref{cond-inv-1a}), (\ref{cond-inv-2a}) are the obvious generalizations of
(\ref{cond-inv-1}) and (\ref{cond-inv-2}).

Now we present the full set of the constraints which occur when the conditions
(\ref{cond-inv-1a}), (\ref{cond-inv-2a}) and (\ref{w-00}) are valid.

The first class bosonic constraint (\ref{constr-b-1}) represents the Schr\"{o}dinger equation as in the previous cases.

Fermionic constraints ($D_\xi{}_\alpha^{a}$, $D_\xi{}_\alpha^{\tilde a}$; $D_\theta{}_\alpha^{a}$, $D_\theta{}_\alpha^{\tilde a}$)
are equivalent to the set ($F_\xi{}_\alpha^{a}$, $F_\xi{}_\alpha^{\tilde a}$; $D_\theta{}_\alpha^{a}$, $D_\theta{}_\alpha^{\tilde a}$),
where ($F_\xi{}_\alpha^{a}$, $F_\xi{}_\alpha^{\tilde a}$) are defined by the following expressions
\bea
\label{constr-f-f1-2}
F_\xi{}_\alpha^{a} &=& D_\xi{}_\alpha^{a} +
\frac{1}{2m_0}\Big( p_{x\,}{}_{\alpha}{}^{\beta}  -i\mu_1\delta_\alpha^\beta\Big) D_\theta{}_\beta^{a} +
\frac{i}{2m_0}\,\nu^{a}{}_{\tilde b} D_\theta{}_\alpha^{\tilde b} \ \approx 0 \,,
\\
\label{constr-f-f2-2}
F_\xi{}_\alpha^{\tilde a} &=& \ D_\xi{}_\alpha^{\tilde a} +
\frac{1}{2m_0}\Big( p_{x\,}{}_{\alpha}{}^{\beta} -i\mu_2\delta_\alpha^\beta\Big) D_\theta{}_\beta^{\tilde a} -
\frac{i}{2m_0}\,\nu_{b}{}^{\tilde a} D_\theta{}_\alpha^{b} \ \approx 0\,.
\eea

The complete set of non-vanishing Poisson brackets for the constraints (\ref{constr-f-f1-2}), (\ref{constr-f-f2-2})
reads
\bea
\{F_\xi{}_\alpha^{a}, D_\xi{}_\beta^{b}\}&=&
-2i\epsilon_{\alpha\beta}\epsilon^{ab}\left( T+m_1 +\frac{(\mu_1)^2+\hat\nu}{2m_0}\right)\,,\qquad
\{ F_\xi{}_\alpha^{a}, D_\theta{}_\beta^{\tilde b} \} \ = \ -2i\epsilon_{\alpha\beta}w^{a\tilde b} \,, \nn
\{F_\xi{}_\alpha^{\tilde a}, D_\xi{}_\beta^{\tilde b}\}&=&
-2i\epsilon_{\alpha\beta}\epsilon^{\tilde a\tilde b}\left( T+m_2 +\frac{(\mu_2)^2+\hat\nu}{2m_0}\right)\,,\qquad
\{ F_\xi{}_\alpha^{\tilde a}, D_\theta{}_\beta^{b} \} \ = \ -2i\epsilon_{\alpha\beta}w^{b\tilde a} \,, \nn
\{F_\xi{}_\alpha^{a}, F_\xi{}_\beta^{b}\}&=&
-2i\epsilon_{\alpha\beta}\epsilon^{ab}\left( T+m_1 +\frac{(\mu_1)^2+\hat\nu}{2m_0}\right)\,, \qquad
\{F_\xi{}_\alpha^{a}, F_\xi{}_\beta^{\tilde b}\} \ = \
-2i\epsilon_{\alpha\beta}w^{a\tilde b}\,.\nn
\{F_\xi{}_\alpha^{\tilde a}, F_\xi{}_\beta^{\tilde b}\}&=&
-2i\epsilon_{\alpha\beta}\epsilon^{\tilde a\tilde b}\left( T+m_2 +\frac{(\mu_2)^2+\hat\nu}{2m_0}\right)\,.
\label{PB-2f-constr-2}
\eea
We see that if the conditions (\ref{cond-inv-1a}), (\ref{cond-inv-1a}) and (\ref{w-00}) are valid,
the constraints ($D_\theta{}_\alpha^{a}$, $D_\theta{}_\alpha^{\tilde a}$) are second class, while  the constraints
($F_\xi{}_\alpha^{a}$, $F_\xi{}_\alpha^{\tilde a}$) are first class and generate $\kappa$-symmetries.

Substituting the expressions (\ref{constr-f-2}) into the $\kappa$-symmetry generators (\ref{constr-f-f1-2}),  (\ref{constr-f-f2-2}),
we obtain them in the following explicit form
\bea
F_\xi{}_\alpha^{a}&=& p_\xi{}_\alpha^{a}
-i\left(p_{\,t}+m_1\right)\xi_\alpha^{a} -i n^{a\tilde b}\xi_{\alpha\tilde b}  \nn
&&
+
\frac{1}{2m_0}\Big( p_{x\,}{}_{\alpha}{}^{\beta} -i\mu_1\delta_\alpha^\beta\Big)
\Big( p_\theta{}_\beta^{a}+2m_0 i\,\theta_\beta^{a} \Big)
+
\frac{i}{2m_0}\,\nu^{a}{}_{\tilde b}
\Big( p_\theta{}_\alpha^{\tilde b}+2m_0 i\,\theta_\alpha^{\tilde b} \Big) \ \approx \ 0 \,,
\nn
F_\xi{}_\alpha^{\tilde a}&=& p_\xi{}_\alpha^{\tilde a}
-i\left(p_{\,t}+m_2\right)\xi_\alpha^{\tilde a} +i n^{b\tilde a}\xi_{\alpha b} \nn
&&
+\frac{1}{2m_0}\Big( p_{x\,}{}_{\alpha}{}^{\beta} -i\mu_2\delta_\alpha^\beta\Big)
\Big( p_\theta{}_\beta^{\tilde a}+2m_0 i\,\theta_\beta^{\tilde a} \Big) -
\frac{i}{2m_0}\,\nu_{b}{}^{\tilde a}
\Big( p_\theta{}_\alpha^{b}+2m_0 i\,\theta_\alpha^{b} \Big) \ \approx \ 0 \,.
\label{F-constr-2}
\eea
As opposed to the constraints (\ref{F-constr}) in Sect.\,4.2, in the considered case the constraints (\ref{F-constr-2})
mix two $\mathrm{USp}(2)$ sectors characterized  by untilded and tilded $\mathrm{USp}(2)$-indices.

It turns out, however, that this model for $m_1 \neq m_2$ and $\mu_1 \neq \mu_2$ is just the model of Sect. 4.2 in disguise.
To show this, recall
that the full Lagrangian (\ref{Lagr-t}) is formally invariant under the simultaneous $\mathrm{O}(5)$ rotation of the $d=1$ fields
and the set of coupling constants $m_1, m_2, \mu_1, \mu_2, n^{a\tilde b}, \nu^{a\tilde b}$, which gives an opportunity to pass to
$\mathrm{O}(5)$ frame where these
constants are reduced to some minimal set.\footnote{To avoid a possible confusion, we point out that this ``invariance'' is
provided if we transform as well the constant parameters in the model;
it does not give rise to any Noether charges and serves just to choose the convenient $\mathrm{O}(5)$ frame.}
It is important that the coupling constants are divided into the $O(5)$ singlets
$m_1 + m_2, \; \mu_1 + \mu_2$ and $\mathrm{O}(5)$ vectors $(n^{a\tilde b}, m_1 - m_2)$ and $(\nu^{a\tilde b}, \mu_1 -\mu_2)$;
then the condition \p{w-00} means the vanishing of some particular linear combination of the $\mathrm{O}(4)$ vector components of these two
$\mathrm{O}(5)$ vectors.
The fifth component of the $\mathrm{O}(5)$ vector containing the $\mathrm{O}(4)$ vector \p{w-00}, is
\be
m_ 1 - m_2 +  \frac{\mu_1 + \mu_2}{2m_0}\,(\mu_1 - \mu_2) = m_1 - m_2 + \frac{(\mu_1)^2  - (\mu_2)^2}{2m_0}\,.
\ee
This quantity is zero just as a difference of the conditions (\ref{cond-inv-1a}), and(\ref{cond-inv-2a})!
The rest of these conditions, their sum, expresses the particular $\mathrm{O}(5)$ invariant $(m_1 + m_2)$ through other one,
\be
m_1 + m_2 = -\frac1{2 m_0}[{\cal A} + \frac12 (\mu_1 + \mu_2)^2 ]\,, \quad
{\cal A} :=\nu^{a\tilde b}\nu_{a\tilde b} + \frac12(\mu_1 -\mu_2)^2 \,.
\ee
Thus it follows that the conditions \p{w-00} and (\ref{cond-inv-1a}), (\ref{cond-inv-2a}) lead to the vanishing of one out of two
independent $\mathrm{O}(5)$ vectors  in the space of coupling constants and as well relate with each other some $\mathrm{O}(5)$ invariant combinations
of these constants.  Thus these conditions preserve $\mathrm{O}(5)$ covariance, and one can still
use the $\mathrm{O}(5)$ rotations in order to choose the frame where
$n^{a\tilde b}$ (or $\nu^{a\tilde b}$) are zero. In such a frame, due to \p{w-00}, both $\mathrm{O}(4)$ vectors are zero and we are
left with the constraints \p{cond-inv-1}, \p{cond-inv-2} as the only remaining ones. So
for the off-diagonal central charges satisfying the conditions \p{w-00}
our model becomes identical
to the one considered in  Sect.\,4.2.

To construct a system with eight
first-class constraints, which would be non-equivalent to the system of Sect. 4.2 and involve the constants
$n^{a \tilde b}$, $\nu^{a\tilde b}$ which cannot be removed, one needs to break explicitly the $\mathrm{USp}(4) \simeq \mathrm{O}(5)$ covariance
in the space of coupling constants. The simplest option is to assume
\be
m_1 = m_2 \quad \Leftrightarrow \quad \mu_1 = \pm \mu_2.\lb{BreakO5}
\ee

The $\mathrm{USp}(4) \sim \mathrm{O}(5)$ covariance is also explicitly broken in a system with four first-class constraints corresponding to
$1/4$ BPS states. We will consider it as the second example.

\subsubsection{The case when quarter of odd constraints is  first class}

Our second example is characterized by non-vanishing off-diagonal central charges, with all
quasi-diagonal ones vanishing:
\begin{equation}\label{m-mu-0}
m_1 =m_2=0\,,\qquad \mu_1 =\mu_2=0\,.
\end{equation}
In this case it follows from (\ref{def-w}) that $w^{a\tilde b}=n^{a\tilde b}$. Further, vanishing of the quantity
(\ref{det-PB-2}) required for the presence of odd first class constraints leads to the condition
\begin{equation}\label{constr-3}
(\hat \nu)^{2}=4(m_0)^2\hat n \,.
\end{equation}
If  we wish to have eight odd first class constraints with the conditions (\ref{m-mu-0}),
the relations (\ref{12-1}) and (\ref{constr-3}) (as consequences of the condition $A=0$) are valid, and they imply
that $\hat n=0$, $\hat\nu=0\,$, whence $\hat w=0\,$ and, further, $n^{a\tilde b} = \nu^{a\tilde b} =0\,$.
Thus, if non-vanishing central charges $n^{a\tilde b}$, $\nu^{a\tilde b}$ are present
and $m_0\neq 0$, we can only obtain four odd first class constraints.

Let us separate now odd first and second class constraints.

The initial fermionic constraints ($D_\xi{}_\alpha^{a}$, $D_\xi{}_\alpha^{\tilde a}$;
$D_\theta{}_\alpha^{a}$, $D_\theta{}_\alpha^{\tilde a}$)
are equivalent to the set ($G{}_\alpha^{a}$, $F_\alpha^{\tilde a}$; $D_\theta{}_\alpha^{a}$, $D_\theta{}_\alpha^{\tilde a}$),
where ($G{}_\alpha^{a}$, $F_\alpha^{\tilde a}$) are defined by the following expressions
\bea
\label{constr-f-f1-3}
G_\alpha^{a} &:=& D_\xi{}_\alpha^{a} +
\frac{1}{2m_0}\, p_{x\,}{}_{\alpha}{}^{\beta}  D_\theta{}_\beta^{a} +
\frac{i}{2m_0}\,\nu^{a}{}_{\tilde b} D_\theta{}_\alpha^{\tilde b} \ \approx 0 \,,
\\
\label{constr-f-f2-3}
F_\alpha^{\tilde a} &:=& \ D_\xi{}_\alpha^{\tilde a} +
\frac{1}{2m_0}\, p_{x\,}{}_{\alpha}{}^{\beta}  D_\theta{}_\beta^{\tilde a} -
\frac{i}{2m_0}\,\nu_{b}{}^{\tilde a} D_\theta{}_\alpha^{b} -
\frac{2m_0}{\hat\nu}\,n_{b}{}^{\tilde a} G_\alpha^{b}\ \approx 0
\eea
(cf. (\ref{constr-f-f1-2})).
The complete set of non-vanishing Poisson brackets for the constraints (\ref{constr-f-f1-3}), (\ref{constr-f-f2-3})
and $D_\theta{}_\alpha^{a}$, $D_\theta{}_\alpha^{\tilde a}$
looks as follows
\bea
\{F_\alpha^{\tilde a}, F_\beta^{\tilde b}\}&=&
-\,2i\left[ 1 +\hat n\left(\frac{2m_0}{\hat\nu}\right)^2\right]
\epsilon_{\alpha\beta}\epsilon^{\tilde a\tilde b}\,T
\ + \ \frac{4im_0}{\hat\nu}\,\epsilon_{\alpha\beta}\epsilon^{\tilde a\tilde b}\left[ \hat n - \left(\frac{\hat\nu}{2m_0}\right)^2\right] \,, \nn
\{F_\alpha^{\tilde a}, G_\beta^{b}\}&=&
-\,\frac{4im_0}{\hat\nu}\,\epsilon_{\alpha\beta}n^{b\tilde a}\,T \,, \nn
\{G_\alpha^{a}, G_\beta^{b}\}&=&
-\,2i\,\epsilon_{\alpha\beta}\epsilon^{ab}\left( T+ \frac{\hat\nu}{2m_0}\right)\,,\nn
\{D_\theta{}_\alpha^{a}, D_\theta{}_\beta^{b}\}&=&
-\,4i\,m_0\epsilon_{\alpha\beta}\epsilon^{ab} \,,\qquad\quad
\{D_\theta{}_\alpha^{\tilde a}, D_\theta{}_\beta^{\tilde b}\} \ = \
-\,4i\,m_0\epsilon_{\alpha\beta}\epsilon^{\tilde a\tilde b}\,.
\label{PB-2f-constr-3}
\eea
Thus, if the condition (\ref{constr-3}) is valid, four constraints $F_\alpha^{\tilde a}$, defined in (\ref{constr-f-f2-3})
are first class. The constraints $D_\theta{}_\alpha^{a}$, $D_\theta{}_\alpha^{\tilde a}$ and $G_\alpha^{a}$,
defined in (\ref{constr-f-f1-3}) are second class.

\setcounter{equation}{0}
\section{Quantization of the model and $\mathcal{N}=4,$ $d=3$ Galilean superfields}

\quad\, In this section we present the canonical operator quantization of our model.
We will introduce the (super)Schr\"{o}dinger realization of quantum phase coordinates and obtain
the superfield description of $\mathcal{N}=4,$ $d=3$  Galilean states.
For this purpose we will quantize the second class constraints by the Gupta-Bleuler (GB) procedure \cite{AzLuk88,HasKL},
without introducing for them the Dirac brackets.

We will consider two versions of our model:
the first one with eight first class constraints (introducing $\frac12$ BPS states or  fraction $\frac12$ of unbroken supersymmetry)
and the second one with four first class constraints (introducing $\frac14$ BPS states or  fraction $\frac14$ of unbroken supersymmetry).

We will use, instead of the symplectic-Majorana real quantities, the complex Hermitian conjugate Grassmann coordinates,
which are defined by (\ref{MW-cond-G-c}) in the following way
\bea
&
\xi^{\alpha} := \xi^{\alpha}_{1} \,,\quad \bar\xi_{\alpha} := (\xi^{\alpha})^\dagger\,,\quad
\xi^{\alpha}_{2}= -\epsilon^{\alpha\beta}\bar\xi_{\beta} \,,\,\,&\,\,
\tilde\xi^{\alpha} := \xi^{\alpha}_{\tilde 1} \,,\quad \bar{\tilde\xi}_{\alpha} := (\tilde\xi^{\alpha})^\dagger\,,\quad
\xi^{\alpha}_{\tilde 2}= -\epsilon^{\alpha\beta}\bar{\tilde\xi}_{\beta} \,, \nn
&
\theta^{\alpha} := \theta^{\alpha}_{1} \,,\quad \bar\theta_{\alpha} := (\theta^{\alpha})^\dagger\,,\quad
\theta^{\alpha}_{2}= \epsilon^{\alpha\beta}\bar\theta_{\beta} \,,\,\,&\,\,
\tilde\theta^{\alpha} := \theta^{\alpha}_{\tilde 1} \,,\quad \bar{\tilde\theta}_{\alpha} := (\tilde\theta^{\alpha})^\dagger\,,\quad
\theta^{\alpha}_{\tilde 2}= \epsilon^{\alpha\beta}\bar{\tilde\theta}_{\beta} \,.
\label{C-cond-G-c}
\eea
The corresponding complex momenta, which are explicit solutions of the conditions (\ref{MW-cond-G-p}), are
\bea
&
p_\xi{}_{\alpha} := p_\xi{}^{1}_{\alpha} ,\  \bar p_\xi{}^{\alpha} := -(p_\xi{}_{\alpha})^\dagger,\
p_\xi{}^{2}_{\alpha}= \epsilon_{\alpha\beta}\bar p_\xi{}^{\beta} ,\,&\,
p_{\tilde\xi}{}_{\alpha} := p_{\xi}{}^{\tilde 1}_{\alpha} ,\  \bar p_{\tilde\xi}{}^{\alpha} := -(p_{\tilde\xi}{}_{\alpha})^\dagger,\
p_{\xi}{}^{\tilde 2}_{\alpha}= \epsilon_{\alpha\beta}\bar p_{\tilde\xi}{}^{\beta} , \nn
&
p_\theta{}_{\alpha} := p_\theta{}^{1}_{\alpha} ,\  \bar p_\theta{}^{\alpha} := -(p_\theta{}_{\alpha})^\dagger,\
p_\theta{}^{2}_{\alpha}= -\epsilon_{\alpha\beta}\bar p_\theta{}^{\beta} ,\,&\,
p_{\tilde\theta}{}_{\alpha} := p_\theta{}^{\tilde 1}_{\alpha} ,\  \bar p_{\tilde\theta}{}^{\alpha} := -(p_{\tilde\theta}{}_{\alpha})^\dagger,\
p_\theta{}^{\tilde 2}_{\alpha}= -\epsilon_{\alpha\beta}\bar p_{\tilde\theta}{}^{\beta}.
\label{C-cond-G-p}
\eea
Poisson brackets of these phase superspace variables are given by
\bea
&&\{ \xi^\alpha, p_\xi{}_\beta \}=\delta^\alpha_\beta \,,\quad
\{ \tilde\xi{}^\alpha, p_{\tilde\xi}{}_\beta \}=\delta^\alpha_\beta \,,\qquad
\{ \theta^\alpha, p_\theta{}_\beta \}=\delta^\alpha_\beta \,,\quad
\{ \tilde\theta{}^\alpha, p_{\tilde\theta}{}_\beta \}=\delta^\alpha_\beta\,, \nn
&&\{ \bar\xi_\alpha, \bar p_\xi{}^\beta \}=\delta_\alpha^\beta \,,\quad
\{ \bar{\tilde\xi}{}_\alpha, \bar p_{\tilde\xi}{}^\beta \}=\delta_\alpha^\beta \,,\qquad
\{ \bar\theta_\alpha, \bar p_\theta{}^\beta \}=\delta_\alpha^\beta \,,\quad
\{ \bar{\tilde\theta}{}_\alpha, \bar p_{\tilde\theta}{}^\beta \}=\delta_\alpha^\beta\,.
\label{PB-c-1}
\eea

In the quantization procedure, we will use the graded coordinate representation with the following
super-Schr\"{o}dinger realization for the momenta:
\bea
&&p_t=-i\frac{\partial}{\partial t}\,, \quad
p_{x\, i} =-i\frac{\partial}{\partial x_i} \,, \nn
&& p_\xi{}_\alpha=i\frac{\partial}{\partial \xi^\alpha}\,, \quad
p_{\tilde\xi}{}_\alpha =i\frac{\partial}{\partial \tilde\xi{}^\alpha} \,, \qquad
p_\theta{}_\alpha=i\frac{\partial}{\partial \theta^\alpha}\,, \quad
p_{\tilde\theta}{}_\alpha =i\frac{\partial}{\partial \tilde\theta{}^\alpha} \,, \nn
&& \bar p_\xi{}^\alpha=i\frac{\partial}{\partial \bar \xi_\alpha}\,, \quad
\bar p_{\tilde\xi}{}^\alpha =i\frac{\partial}{\partial \bar{\tilde\xi}{}_\alpha} \,, \qquad
\bar p_\theta{}^\alpha=i\frac{\partial}{\partial \bar\theta_\alpha}\,, \quad
\bar p_{\tilde\theta}{}^\alpha =i\frac{\partial}{\partial \bar{\tilde\theta}{}_\alpha} \,.
\label{mom-real}
\eea

\subsection{The model with vanishing off-diagonal central charges }

The system is described by
even first class constraint (\ref{constr-b-1})
\begin{equation}\label{constr-b-c}
T \ = \ p_{\,t}+ \frac{p_{\,x}{}_i p_{\,x}{}_i}{2m_0} \ \approx \  0\,,
\end{equation}
odd first class constraints (\ref{F-constr}) and odd second class constraints (\ref{constr-f-1}).
In terms of complex variables (\ref{C-cond-G-c}), (\ref{C-cond-G-p}) odd first class constraints (\ref{F-constr})
are written as
\bea
F_\xi{}_\alpha =& \!\!\!F_\xi{}_\alpha^{1}\!\!\! & =  p_\xi{}_\alpha
-i\left(p_{\,t}+m_1\right)\bar\xi_\alpha +
\frac{1}{2m_0}\Big( p_{x\,}{}_{\alpha}{}^{\beta} -i\mu_1\delta_\alpha^\beta\Big)
\Big( p_\theta{}_\beta -2m_0 i\,\bar\theta_\beta  \Big)  \approx  0 \,, \nn
\bar F_\xi{}^\alpha =& \!\!\!\epsilon^{\alpha\beta}F_\xi{}_\beta^{2}\!\!\! & = \bar p_\xi{}^\alpha
-i\left(p_{\,t}+m_1\right)\xi^\alpha +
\frac{1}{2m_0}\Big(\bar p_\theta{}^\beta -2m_0 i\,\theta^\beta \Big)
\Big( p_{x\,}{}_{\beta}{}^{\alpha} +i\mu_1\delta_\beta^\alpha\Big)
\approx  0 \,, \nn
F_{\tilde\xi}{}_\alpha =& \!\!\!F_{\xi}{}_\alpha^{\tilde1}\!\!\! & = p_{\tilde\xi}{}_\alpha
-i\left(p_{\,t}+m_2\right)\bar{\tilde\xi}_\alpha +
\frac{1}{2m_0}\Big( p_{x\,}{}_{\alpha}{}^{\beta} -i\mu_2\delta_\alpha^\beta\Big)
\Big( p_{\tilde\theta}{}_\beta -2m_0 i\,\bar{\tilde\theta}_\beta  \Big)  \approx  0 \,, \nn
\bar F_{\tilde\xi}{}^\alpha =& \!\!\!\epsilon^{\alpha\beta}F_\xi{}_\beta^{\tilde 2}\!\!\! & = \bar p_{\tilde\xi}{}^\alpha
-i\left(p_{\,t}+m_2\right){\tilde\xi}^\alpha +
\frac{1}{2m_0}\Big(\bar p_{\tilde\theta}{}^\beta -2m_0 i\,{\tilde\theta}^\beta \Big)
\Big( p_{x\,}{}_{\beta}{}^{\alpha} +i\mu_2\delta_\beta^\alpha\Big) \approx  0 \,.
\label{F-constr-c}
\eea
Odd second class constraints (\ref{constr-f-1}) read
\bea
D_{\theta}{}_\alpha \  =& D_{\theta}{}^{1}_\alpha & = \ p_{\theta}{}_\alpha +2m_0i\,\bar\theta_\alpha \ \approx \ 0\,, \nn
\bar D_{\theta}{}^\alpha \  =& -\epsilon^{\alpha\beta}D_{\theta}{}^{2}_\beta & = \  \bar p_{\theta}{}^\alpha +2m_0i\, \theta^\alpha \ \approx \ 0\,, \nn
D_{\tilde \theta}{}_\alpha \  =& D_{\theta}{}^{\tilde 1}_\alpha & = \  p_{\tilde \theta}{}_\alpha +2m_0i\,\bar{\tilde \theta}_\alpha \ \approx \ 0\,, \nn
\bar D_{\tilde \theta}{}^\alpha \  =& -\epsilon^{\alpha\beta}D_{\theta}{}^{\tilde 2}_\beta & = \
\bar p_{\tilde \theta}{}^\alpha +2m_0i\, \tilde \theta^\alpha \ \approx \ 0\,.
\label{2nd-1}
\eea

Second class constraints (\ref{2nd-1}) form the Hermitian conjugate pairs $\big(D_{\theta}{}_\alpha\,,\, \bar D_{\theta}{}^\alpha\big)$
and $\big(D_{\tilde \theta}{}_\alpha\,,\, \bar D_{\tilde \theta}{}^\alpha\big)$, what permits us to apply Gupta-Bleuler quantization.
In accord with this quantization technique we impose on wave function half of second class constraints, {\it i.e.},
$\bar D_{\theta}{}^\alpha$ and $\bar D_{\tilde \theta}{}^\alpha\,$.

Thus the wave function is the superfield
\begin{equation}\label{wf-1}
\Psi \ = \ \Psi(t,x_i;\, \xi^{\alpha}, \bar\xi_{\alpha};\,
\tilde\xi^{\alpha}, \bar{\tilde\xi}_{\alpha}; \,
\theta^{\alpha}, \bar\theta_{\alpha};\,
\tilde\theta^{\alpha}, \bar{\tilde\theta}_{\alpha})\,,
\end{equation}
which satisfies the conditions
\begin{equation}\label{constr-Sh}
T \, \Psi \ = \  0\,,
\end{equation}
\begin{equation}\label{constr-1st}
F_\xi{}_\alpha \, \Psi \ = \  0\,,\qquad
\bar F_\xi{}^\alpha \, \Psi \ = \  0\,,\qquad
F_{\tilde\xi}{}_\alpha \, \Psi \ = \  0\,,\qquad
\bar F_{\tilde\xi}{}^\alpha \, \Psi \ = \  0\,,
\end{equation}
\begin{equation}\label{constr-2nd}
\bar D_{\theta}{}^\alpha \, \Psi \ = \  0\,,
\qquad
\bar D_{\tilde \theta}{}^\alpha \, \Psi \ = \  0\,.
\end{equation}

The solution of second class constraints (\ref{constr-2nd}) is the following
\begin{equation}\label{wf-sol0-1}
\Psi \ = \ \exp\left\{ 2m_0\left({\theta}^\alpha\bar{\theta}_\alpha+{\tilde \theta}^\alpha\bar{\tilde \theta}_\alpha \right)\right\}\Phi\,,
\end{equation}
where the superfield
\begin{equation}\label{wf-red-1}
\Phi \ = \ \Phi(t,x_i;\, \xi^{\alpha}, \bar\xi_{\alpha};\,
\tilde\xi^{\alpha}, \bar{\tilde\xi}_{\alpha}; \,
\theta^{\alpha};\,\tilde\theta^{\alpha})
\end{equation}
is chiral with respect to the variables ($\theta^\alpha$, $\tilde\theta^\alpha$).

Even first class constraint (\ref{constr-Sh}) yields the Schr\"{o}dinger equations for all component fields.

Odd first class constraint (\ref{constr-1st}) provide  the following  four superwave equations for the superfield (\ref{wf-red-1})
\bea
&\left[\mathcal{D}_\xi{}_\alpha -\Delta_\theta{}_\alpha \right] \Phi :=&
\left[\mathcal{D}_\xi{}_\alpha +
\frac{1}{2m_0}\Big( p_{x\,}{}_{\alpha}{}^{\beta} -i\mu_1\delta_\alpha^\beta\Big)
\frac{\partial}{\partial\theta^\beta}  \right] \Phi =  0 \,, \nn
&\left[\bar{\mathcal{D}}_\xi{}^\alpha -\bar{\Delta}_\theta{}^\alpha \right] \Phi :=&
\left[\bar{\mathcal{D}}_\xi{}^\alpha -2\,\theta^\beta
\Big( p_{x\,}{}_{\beta}{}^{\alpha} +i\mu_1\delta_\beta^\alpha\Big)
\right] \Phi =  0 \,, \nn
&\left[\mathcal{D}_{\tilde\xi}{}_\alpha -\tilde\Delta_{\theta}{}_\alpha \right] \Phi :=&
\left[\mathcal{D}_{\tilde\xi}{}_\alpha +
\frac{1}{2m_0}\Big( p_{x\,}{}_{\alpha}{}^{\beta} -i\mu_2\delta_\alpha^\beta\Big)
\frac{\partial}{\partial{\tilde\theta}^\beta}  \right] \Phi =  0 \,, \nn
&\left[\bar{\mathcal{D}}_{\tilde\xi}{}^\alpha -\bar{\tilde\Delta}_{\theta}{}^\alpha \right] \Phi :=&
\left[\bar{\mathcal{D}}_{\tilde\xi}{}^\alpha -2\,\tilde\theta^\beta
\Big( p_{x\,}{}_{\beta}{}^{\alpha} +i\mu_2\delta_\beta^\alpha\Big)
\right] \Phi =  0  \,.
\label{F-constr-Phi}
\eea
Here we have introduced the operators
$\Delta_{\theta}{}_\alpha$, $\bar{\Delta}_{\theta}{}^\alpha$, $\tilde\Delta_{\theta}{}_\alpha$, $\bar{\tilde\Delta}_{\theta}{}^\alpha$ which do not depend on $\xi$-variables, and the  covariant derivatives for $\mathcal{N}{=}\,8$ extended one-dimensional supersymmetry
\bea
&&
\mathcal{D}_\xi{}_\alpha := \frac{\partial}{\partial\xi^\alpha}
-\left(p_{\,t}+m_1\right)\bar{\xi}_\alpha\,,\qquad
\bar{\mathcal{D}}_\xi{}^\alpha := \frac{\partial}{\partial\bar\xi_\alpha}
-\left(p_{\,t}+m_1\right)\xi^\alpha  \,, \nn
&&
\mathcal{D}_{\tilde\xi}{}_\alpha := \frac{\partial}{\partial\tilde\xi^\alpha}
-\left(p_{\,t}+m_2\right)\bar{\tilde\xi}_\alpha \,, \qquad
\bar{\mathcal{D}}_{\tilde\xi}{}^\alpha := \frac{\partial}{\partial\bar{\tilde\xi}_\alpha}
-\left(p_{\,t}+m_2\right){\tilde\xi}^\alpha\,,
\label{cov-D}
\eea
which form two mutually anticommuting $\mathcal{N}{=}\,4$ $d\,{=}1$ superalgebras
\begin{equation}\label{1d-salgebra}
\{\mathcal{D}_\xi{}_\alpha, \bar{\mathcal{D}}_\xi{}^\beta\}=-2\left(p_{\,t}+m_1\right)\delta_\alpha^\beta\,,\qquad
\{\mathcal{D}_{\tilde\xi}{}_\alpha, \bar{\mathcal{D}}_{\tilde\xi}{}^\beta\}=-2\left(p_{\,t}+m_1\right)\delta_\alpha^\beta\,,
\end{equation}
with constants $m_1$, $m_2$ playing the role of central charges. It is straightforward to check that
the integrability condition for the equations (\ref{F-constr-Phi}) is just the Schr\"{o}dinger equations  (\ref{constr-Sh})
(we mention that the conditions (\ref{cond-inv-1}), (\ref{cond-inv-2}) are valid).

It is easy to find the explicit solution of equations (\ref{F-constr-Phi}). Consider the expansion
of superfield (\ref{wf-red-1}) over the Grassmann coordinates ${\xi}^\alpha$, $\bar{\xi}_\alpha$, ${\tilde\xi}^\alpha$,
$\bar{\tilde\xi}_\alpha$
\begin{equation}\label{Phi-exp-xi}
\Phi \ = \ \Phi_0 + {\xi}^\alpha\Phi_\alpha +\bar{\xi}_\alpha\bar\Phi^\alpha +{\tilde\xi}^\alpha\tilde\Phi_\alpha +
\bar{\tilde\xi}_\alpha\bar{\tilde\Phi}^\alpha+{\xi}^\alpha{\xi}^\beta\Phi_{\alpha\beta}+
{\tilde\xi}^\alpha{\tilde\xi}^\beta\tilde\Phi_{\alpha\beta}+\cdots\,,
\end{equation}
where all components $\Phi_0$, $\Phi_\alpha$, $\bar\Phi^\alpha$, etc,  are ``two-chiral'' superfields dependent
on ($t$, $x_i$) and two chiral spinorial coordinates, ($\theta^\alpha$, $\tilde\theta^\alpha$).
The equations (\ref{F-constr-Phi}) express all the component superfields in the expansion (\ref{Phi-exp-xi})
in terms of the first component, the superfield $\Phi_0(t, x_i, \theta^\alpha, \tilde\theta^\alpha)$.
It follows from (\ref{F-constr-Phi}), that all covariant derivatives $\mathcal{D}_\xi{}_\alpha\Phi$, $\bar{\mathcal{D}}_\xi{}^\alpha\Phi$,
$\mathcal{D}_{\tilde\xi}{}_\alpha\Phi$, $\bar{\mathcal{D}}_{\tilde\xi}{}^\alpha\Phi$ and all their powers are expressed
in terms of powers of the $\Delta$-operators acting on $\Phi$:
$\mathcal{D}_{\xi}{}_\alpha\bar{\mathcal{D}}_{\xi}{}^\beta\Phi=
-\bar{\Delta}_{\theta}{}^\beta\Delta_{\theta}{}_\alpha\Phi$ etc.
So, being taken at ${\xi}^\alpha=\bar{\xi}_\alpha={\tilde\xi}^\alpha=\bar{\tilde\xi}_\alpha=0$,
all superfields in the expansion (\ref{Phi-exp-xi}) are expressed through derivatives of a single superfield
$\Phi_0(t, x_i, \theta^\alpha, \tilde\theta^\alpha)$.

To summarize, physical states of the considered model are described by the two-chiral
superfield $\Phi_0(t, x_i, \theta^\alpha, \tilde\theta^\alpha)$ with the following component expansion
\bea
\Phi_0 &=& A \ + \ \theta^{\alpha}\,\Omega_\alpha  \ + \ \tilde\theta^{\alpha}\,\tilde\Omega_\alpha  \ + \ \theta^{\alpha}\theta_{\alpha}\,B +
\tilde\theta^{\alpha}\tilde\theta_{\alpha}\,\tilde B  \ + \ \theta^{\alpha}\tilde\theta^{\beta}\, B_{\alpha\beta} \nn
&&  + \
\tilde\theta^{\beta}\tilde\theta_{\beta}\theta^{\alpha}\,\Lambda_\alpha  \ + \ \theta^{\beta}\theta_{\beta}\tilde\theta^{\alpha}\,\tilde\Lambda_\alpha  \ + \
\theta^{\alpha}\theta_{\alpha}\tilde\theta^{\beta}\tilde\theta_{\beta}\,C\,.\label{wf-red-1-b}
\eea
In  (\ref{wf-red-1-b}) all component fields are complex functions of $t$ and $x_i$ which satisfy
the Schr\"{o}dinger equation  (\ref{constr-Sh}).

Thus the presented model results in five complex scalar fields $A(t, x_i)$, $B(t, x_i)$, $\tilde B(t, x_i)$,
$B_{[\alpha\beta]}(t, x_i)$, $C(t, x_i)$ which describe spin $0$ states;
one complex vectorial field $B_{(\alpha\beta)}(t, x_i)$, which accommodates  spin $1$ states; and four spinorial fields
$\Omega_\alpha(t, x_i)$, $\tilde\Omega_\alpha(t, x_i)$, $\Lambda_\alpha(t, x_i)$, $\tilde\Lambda_\alpha(t, x_i)$ corresponding to spin $1/2$.
It is easy to see that we obtain in such a way equal number of 8 bosonic and 8 fermionic component fields.

It is worth to emphasize that the description of physical states by the two-chiral superfield (\ref{wf-red-1-b}) is consistent
with the possibility of imposing the gauges $\xi_\alpha^a=0$, $\xi_\alpha^{\tilde a}=0$ on the local transformations (\ref{kappa-tr1}),
as was mentioned in Sect.\,4.2.

\subsection{The models with non-vanishing off-diagonal central charges }

\quad\, Like in the previous subsection, we will use the complex variables (\ref{C-cond-G-c}), (\ref{C-cond-G-p}).

\subsubsection{The models with eight odd first class constraints}

\quad\, We consider firstly the case with condition (\ref{w-00}).

The wave function has the form (\ref{wf-1}) and the wave equations are derived from
even first class constraint (\ref{constr-Sh}) as well as odd first and second class constraints,
the last ones quantized by the Gupta-Bleuler quantization method.

Second class constraints have the same form as in (\ref{2nd-1}), while the solution of the  constraints
(\ref{constr-2nd}) leads to the reduced superwave function (\ref{wf-sol0-1}), (\ref{wf-red-1}).

{}For simplicity we will restrict our study to the case of parallel four-vectors $n_M$, $\nu_M$
\begin{equation}\label{n4}
n^{a\tilde b}=n\epsilon^{a\tilde b}\,,\qquad \nu^{a\tilde b}=\nu\epsilon^{a\tilde b}\,,\qquad \hat\nu =\nu^2
\end{equation}
where (see (\ref{w-00}))
\begin{equation}\label{n4-nu4}
n=-\, \frac{\mu_1+\mu_2}{2m_0}\, \nu\,.
\end{equation}
Other choices of the constants $n^{a\tilde b}$ and $\nu^{a\tilde b}$ satisfying (\ref{w-00}) lead to the same set of
physical states.

Using complex variables (\ref{C-cond-G-c}), (\ref{C-cond-G-p}), the odd first class constraints (\ref{F-constr-2})
can be brought to the form
\bea
F_\xi{}_\alpha &=&   p_\xi{}_\alpha
-i\left(p_{\,t}+m_1\right)\bar\xi_\alpha -i n \bar{\tilde\xi}_\alpha \nn
&& +
\frac{1}{2m_0}\,\Big( p_{x\,}{}_{\alpha}{}^{\beta} -i\mu_1\delta_\alpha^\beta\Big)
\Big( p_\theta{}_\beta -2m_0 i\,\bar\theta_\beta  \Big) -
\frac{i}{2m_0}\,\nu
\Big( p_{\tilde\theta}{}_\alpha -2m_0 i\,\bar{\tilde\theta}_\alpha  \Big) \ \approx \ 0 \,, \nn
\bar F_\xi{}^\alpha &=& \bar p_\xi{}^\alpha
-i\left(p_{\,t}+m_1\right)\xi^\alpha -i n {\tilde\xi}^\alpha \nn
&& +
\frac{1}{2m_0}\,\Big(\bar p_\theta{}^\beta -2m_0 i\,\theta^\beta \Big)
\Big( p_{x\,}{}_{\beta}{}^{\alpha} +i\mu_1\delta_\beta^\alpha\Big) +
\frac{i}{2m_0}\,\nu
\Big( \bar p_{\tilde\theta}{}^\alpha -2m_0 i\,{\tilde\theta}^\alpha  \Big)
\ \approx \ 0 \,,
\label{F-constr-c2a}
\\
F_{\tilde\xi}{}_\alpha &=& p_{\tilde\xi}{}_\alpha
-i\left(p_{\,t}+m_2\right)\bar{\tilde\xi}_\alpha -i n \bar{\xi}_\alpha\nn
&& +
\frac{1}{2m_0}\,\Big( p_{x\,}{}_{\alpha}{}^{\beta} -i\mu_2\delta_\alpha^\beta\Big)
\Big( p_{\tilde\theta}{}_\beta -2m_0 i\,\bar{\tilde\theta}_\beta  \Big)-
\frac{i}{2m_0}\,\nu
\Big( p_{\theta}{}_\alpha -2m_0 i\,\bar{\theta}_\alpha  \Big) \ \approx \ 0 \,, \nn
\bar F_{\tilde\xi}{}^\alpha &=& \bar p_{\tilde\xi}{}^\alpha
-i\left(p_{\,t}+m_2\right){\tilde\xi}^\alpha -i n {\xi}^\alpha\nn
&& +
\frac{1}{2m_0}\,\Big(\bar p_{\tilde\theta}{}^\beta -2m_0 i\,{\tilde\theta}^\beta \Big)
\Big( p_{x\,}{}_{\beta}{}^{\alpha} +i\mu_2\delta_\beta^\alpha\Big) +
\frac{i}{2m_0}\,\nu
\Big( \bar p_{\theta}{}^\alpha -2m_0 i\,{\theta}^\alpha  \Big) \ \approx \ 0 \,.
\label{F-constr-c2b}
\eea
These expressions generalize the ones given in  (\ref{F-constr-c}).
Imposing the constraints (\ref{constr-1st}) on the superwave function (\ref{wf-sol0-1}),
we obtain that the superfield (\ref{wf-red-1}) satisfies the following
generalization of four superwave equations (\ref{F-constr-Phi})
\bea
&&
\left[\nabla_\xi{}_\alpha  +
\frac{1}{2m_0}\Big( p_{x\,}{}_{\alpha}{}^{\beta} -i\mu_1\delta_\alpha^\beta\Big)
\frac{\partial}{\partial\theta^\beta} -
\frac{i\nu}{2m_0}\,
\frac{\partial}{\partial\tilde\theta^\alpha} \right] \Phi =  0 \,, \nn
&&
\left[\bar{\nabla}_\xi{}^\alpha   -2\,\theta^\beta
\Big( p_{x\,}{}_{\beta}{}^{\alpha} +i\mu_1\delta_\beta^\alpha\Big) - 2i\,\nu\tilde\theta^\alpha
\right] \Phi =  0 \,,
\label{F-constr-Phi-2a}
\\
&&
\left[\nabla_{\tilde\xi}{}_\alpha  +
\frac{1}{2m_0}\Big( p_{x\,}{}_{\alpha}{}^{\beta} -i\mu_2\delta_\alpha^\beta\Big)
\frac{\partial}{\partial{\tilde\theta}^\beta}  -
\frac{i\nu}{2m_0}\,
\frac{\partial}{\partial\theta^\alpha}\right] \Phi =  0 \,, \nn
&&
\left[\bar{\nabla}_{\tilde\xi}{}^\alpha   -2\,\tilde\theta^\beta
\Big( p_{x\,}{}_{\beta}{}^{\alpha} +i\mu_2\delta_\beta^\alpha\Big) - 2i\,\nu\theta^\alpha
\right] \Phi =  0  \,,
\label{F-constr-Phi-2b}
\eea
where
\bea
&&
\nabla_\xi{}_\alpha := \mathcal{D}_\xi{}_\alpha
-n\bar{\tilde\xi}_\alpha\,,\qquad
\bar{\nabla}_\xi{}^\alpha := \bar{\mathcal{D}}_\xi{}^\alpha
-n{\tilde\xi}^\alpha   \,, \nn
&&
\nabla_{\tilde\xi}{}_\alpha := \mathcal{D}_{\tilde\xi}{}_\alpha
-n\bar{\xi}_\alpha \,, \qquad
\bar{\nabla}_{\tilde\xi}{}^\alpha := \bar{\mathcal{D}}_{\tilde\xi}{}^\alpha
-n \xi^\alpha
\label{cov-nabla}
\eea
and $\mathcal{D}_\xi{}_\alpha$, $\bar{\mathcal{D}}_\xi{}^\alpha$, $\mathcal{D}_{\tilde\xi}{}_\alpha$,
$\bar{\mathcal{D}}_{\tilde\xi}{}^\alpha$ are defined in (\ref{cov-D}).
If the conditions
(\ref{cond-inv-1a}), (\ref{cond-inv-2a}) and (\ref{w-00}) are valid once again, the integrability condition for the system of equations (\ref{F-constr-Phi-2a}), (\ref{F-constr-Phi-2b})
is the free Schr\"{o}dinger equations  (\ref{constr-Sh}).

The unconstrained superfield can be found as in the previous subsection:
we apply the expansion (\ref{Phi-exp-xi}) and again find that all superfield components in this expansion
are expressed through the derivatives of single two-chiral superfield $\Phi_0(t, x_i, \theta^\alpha, \tilde\theta^\alpha)$
(see (\ref{wf-red-1-b})). Therefore, the considered model with off-diagonal central charges has the same physical fields content
as the previously studied model with the diagonal central charges only.

\subsubsection{The models with four odd first class constraints}

\quad\, Now we will consider the case when central charges take the values (\ref{m-mu-0}), (\ref{constr-3}).
Also, for definiteness, we choose the values (\ref{n4}), (\ref{n4-nu4}) for off-shell central charges.
Imposing on the wave function (\ref{wf-1}) the second class constraints (\ref{2nd-1}), (\ref{constr-2nd})
we obtain the same superfield solution (\ref{wf-sol0-1}), (\ref{wf-red-1}) as in the previous cases.

The remaining odd constraints are second class constraints $G{}_\alpha^{a}$, defined in (\ref{constr-f-f1-3}),
and first class constraints $F_\alpha^{\tilde a}$, defined in  (\ref{constr-f-f2-3}).
The second class constraints (\ref{constr-f-f1-3}) coincide with the constraints
(\ref{F-constr-c2a}), (\ref{F-constr-Phi-2a}) taken at $m_1 =m_2=0$, $\mu_1 =\mu_2=0$:
\begin{equation}\label{F-G}
G{}_\alpha^{a}=(F_\xi{}_\alpha, \bar F_\xi{}^\alpha)\,.
\end{equation}
We quantize these constraints by the Gupta-Bleuler procedure, imposing on the wave function the
half of them:
\begin{equation}\label{G-GB}
\bar F_\xi{}^\alpha\,\Psi=0\quad\Rightarrow\quad\bar F_\xi{}^\alpha\,\Phi=0\,.
\end{equation}
Using (\ref{F-constr-Phi-2a}) at $m_1 =m_2=0$, $\mu_1 =\mu_2=0$, we find the solution of (\ref{G-GB})
\begin{equation}\label{wf-sol0-g}
\Phi \ = \ \exp\left\{ -p_t{\xi}^\alpha\bar{\xi}_\alpha
-n\tilde{\xi}^\alpha\bar{\xi}_\alpha-2\theta^\alpha p_x{}_\alpha{}^\beta\bar{\xi}_\beta
-2i\nu\tilde{\theta}^\alpha\bar{\xi}_\alpha \right\}\Omega\,,
\end{equation}
where
\begin{equation}\label{wf-red-g}
\Omega \ = \ \Omega(t,x_i;\, \xi^{\alpha};\,
\tilde\xi^{\alpha}, \bar{\tilde\xi}_{\alpha}; \,
\theta^{\alpha};\,\tilde\theta^{\alpha})
\end{equation}
is a three-chiral superfield with respect to $\xi^{\alpha}$ and ($\theta^{\alpha}$, $\tilde\theta^{\alpha}$) variables.

The last constraints which we need to solve are the first class constraints rewritten in terms of the operators
(\ref{F-constr-c2a}), (\ref{F-constr-c2b}) and (\ref{F-constr-Phi-2a}), (\ref{F-constr-Phi-2b})
(at $m_1 =m_2=0$, $\mu_1 =\mu_2=0$) in the following form
\bea
F_\alpha \Phi&=& \left( F_{\tilde\xi}{}_\alpha  -\frac{2m_0 n}{\hat\nu}\, F_\xi{}_\alpha\right) \Phi \ = \ 0 \,,
\label{F-constr-1Phi}
\\
\bar F{}^\alpha \Phi &=& \left( \bar F_{\tilde\xi}{}^\alpha-\frac{2m_0 n}{\hat\nu}\, \bar F_\xi{}^\alpha\right) \Phi \ = \
\bar F_{\tilde\xi}{}^\alpha \,   \Phi \ = \ 0 \,.
\label{F-constr-2Phi}
\eea
Using (\ref{F-constr-Phi-2a}), (\ref{F-constr-Phi-2b}),  we obtain that the equations (\ref{F-constr-1Phi}), (\ref{F-constr-2Phi})
amount to the following pair of equations for the superfield (\ref{wf-red-g})
\bea
&&
\left(\mathcal{D}_{\tilde\xi}{}_\alpha  +
\frac{1}{2m_0}\, p_{x\,}{}_{\alpha}{}^{\beta} \frac{\partial}{\partial{\tilde\theta}^\beta}  -
\frac{i\nu}{2m_0}\,\frac{\partial}{\partial\theta^\alpha}
-\frac{2m_0n}{\hat\nu} \frac{\partial}{\partial\xi^\alpha}  -
\frac{n}{\hat\nu}\, p_{x\,}{}_{\alpha}{}^{\beta} \frac{\partial}{\partial\theta^\beta} +
\frac{i n}{\nu}\,
\frac{\partial}{\partial\tilde\theta^\alpha}\right) \Omega =  0 \,,
\label{F-cons-Om-2a}
\\
&&
\left(\bar{\mathcal{D}}_{\tilde\xi}{}^\alpha  -n \xi^\alpha -2\,\tilde\theta^\beta
p_{x\,}{}_{\beta}{}^{\alpha} - 2i\,\nu\theta^\alpha
\right) \Omega =  0  \,.
\label{F-cons-Om-2b}
\eea
One can consider its  general expansion
with respect to Grassmann coordinates ${\tilde\xi}^\alpha$, $\bar{\tilde\xi}_\alpha$
\begin{equation}\label{Om-exp-xi}
\Omega \ = \ \Omega_0 + \tilde{\xi}^\alpha\Omega_\alpha +\bar{\tilde\xi}_\alpha\bar\Omega^\alpha +
{\tilde\xi}^\alpha{\tilde\xi}_\alpha \Omega_1 +
\bar{\tilde\xi}_\alpha\bar{\tilde\xi}^\alpha\bar{\Omega}_1+{\tilde\xi}^\alpha\bar{\tilde\xi}^\beta\Omega_{\alpha\beta}+
\cdots\,,
\end{equation}
where all components $\Omega_0$, $\Omega_\alpha$, $\bar\Omega^\alpha$, etc., are three-chiral superfields depending
on $t$, $x_i$ and three chiral Weyl spinors $\theta^\alpha$, $\tilde\theta^\alpha$, $\xi^\alpha$.
As in the previous subsections, we obtain that  due to the equations (\ref{F-cons-Om-2a}), (\ref{F-cons-Om-2b}), all three-chiral
superfields in \p{Om-exp-xi} (except $\Omega_0$) are represented as derivatives of this lowest  component superfield $\Omega_0$.

Thus, the solutions of the model in this subsection are
parametrized by a three-chiral superfield
\begin{equation}\label{Om-sol}
\Omega_0 \ = \ \Omega_0 (t, x_i, \theta^\alpha, \tilde\theta^\alpha, \xi^\alpha)\,.
\end{equation}
Furthermore we recall that all component fields of the superfield (\ref{Om-sol}) satisfy the Schr\"{o}dinger equation.

Finally, we would like to notice that both types of chiral superfields
describing superwave functions
in this Section are essentially
on-shell since they require Schr\"{o}dinger equation as the integrability condition of the relevant odd first class constraints.

\setcounter{equation}{0}
\section{Conclusions}

\quad\, In this paper we considered the $\mathcal{N}{=}\,4,$ $d{=}\,3$ NR superparticle models with twelve constant central charges
transforming in certain representations of  $\mathrm{USp}(4) \sim \mathrm{O}(5) $ and
$\mathrm{USp}(2){\otimes} \mathrm{USp}(2)\simeq \mathrm{O}(4) \subset \mathrm{O}(5)$ internal R-symmetry groups.
The maximal $\mathrm{U}(4)$ R-symmetry group of relativistic ${\cal N}=4\,, \,D=4$ superalgebra in the NR contraction limit
is reduced to a semi-direct product of the compact R-symmetry group  $\mathrm{USp}(4)\simeq \mathrm{O}(5)$ and some
abelian six-dimensional commutative ideal.  In the dynamical framework of our superparticle model, after quantization
the central charges are identified with constant parameters of the underlying world-line
Lagrangian. Depending on the specific non-vanishing values of these central charges, we are left, before any Hamiltonian analysis,
with different fractions of unbroken internal symmetry,  $G_{int}\subset \mathrm{USp}(4)\simeq \mathrm{O}(5)$, namely
\begin{description}
\item[a)]
If only  one central charge $\mathcal{Z}$ (see (\ref{Z-1})) is present, the maximal R-symmetry
$\mathrm{O}(5)$ of NR $\mathcal{N}{=}\,4$ superalgebra remains in the model; the central charge is $\mathrm{O}(5)$ singlet;
\item[b)]
If we have two quasi-diagonal central charges (see (\ref{eqfli1.5})) the internal symmetry is broken to
$G_{int}=\mathrm{USp}(2){\otimes} \mathrm{USp}(2)\simeq \mathrm{SU}(2){\otimes} \mathrm{SU}(2)\simeq \mathrm{O}(4)$.
The central charges are presented by four $\mathrm{USp}(2)$ singlets $m_1, m_2, \mu_1$ and $\mu_2$;
\item[c)]
Adding the off-diagonal central charges described by two arbitrary constant $\mathrm{O}(4)$ isovectors $n^{a\tilde b}$ and
$\nu^{a\tilde b}$ (see (\ref{nnu-sigma})) radically changes the situation. At $m_1 \neq m_2$ and $\mu_1 \neq \mu_2$ the set
of twelve constant central charges determines two $\mathrm{O}(5)$ vectors $(n^{a\tilde b}\,, \;m_1 -m_2)$ and
$(\nu^{a\tilde b}\,, \; \mu_1 -\mu_2)$ and two $O(5)$ singlets $(m_1 + m_2)$ and $(\mu_1 + \mu_2)$. The $\mathrm{O}(5)$ frame can be fixed
so that one of these $\mathrm{O}(5)$ vectors carries only one non-zero component, say $m_1 -m_2$. There still remains $\mathrm{O}(4)$ covariance
which can
be further restricted in such a way that another  $\mathrm{O}(4)$ vector $\nu^{a\tilde b}$
(see (\ref{nnu-sigma})) will have only one non-zero
component, $\nu^{a\tilde b} = \epsilon^{a\tilde b}\nu \,$. Thus in such R-symmetry frame we end up with five independent constant central charges
and $\mathrm{O}(3)$ as the residual R-symmetry  group;
\item[d)] If $m_1 = m_2$ and/or $\mu_1 = \mu_2$, the $\mathrm{O}(5)$  covariance is reduced to $\mathrm{O}(4)$.
Hence we can choose the frame where $\mathrm{O}(4)$ isovector $n^{a\tilde b}$ contains only one non-zero component,
$n^{a\tilde b}  = \epsilon^{a\tilde b}n \,$, and the residual R-symmetry group is $\mathrm{O}(3)$. The second non-parallel $\mathrm{O}(4)$
vector $\nu^{a\tilde b}$ can be split into $\mathrm{O}(3)$ singlet and vector parts,
with only one
non-zero vector component, $\nu^{a\tilde b} \rightarrow (\epsilon^{a\tilde b}\,\nu, \;\delta^{a\tilde b} \nu_2)$, that reduces
R-symmetry $\mathrm{O}(3)$ to the  minimally possible one, given by $\mathrm{O}(2)$.
Thus in this particular frame we end up with six (or less) independent constant central charges and
$\mathrm{O}(2)$ as the minimal exact internal symmetry in our model.
\end{description}

It should be added that rest mass $m_0$, describing the Bargmann central charge in Galilean sector,
can be treated as the thirteenth central charge, which does not break any R-symmetry.

The superparticle action, constructed in Sect.\,3 and Sect.\,4 of the present paper, is linear in the MC one-forms
associated with central charges. The numerical coefficients in front of the central charge  MC one-forms
provide  the numerical values of central charges. In our further work we plan to consider also alternative
action densities as nonlinear functions of MC forms, which would permit, e.g., a generalization of (\ref{act-2})
to the action  manifestly invariant under the $\mathrm{O}(4)$ internal symmetries.
In particular, following the construction of the model for free relativistic massive particle
\footnote{In the case of relativistic particle in a flat space the action is equal to
$m\int (\omega^\mu \omega_\mu)^{1/2}$, where $\omega^\mu=dx^\mu$.
In NR theories we are interested in such type of actions will be employed in the internal symmetry sector only.},
one can replace the action (\ref{act-2}) by the $\mathrm{USp}(2){\otimes}\mathrm{USp}(2)$-invariant action depending on
all eight off-diagonal central charges
\begin{equation}\label{act-2-a}
S_2^\prime\ =\  \int \left( \ k_1\,\sqrt{\hat\omega_{(X)}^{\,a\tilde b}\hat\omega_{(X)\,a\tilde b}} +
k_2\,\sqrt{\hat\omega_{(Y)}^{\,a\tilde b}\hat\omega_{(Y)\,a\tilde b}} +
k_3\,\sqrt{\hat\omega_{(X)}^{\,a\tilde b}\hat\omega_{(Y)\,a\tilde b}} \ \right)\,,
\end{equation}
where $k_1$, $k_2$, $k_3$ are constant and $\hat\omega_{(X)}^{\,a\tilde b}$, $\hat\omega_{(Y)}^{\,a\tilde b}$
are defined in  (\ref{Car-f-hat}). The system described by the action $S_1 + S_2^\prime$
produces the same fermionic constraints (\ref{constr-f-2}) where, however, $n^{a\tilde b}$ and $\nu^{a\tilde b}$ are not
constant anymore: they become the canonical momenta for the tensorial central charge coordinates
$h_{a\tilde b}$ and $f_{a\tilde b}$ (see Sect.\,3, eqs. (\ref{factor}) and (\ref{Car-f-hat})).
So, although the fermionic constraints have basically the same form in both models
(see (\ref{act-2-a}) and (\ref{act-2})),
in the case of the action (\ref{act-2-a})
the group parameters $h_{a\tilde b}$ and $f_{a\tilde b}$
are introduced as the dynamical tensorial central charges coordinates. In such a way we deal with an extension of the bosonic target space sector
$(t,x_i)$ describing NR space-time to an extended
target space with auxiliary central charge coordinates $(t,x_i;h_{a\tilde b},f_{a\tilde b})$.
Additional coordinates $h_{a\tilde b}$, $f_{a\tilde b}$ enter into new three bosonic constraints
which fix $\hat n=n_M n_M$, $\hat \nu=\nu_M \nu_M$, $n_{M}\nu_{M}$ (see (\ref{def-scalars})) by $(k_1)^2$, $(k_2)^2$, $(k_3)^2$.
In such a way we obtain a sort of Kaluza-Klein (KK) extension of the superparticle model, with auxiliary KK bosonic dimensions represented by
central charge coordinates.  Analysis of this modified ${\cal N}=4$ NR
superparticle model will be given elsewhere (for an early attempt in this direction see \cite{LukAzc86}).

In the future we plan also to examine another way of  preserving internal symmetry by using
the harmonic type variables $u_a^b$ and $u_{\tilde a}^{\tilde b}$ which occur in the coset space parametrization
(\ref{factor}), as well as the ``genuine'' harmonic variables defined for the R-symmetry group $\mathrm{USp}(4)$.\footnote{The description
of various harmonic  superspaces for the relativistic ${\cal N}=4, D=4$ superalgebra and their implications in studying the quantum
structure of  ${\cal N}=4$ SYM theory can be found in a recent review \cite{BIS}.}
Further direction for the future study is to couple the NR superparticle presented here to electromagnetic, YM
and supergravity backgrounds.
It can be important for the following reason.
The energy momentum dispersion relations in our model for arbitrary spin states are described
by the free Schr\"{o}dinger equation depending on the same mass parameter $m_0\,$. One can argue that,
after switching on the background fields,
other central charges will also become dynamically active and will contribute to the modification of   Schr\"{o}dinger equation.

\section*{Acknowledgements}

\noindent
This research was supported by Polish National Science Center (NCN),
project \break $2014/13/\mathrm{B}/\mathrm{ST}2/04043$ (J.L.), and
partially supported by  the JINR-Polish Bogoliubov-Infeld program (all the authors).
The work of S.F. \&  E.I.
was partially supported by the RFBR Grant No.\,18-02-01046,  the DFG project LE 838/12-2,
Russian Science Foundation Grant No.\,16-12-10306
and Russian Ministry of Education and Science grant, project No.\,3.1386.201.
S.F. and E.I. would like to thank Institute for Theoretical Physics, Wroclaw University,
for the warm hospitality during this study.

\end{document}